\documentclass[
		twoside,openright,titlepage,numbers=noenddot,headinclude,%1headlines,
                footinclude=true,cleardoublepage=empty,
                BCOR=5mm,paper=a4,fontsize=10pt, % Binding correction, paper type and font size
                ngerman,american, % Languages
                ]{scrreprt} 
                
\PassOptionsToPackage{eulerchapternumbers,listings, pdfspacing, subfig,beramono,eulermath,parts}{classicthesis}
\newcommand{\myTitle}{Brain as a Complex System}
\newcommand{\mySubtitle}{harnessing systems neuroscience tools \& notions for an empirical approach}

\newcommand{\myName}{Shervin safavi\xspace}

\newcommand{\myInst}{Max Planck Institute for Biological Cybernetics\xspace}

\newcommand{\myInstDeptLong}{Department of Physiology of Cognitive Processes\xspace}

\newcommand{\ie}{i.\,e.\xspace}

\newcommand{\eg}{e.\,g.\xspace}

\newcommand{\seealso}{also can refer to the corresponding summary}
\newcommand{\seealsos}{also can refer to the corresponding summaries}

\makeatletter
\newcommand*{\rom}[1]{\textsf{\expandafter\@slowromancap #1@}}
\makeatother
\newcommand{\matmet}{``Material and Methods''}

\newcounter{dummy} % Necessary for correct hyperlinks (to index, bib, etc.)
\providecommand{\mLyX}{L\kern-.1667em\lower.25em\hbox{Y}\kern-.125emX\@}

\newcommand{\PRLsep}{\noindent\makebox[\linewidth]{\resizebox{0.3333\linewidth}{1pt}{$\bullet$}}\bigskip}

\usepackage{lipsum} % Used for inserting dummy 'Lorem ipsum' text into the template

\PassOptionsToPackage{latin9}{inputenc} % latin9 (ISO-8859-9) = latin1+"Euro sign"
\usepackage{inputenc}
 
\PassOptionsToPackage{ngerman,american}{babel}  % Change this to your language(s)
\usepackage{babel}

\PassOptionsToPackage{comma,square,numbers}{natbib}

\newcommand\citeAYt[1]{\citeauthor{#1} (\citeyear{#1})}

\usepackage[			% use biblatex for bibliography
	backend=biber,      % use biber or bibtex backend
        style=numeric,   % choose style
        citestyle=numeric-comp,
	natbib=true,		% allow natbib commands
	hyperref=true,	    % activate hyperref support
	alldates=year,      % only show year (not month)
        mincitenames=1,
        maxcitenames=1,
        backref=true,
        doi=true,
        url=false,
        isbn=false,
        eprint=false,
        sorting=none
        ]{biblatex}

\renewbibmacro{in:}{}

\IfFileExists{/home/ssafavi/Nextcloud/libraries/zotlib.bib}{\bibliography{/home/ssafavi/Nextcloud/libraries/zotlib,locallib,tmplib}}{\bibliography{locallib,tmplib}}

\PassOptionsToPackage{fleqn}{amsmath} % Math environments and more by the AMS 
 \usepackage{amsmath}
 
\PassOptionsToPackage{T1}{fontenc} % T2A for cyrillics
\usepackage{fontenc}

\usepackage{xspace} % To get the spacing after macros right

\usepackage{mparhack} % To get marginpar right

\usepackage{fixltx2e} % Fixes some LaTeX stuff 

\PassOptionsToPackage{smaller}{acronym} % Include printonlyused in the first bracket to only show acronyms used in the text
\usepackage{acronym} % nice macros for handling all acronyms in the thesis

\PassOptionsToPackage{pdftex}{graphicx}
\usepackage{graphicx} 

\usepackage{tabularx} % Better tables
\usepackage{caption}
\usepackage{subfig}  
\usepackage{epstopdf} % eps conversion

\usepackage{listings} 
\usepackage[dvipsnames]{xcolor}
\definecolor{grayForLinks}{gray}{0.40} %  %
\definecolor{webgreen}{rgb}{0,.5,0}

\PassOptionsToPackage{pdftex,hyperfootnotes=false,pdfpagelabels}{hyperref}
\usepackage{hyperref}  % backref linktocpage pagebackref
\hypersetup{
colorlinks=true, linktocpage=true, pdfstartpage=1, pdfstartview=Fit,
breaklinks=true, pdfpagemode=UseNone, pageanchor=true, pdfpagemode=UseOutlines,
plainpages=false, bookmarksnumbered, bookmarksopen=true, bookmarksopenlevel=1,
hypertexnames=true, pdfhighlight=/O, urlcolor=grayForLinks, linkcolor=RoyalBlue, citecolor=webgreen,
pdftitle={\myTitle},
pdfauthor={\textcopyright\ \myName, \myInst},
pdfsubject={},
pdfkeywords={},
pdfcreator={pdfLaTeX},
pdfproducer={LaTeX with hyperref and classicthesis}
}   

\usepackage{ifthen} % Allows the user of the \ifthenelse command
\makeatletter
\@ifpackageloaded{babel}
{
\addto\extrasamerican{

}
\addto\extrasngerman{

}
}{\relax}
\makeatother

\usepackage{classicthesis} 

\usepackage{comment}
\numberwithin{equation}{chapter}
\numberwithin{figure}{chapter}

\usepackage{pdfpages} % for adding PDF files

\usepackage{nameref}

\usepackage{csquotes}

\usepackage{chemmacros}
\chemsetup[phases]{pos=sub}

\usepackage{animate}             % for animations
\def \fmisc {./gfx/} % address for addiional figures
\usepackage{microtype}

\usepackage{bibentry}

\definecolor{pantone328}{cmyk}{1,0,0.57,0.30}

\colorlet{CTtitle}{pantone328}
\colorlet{CTurl}{grayForLinks}

\begin{document}

\frenchspacing % Reduces space after periods to make text more compact

\raggedbottom % Makes all pages the height of the text on that page

\selectlanguage{american} % Select your default language - e.g. american or ngerman

%\renewcommand*{\bibname}{new name} % Uncomment to change the name of the bibliography
%\setbibpreamble{} % Uncomment to include a preamble to the bibliography - some text before the reference list starts

\pagenumbering{roman} % Roman page numbering prior to the start of the thesis content (i, ii, iii, etc)

\pagestyle{plain} % Suppress headers for the pre-content pages

%----------------------------------------------------------------------------------------
%	PRE-CONTENT THESIS PAGES
%----------------------------------------------------------------------------------------
\begin{titlepage}
% \begin{center}
  \large
  % \begingroup
  \noindent {\LARGE{\color{pantone328}\spacedallcaps{\myTitle}}} \\ %\bigskip % Thesis title
  \emph{\Large \mySubtitle} \\ \medskip % Thesis subtitle

  % \spacedlowsmallcaps
  \noindent {\Large \myName}\\[5pt] % Your name
  \emph{\myInst} \\
  \emph{\myInstDeptLong}\\[1pt] %\bigskip

  \noindent
  \emph{IMPRS for Cognitive and Systems Neuroscience} \\
  \emph{School of Neural Information Processing }\\[20pt] %\bigskip

  \noindent
  For the full version of this PhD thesis please refer to the original reference \cite{safaviBrainComplexSystem2022}:\\[5pt]

  \fullcite{safaviBrainComplexSystem2022}
  % \endgroup
  % \end{center}
\end{titlepage}
\newpage
\begin{titlepage}
  \begin{addmargin}[-1cm]{-3cm}

    \begin{center}

      \vspace{9.5pt}

      \begingroup
      {\LARGE{\color{black}{\myTitle}}} \\ [5pt]%\bigskip % Thesis title
      \emph{\Large \mySubtitle} \\[4cm]%\medskip % Thesis subtitle
      \endgroup

      \vspace{7.5pt}

      \begin{minipage}{0.8\textwidth}

        \begin{center}

          \Large Dissertation\\

          \vspace{7.5pt}

          zur Erlangung des Grades eines\\

          Doktors der Naturwissenschaften\\

        \end{center}

      \end{minipage}
      \\[2cm]

      % Year

      {\Large der Mathematisch-Naturwissenschaftlichen Fakult\"at\\

        und\\

        der Medizinischen Fakult\"at\\[5pt]

        der Eberhard-Karls-Universit\"at T\"ubingen}
      \\[3cm]

      {\Large vorgelegt\\

        von\\[1cm]

        Shervin Safavi \\[5pt]

        aus Tehran, Iran}
      \\[1.6cm]

      % {\Large December - 2020}
      {\Large 2021}

    \end{center} 
  \end{addmargin}
\end{titlepage}

%%% Local Variables:
%%% mode: latex
%%% TeX-master: "../phdThesis_csb"
%%% End:
 % Main title - page GTC format

\cleardoublepage\vspace*{\fill}

\begin{table}[h!]
{\begin{tabular}{ll}

Tag der m\"undlichen Pr\"ufung: & 2021-10-20 \\
&  \\
&  \\
Dekan der Math.-Nat. Fakult\"at: & Prof. Dr. Thilo Stehle \\
Dekan der Medizinischen Fakult\"at: & Prof. Dr. Bernd Pichler  \\
&  \\
&  \\
1. Berichterstatter:  & Prof. Dr. Nikos K. Logothetis  \\
2. Berichterstatter:  & Prof. Dr. Anna Levina  \\
3. Berichterstatter:  & Prof. Dr. Sonja Gr\"un   \\
&  \\
   Pr\"ufungskommission:  & Prof. Dr. Nikos K. Logothetis  \\
  & Prof. Dr. Martin Giese \\
  & Prof. Dr. Anna Levina \\
  & Prof. Dr. Gustavo Deco \\

\end{tabular}}{}

\end{table}

%%% Local Variables:
%%% mode: latex
%%% TeX-master: "../phdThesis_csb"
%%% End:
 % Supervisors + 2nd Reader and their affilations

\cleardoublepage% Declaration

\refstepcounter{dummy}
\pdfbookmark[0]{Declaration}{declaration} % Bookmark name visible in a PDF viewer

\thispagestyle{empty}

\vspace*{\fill}

Erkl\"arung / Declaration:
Ich erkl\"are, dass ich die zur Promotion eingereichte Arbeit mit dem Titel:

\begin{center}
\textbf{"\myTitle, \mySubtitle"}
\end{center}

selbst\"andig verfasst, nur die angegebenen Quellen und Hilfsmittel benutzt und w\"ortlich oder inhaltlich \"ubernommene Stellen als solche gekennzeichnet habe. Ich versichere an Eides statt, dass diese Angaben wahr sind und dass ich nichts verschwiegen habe. Mir ist bekannt, dass die falsche Abgabe einer Versicherung an Eides statt mit Freiheitsstrafe bis zu drei Jahren oder mit Geldstrafe bestraft wird.

I hereby declare that I have produced the work entitled \textbf{\textit{"\myTitle, \mySubtitle"}}, submitted for the award of a doctorate, on my own (without external help), have used only the sources and aids indicated and have marked passages included from other works, whether verbatim or in content, as such. I swear upon oath that these statements are true and that I have not concealed anything. I am aware that making a false declaration under oath is punishable by a term of imprisonment of up to three years or by a fine.
\vspace{25pt}
\begin{table}[h!]

\centering

{\begin{tabular}{c c c}

T\"ubingen, den & ..................................................... & .....................................................\\

& Datum / Date	& Unterschrift / Signature \\

\end{tabular}}{}
\end{table}

%%% Local Variables:
%%% mode: latex
%%% TeX-master: "../phdThesis_csb"
%%% End:
 % Declaration

\thispagestyle{empty}

\hfill

\vfill

%\textcopyright\ \myTime
\noindent\myName:\\[2pt]
\textit{\myTitle}\\[-3.5pt]{\small\mySubtitle}\\[4pt]%\myDegree, 
\includegraphics[width=2cm]{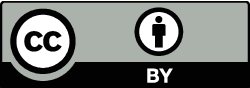}\\[-2.5pt]
{\footnotesize Content of this thesis is licensed under a \href{https://creativecommons.org/licenses/by/3.0/}{Creative Commons Attribution 3.0},
  except the scientific papers reprinted in the thesis (see part \ref{part:manuscripts}),
  that are subject to their own copyright protection.
}

%%% Local Variables:
%%% mode: latex
%%% TeX-master: "../phdThesis_csb"
%%% End:
 % Back of the title page

\cleardoublepage% Dedication

\thispagestyle{empty}
\refstepcounter{dummy}
\pdfbookmark[0]{Dedication}{Dedication} % Bookmark name visible in a PDF viewer

%%%%%%%%%%%%%%%%%%%%%%%%%%%%%%%%%%%%%%%%%%%%%%%%%%%%%%%%
% v2
\vspace*{3cm}

\begin{center}
  Dedicated to all  the nurses, doctors, clinicians and scientists \dots \\
  who sacrifice their lives
  to save ours
  during the COVID-19 pandemic.\\ \smallskip
\end{center}

\PRLsep

\begin{center}
  Dedicated to loving Farhad Meysami \\
  who has an important contribution in shaping my mindset.
\end{center}

\vfill

%%% Local Variables:
%%% mode: latex
%%% TeX-master: "../phdThesis_csb"
%%% End:
 % Dedication page
\cleardoublepage\thispagestyle{empty}
\refstepcounter{dummy}

\vfill

\begin{figure}
  \centering
    \includegraphics[width = .5\linewidth]{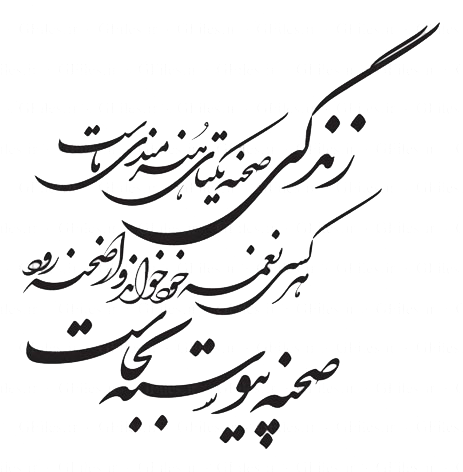}
\end{figure}

\vfill

% potential source:
% http://www.gfiles.ir/2273/%D9%88%DA%A9%D8%AA%D9%88%D8%B1-%D8%B4%DA%A9%D8%B3%D8%AA%D9%87-%D9%86%D8%B3%D8%AA%D8%B9%D9%84%DB%8C%D9%82-%D8%B4%D8%B9%D8%B1-%D8%B2%D9%86%D8%AF%DA%AF%DB%8C-%D8%B5%D8%AD%D9%86%D9%87-%DB%8C%DA%A9%D8%AA%D8%A7%DB%8C-%D9%87%D9%86%D8%B1%D9%85%D9%86%D8%AF%DB%8C

%%% Local Variables:
%%% mode: latex
%%% TeX-master: "../phdThesis_csb"
%%% End:
 % The poem 
\pagestyle{empty}
\refstepcounter{dummy}

\hfill

\vfill

\noindent{\footnotesize{\textit{Translation:}}}\\[5pt]
Life is our unique stage of performance!\\
Everyone sing their own song and leave ...\\
Stage remains ... \\
Remembered songs are the delighted ones.\\[5pt]
\begin{flushright}
  \noindent {--- Poem by Zhale Esfehani}\\
  \noindent \footnotesize{\textit{(\footnotesize{Subjectively translated by Shervin Safavi})}}
\end{flushright}

\vfill

%%% TeX-master: "../phdThesis_csb"
%%% End:
                 % credit of the poem

\pagestyle{scrheadings} % Show chapter titles as \cleardoublepage\include{FrontBackMatter/Contents} % Contents, list of figures/tables/listings and acronyms
% \include{Chapters/Chapter00} % Preface

% \cleardoublepage\include{FrontBackMatter/Acknowledgments} % Acknowledgements page

\cleardoublepage% Table of Contents - List of Tables/Figures/Listings and Acronyms

\refstepcounter{dummy}

\pdfbookmark[0]{\contentsname}{tableofcontents} % Bookmark name visible in a PDF viewer

\setcounter{tocdepth}{2} % Depth of sections to include in the table of contents - currently up to subsections

\setcounter{secnumdepth}{3} % Depth of sections to number in the text itself - currently up to subsubsections

\manualmark
\markboth{\spacedlowsmallcaps{\contentsname}}{\spacedlowsmallcaps{\contentsname}}
\tableofcontents 
\automark[section]{chapter}
\renewcommand{\chaptermark}[1]{\markboth{\spacedlowsmallcaps{#1}}{\spacedlowsmallcaps{#1}}}
\renewcommand{\sectionmark}[1]{\markright{\thesection\enspace\spacedlowsmallcaps{#1}}}

\clearpage

\begingroup 
\let\clearpage\relax
\let\cleardoublepage\relax
\let\cleardoublepage\relax

\endgroup % Contents, list of figures/tables/listings and acronyms
\cleardoublepage% Preface
\begingroup
\let\clearpage\relax
\let\cleardoublepage\relax
\let\cleardoublepage\relax

\chapter*{Preface} % Abstract name
\addtocontents{toc}{\protect\vspace{\beforebibskip}} % Place the bibliography slightly below the rest of the document content in the table of contents
\addcontentsline{toc}{chapter}{\tocEntry{Preface}}

Finding general principles underlying brain function has been appealing to scientists.
Indeed, in some branches of science like physics and chemistry (and to some degree biology) a general theory often can capture the essence of a wide range of phenomena.
Whether we can find such principles in neuroscience, and [assuming they do exist]
what those principles are, are important questions.
Abstracting the brain as a complex system is one of the perspectives that
may help us answer this question.

While it is commonly accepted that the brain is a (or even \emph{the}) prominent example of a complex system, 
the far reaching implications of this fact are still arguably overlooked in our approaches to neuroscientific questions.
One of the reasons for the lack of attention could be the apparent difference in foci of investigations in these two fields --- neuroscience and complex systems.
This thesis is an effort toward providing a bridge between systems neuroscience and complex systems
by harnessing systems neuroscience tools \& notions for building empirical approaches toward the brain as a complex system.

Perhaps, in the spirit of \emph{searching for principles},
we should abstract and approach the brain as a complex \emph{adaptive} system as the more complete perspective (rather than just a complex system).
In the end, the brain, even the most ``complex system'', need to survive in the environment.
Indeed, in the field of \emph{complex adaptive systems}, the intention is understanding very similar questions in nature.
As an outlook, we also touch on some research directions pertaining to the adaptivity of the brain as well.

\endgroup			

\vfill

%%% Local Variables:
%%% mode: latex
%%% TeX-master: "../phdThesis_csb"
%%% End:
 % Acknowledgements page
\cleardoublepage\begingroup
\let\clearpage\relax
\let\cleardoublepage\relax
\let\cleardoublepage\relax

\chapter*{Summary} % Abstract name
\addtocontents{toc}{\protect\vspace{\beforebibskip}} % Place the bibliography slightly below the rest of the document content in the table of contents
\addcontentsline{toc}{chapter}{\tocEntry{Summary}}

The brain can be conceived as a complex system,
as it is made up of nested networks of interactions
and moreover,
demonstrates emergent-like behaviors such as oscillations.
Based on this conceptualization,
various tools and frameworks that stem from the field of complex systems
have been adapted to answer neuroscientific questions.
Certainly, using such tools for neuroscientific questions has been insightful for understanding the brain as a complex system.
Nevertheless,
they encounter limitations when they are adapted for the purpose of understanding the brain,
or perhaps better should be stated that,
developing approaches which are closer to the neuroscience side can also be instrumental for approaching the brain as a complex system.

\marginpar{\autoref{cha:brain-as-complex}}

In this thesis, after an elaboration on the motivation of this endeavor in \autoref{cha:brain-as-complex},
we introduce a set of complementary approaches,
with the rationale of exploiting the development in the field of systems neuroscience in order to
be close to the neuroscience side of the problem,
but also still remain connected to the complex systems perspective.
Such complementary approaches can be envisioned through different apertures.
In this thesis, we introduce our complementary approaches,
through the following apertures:
neural data analysis (\autoref{cha:appr-thro-nda}),
neural theories (\autoref{cha:appr-thro-theo}), and
cognition (\autoref{cha:appr-thro-behav}).

In \autoref{cha:appr-thro-nda}, we argue that multi-scale and cross-scale analysis of neural data is one of the important aspects of the neural data analysis from the complex systems perspective toward the brain.
Furthermore, we also elaborate that,
investigating the brain across scales, is not only important from the abstract perspective of complex systems,
but also motivating based on a variety of empirical evidence on coupling between brain activity at different scales, neural coordination and theoretical speculations on neural computation.
\marginpar{\autoref{cha:appr-thro-nda} \\
  \autoref{cha:paper-5}/\nameref{cha:paper-5} \\
  \autoref{cha:paper-gpla}/\nameref{cha:paper-gpla}  \\
  \autoref{cha:paper-besserve2020ned}/\nameref{cha:paper-besserve2020ned}
}
Based on this motivation we first very briefly discuss some of the relevant cross-scale neural data 
analysis methodologies and then introduce two novel methodologies that have been developed as parts of this thesis
(\nameref{cha:paper-5}, \nameref{cha:paper-gpla}, and \nameref{cha:paper-besserve2020ned}).
In \nameref{cha:paper-5} and \nameref{cha:paper-gpla} we introduced a multi-variate methodology for investigating spike-LFP relationship and in \nameref{cha:paper-besserve2020ned} we introduced a methdology for detecting cooperative neural activities (neural events) in local field potentials,
that can be used as a trigger to investigate simultanious activity in larger and smaller scales.
A prominent example of these neural events are sharp wave-ripples that has been shown to co-occur with precise coordination in the spiking activity of individual neurons and the large-scale brain activity as well.

In \autoref{cha:appr-thro-theo}, we introduce a new aperture through neural theories.
One way of approaching the brain as a complex system is seeking for connections between theoretical frameworks that stem from the field of complex systems and the ones established in neuroscience.
On the complex systems side, we consider the \emph{criticality hypothesis of the brain} that has strong roots in the field of complex systems, and on the neuroscience side, we consider the \emph{efficient coding} which is one of the most important theoretical frameworks in systems neuroscience.
We first briefly introduce the background on efficient coding and criticality,
and elaborate further on the motivation behind our integrative approach.
In \nameref{cha:paper-safavi2020cribay}, we present our interim results,
\marginpar{\autoref{cha:appr-thro-theo} \\
  \autoref{cha:paper-safavi2020cribay}/\nameref{cha:paper-safavi2020cribay}
}
which suggests the two influential, and previously disparate fields -- efficient coding, and criticality -- might be intimately related.
We observed that, in the vicinity of the parameters that leads to optimized performance of a network implementing neural coding,
the distribution of avalanche sizes follow a power-law distribution.
In \nameref{cha:paper-safavi2020cribay} we also provide an extensive discussion on the implication of our interim results and its future extensions.
Moreover, in \nameref{cha:paper-safavi2020cribay} we also introduce another perspective which motivates such investigations,
namely seeking for potential bridges between \emph{neural computation} and \emph{neural dynamics}.

In \autoref{cha:appr-thro-behav}, we argue that binocular rivalry,
as a key phenomenon to investigate consciousness,
is particularly relevant for a complex systems perspective toward the brain.
Based on this insight,
we suggest and conduct novel experimental work,
namely, studying this phenomenon at a mesoscopic scale, that has not been done before.
Surprisingly, in the last 30 years, almost all the previous studies on binocular rivalry were either focused on micro-scale (level of an individual neuron) or the macro-scale (level of the whole brain).
Therefore, our work in this domain not only is valuable from the perspective of complex systems,
but also for understanding the neural correlate of visual awareness \emph{per se}.
In \nameref{cha:paper-safavi2014}, \nameref{cha:paper-safavi2018}, \nameref{cha:paper-kapoor2020}, and \nameref{cha:paper-dwarakanath2020} we elaborate on the outcome of this investigation.
\marginpar{\autoref{cha:appr-thro-behav} \\
  \autoref{cha:paper-safavi2014}/\nameref{cha:paper-safavi2014} \\
  \autoref{cha:paper-safavi2018}/\nameref{cha:paper-safavi2018}  \\
  \autoref{cha:paper-kapoor2020}/\nameref{cha:paper-kapoor2020}  \\
  \autoref{cha:paper-dwarakanath2020}/\nameref{cha:paper-dwarakanath2020}
}
\nameref{cha:paper-safavi2014} and \nameref{cha:paper-safavi2018} were prerequisite for the binocular rivalry experiments.
In \nameref{cha:paper-safavi2014} we elaborate on the importance of studying prefrontal cortex (PFC)
(which was the region of interest in our investigation)
for understating the neural correlate of visual awareness.
In \nameref{cha:paper-safavi2018} we investigate
the basic aspects of neural responses (tuning curves and noise correlations) of PFC units to simple visual stimulation
(in a similar setting used for our binocular rivalry experiments).
In \nameref{cha:paper-kapoor2020} and \nameref{cha:paper-dwarakanath2020} we investigate the neural correlate of visual awareness at a mesocopic scale
(which is motivating from the complex system perspective toward the brain).
We show that content of visual awareness is decodable from the population activity of PFC neurons (\nameref{cha:paper-kapoor2020})
and show oscillatory dynamics of PFC (as a reflection of collective neural activity) can be a relevant signature for perceptual switches (\nameref{cha:paper-dwarakanath2020}).
I believe that this is just the very first step toward establishing a connection from a complex systems perspective to cognition and behavior.
Various theoretical and experimental steps need to be taken in the future studies to build a solid bridge between cognition and complex systems perspective toward the brain.

The last chapter, \autoref{cha:brain-as-complex-adaptive}, is dedicated to an outlook, a subjective perspective on how this research line can be proceeded.
In the spirit of this thesis which is \emph{searching for principles},
I believe we are missing an important aspect of the brain which is its \emph{adaptivity}.
At the end, brain, even the most ``complex system'', needs to survive in the environment.
Indeed, in the field of \emph{complex adaptive systems}, the intention is understanding very similar 
\marginpar{\autoref{cha:brain-as-complex-adaptive}}
questions in the nature.
Inspired by ideas discussed in the field of complex adaptive systems,
I introduce a set of new research directions which intend to  incorporate the adaptivity aspect of the brain as one of the principles. 
These research directions also remain close to the neuroscience side, similar to the intention of the research presented in this thesis.

\endgroup			

\vfill

%%% Local Variables:
%%% mode: latex
%%% TeX-master: "../phdThesis_csb"
%%% End:

 % Acknowledgements page
\cleardoublepage

\pagenumbering{arabic} % Arabic page numbering for thesis content (2, 2, 3, etc)

\cleardoublepage % Avoids problems with pdfbookmark

%----------------------------------------------------------------------------------------
%	THESIS CONTENT - CHAPTERS
%----------------------------------------------------------------------------------------

% ------------------------------------------------
% Synopsis

\ctparttext
{
  This part provides a general idea of this thesis.
  % We suggest an important approach toward understanding the brain,
  We suggest an important approach that should be taken toward understanding the brain,
  could be borrowed or inspired from the field of \emph{complex systems}.
  In light of this perspective, new questions can be asked in various domains 
  and moreover, old questions can be revisited based on this perspective. 
  Contents of this thesis, pertain to three different domains, namely
  \textit{methods for  neural data analysis}, \textit{neural theories}, and \textit{cognition}.
  % With this  motivation,
% 1
  In the first domain, 
  we introduce novel statistical methods for multi-scale investigation of neural data
  that we believe should be an important piece in our analysis methods for understanding the brain as a complex system.
% 2
  In the second domain, we first briefly introduce
  \textit{criticality hypothesis of the brain},
  % that has been borrowed from the field of complex systems
  that has been primarily developed based on statistical physics 
  and has been suggested to explain the complex dynamics of the brain activity in different spatial and temporal scales.
  Then we introduce our complementary approach of investigation in this framework,
  and our finding regarding the hypotheses.
  % investigate whether the brain is operating close to a critical is state or not.
  % Then we introduce its potential connection to \textit{efficient coding} 
  % which in contrast to criticality hypothesis of the brain 
  % is a functionally relevant 
  In the third domain,
  we first describe the importance of investigating bistable perception phenomenon from the perspective of complex systems.
  Then we discuss our finding pertaining the mesoscopic neural mechanism underlying this phenomenon.
  % introduce the question we asked and their answers.
  % With this  motivation, we introduce number of questions in three different domains, 
  % \textit{methods for analyis of neural data}, \textit{neural dynamics}, and \textit{behavior and cognition}.
  % We explain our three 
  % apporaches toward understaning the brain 
  % from the perspective of approaching the brain as a complex system.
} % Text on the Part 1 page describing the content in Part 2

%%% Local Variables:
%%% mode: latex
%%% TeX-master: "../phdThesis_csb"
%%% End:

\part{Synopsis} % First part of the thesis
\label{part:synopsis}
\chapter{Brain as a complex system}\label{cha:brain-as-complex}
\section{Complex systems}\label{sec:complex-systems}
Behavior, or better stated \emph{collective} behavior, of wide range of system spanning the scales of movement of atoms to behavior of humans/animals can be studied under an inclusive young framework of sdudying \emph{complex systems}
\cite{bar-yamDynamicsComplexSystems2003,mitchellComplexityGuidedTour2011,hollandComplexityVeryShort2014,bar-yamWhyComplexityDifferent2017}.
\citet[Chapter 1]{mitchellComplexityGuidedTour2011} introduces and defines a complex system as following:
\begin{displayquote}\textsl{
    ``Systems in which organized behavior arises without an internal or external controller
    or leader are sometimes called self-organizing.
    Since simple rules produce complex behavior in hard-to-predict ways,
    the macroscopic behavior of such systems is sometimes called emergent.
    Here is an alternative definition of a complex system:
    a system that exhibits nontrivial emergent and self-organizing behaviors.''
  }
\end{displayquote}

One of the characteristic properties of complex systems are their
emergent properties, or/and their coordinated dynamics.
Interactions between units of the system play a crucial role in the creating its emergent properties.
These two aspects (emergent properties and the underlying interactions) of complex systems is central for the development of the ideas presented in this thesis (also see \autoref{cha:brain-as-complex-adaptive} for the complementary ideas).

To provide an intuition for emergent properties in complex systems and how interaction lead to such emergent properties,
we exploit synchronization phenomena in a system made up of coupled oscillators.
Assume we have $N$ oscillators (indexed by $i$), each oscillates with frequency $\omega_i$,
where oscillation frequencies are drawn from a normal distribution with mean $\overline{\omega}$ and standard deviation $\beta$,
\[
\omega_i \sim \mathcal{N}\left(\overline{\omega}, \beta \right)\,.
\]

In absence of interactions between oscillators, the dynamics of each oscillator
(which is defined based on its phase, $\theta_i$) is governed only by its  oscillation frequency,
\begin{equation}
  \label{eq:kuramotoIndependent}
  \theta'_j = \omega_j\,.
\end{equation}
Whereas, in  presence of interactions between oscillators,
they are allowed to exert forces on each other and therefore the dynamics of each oscillator also depends on the dynamics of other oscillators.
These interactions are incorporated as an interaction term in the differential equation governing the dynamics of each oscillator (second term in \autoref{eq:kuramotoIteracting})
\footnote{
  The particular choice of interaction terms is made to ease the analytical treatment and for purpose of demonstration
(see \cite{kuramotoSelfentrainmentPopulationCoupled1975,kuramotoChemicalOscillationsWaves2003} for more elaborate discussion).
}:
\begin{equation}
  \label{eq:kuramotoIteracting}
  \theta'_j = \omega_j + \kappa \frac{1}{N} \sum_i^N sin(\theta_i - \theta_j)\,,
\end{equation}
where $\kappa$ indicates the strength of these interactions.

The dynamics of system of oscillators described above is illustrated in \autoref{fig:kuramotoDemo_video} (video) and \autoref{fig:kuramotoDemo_snapshors} (snapshots).
Each dot represents an oscillator and colors code for oscillator's intrinsic frequency.
The oscillatory dynamics of the oscillators are represented by the circular motion of the dots.
In the absence of interactions, as is evident in \autoref{eq:kuramotoIndependent},
each oscillator, oscillates independently of the rest of the oscillators
(\autoref{fig:kuramotoDemo_video} and \autoref{fig:kuramotoDemo_snapshors} left).
Nevertheless, in the presence of interactions and if the parameters of the system
are appropriately chosen (in particular,  $\kappa$, to be non-zero),
the oscillators start synchronizing after a certain period
(see \autoref{fig:kuramotoDemo_snapshors} second row, and compare simulations with and without coupling) and ultimatley all oscillators synchronize
(see \autoref{fig:kuramotoDemo_snapshors} third row, and compare simulations with and without coupling).

\begin{figure}
  \centering
%   commented out to be compiled in overleaf
  \animategraphics[autoplay, trim = 3.3cm 1.5cm 2.6cm 0, width=.496\linewidth]{1}
  {\fmisc kuramotoModelSyncDemo/random/kuramotoModel_fNo}{1}{50}
  \animategraphics[autoplay, trim = 3.3cm 1.5cm 2.6cm 0, width=.496\linewidth]{1}
  {\fmisc kuramotoModelSyncDemo/collective/kuramotoModel_fNo}{1}{50}
  % end of comment
  \caption{\textbf{Kuramoto model} (animation, need Okular or Adobe Acrobat Reader)\\ 
    These animation demonstrate the dynamic of Kuramoto model consisting of 100 oscillators.
    Each dot represent an oscillator and the colors code for oscillator's intrinsic frequency.
    On the left, the oscillators do not interact with each other as the coupling parameter is set to zero ($\kappa = 0$).
    On the righ, the oscillators do interact with each other as the coupling parameter is non-zero ($\kappa = 0.5$). 
  }
  \label{fig:kuramotoDemo_video}
\end{figure}

\begin{figure}
  \centering
  \includegraphics[trim = 3.3cm 1.5cm 2.6cm 0, clip, width=.496\linewidth]
  {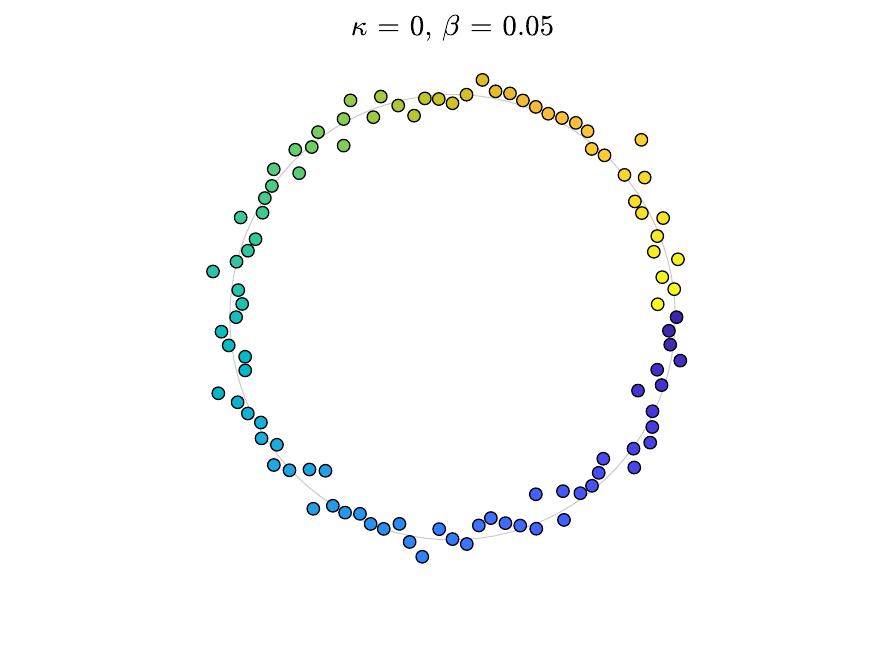}
  \includegraphics[trim = 3.3cm 1.5cm 2.6cm 0, clip, width=.496\linewidth]
  {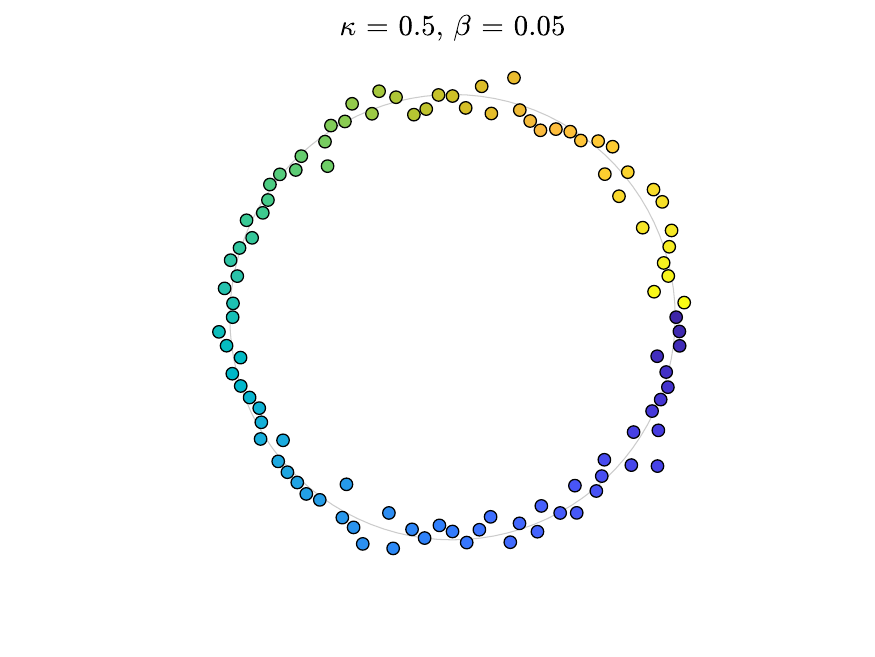}\\
  \includegraphics[trim = 3.3cm 1.5cm 2.6cm 0.7cm, clip, width=.496\linewidth]
  {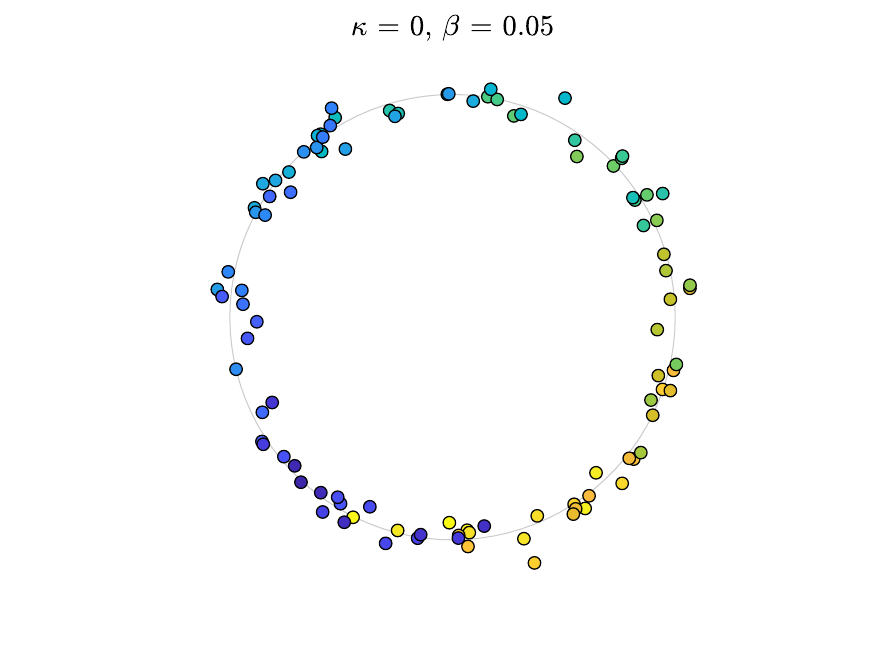}
  \includegraphics[trim = 3.3cm 1.5cm 2.6cm 0.7cm, clip, width=.496\linewidth]
  {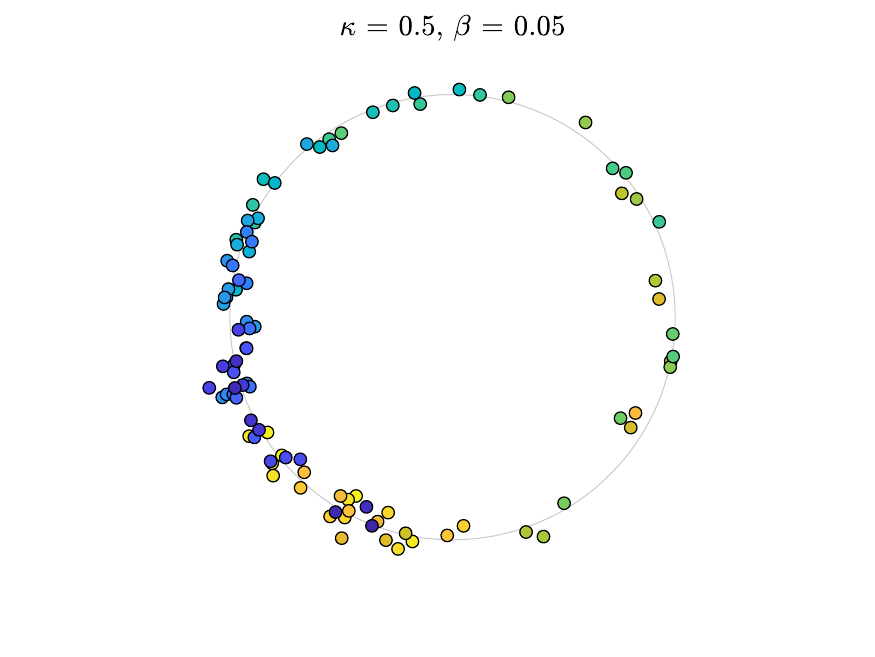}\\
  \includegraphics[trim = 3.3cm 1.5cm 2.6cm 0.7cm, clip, width=.496\linewidth]
  {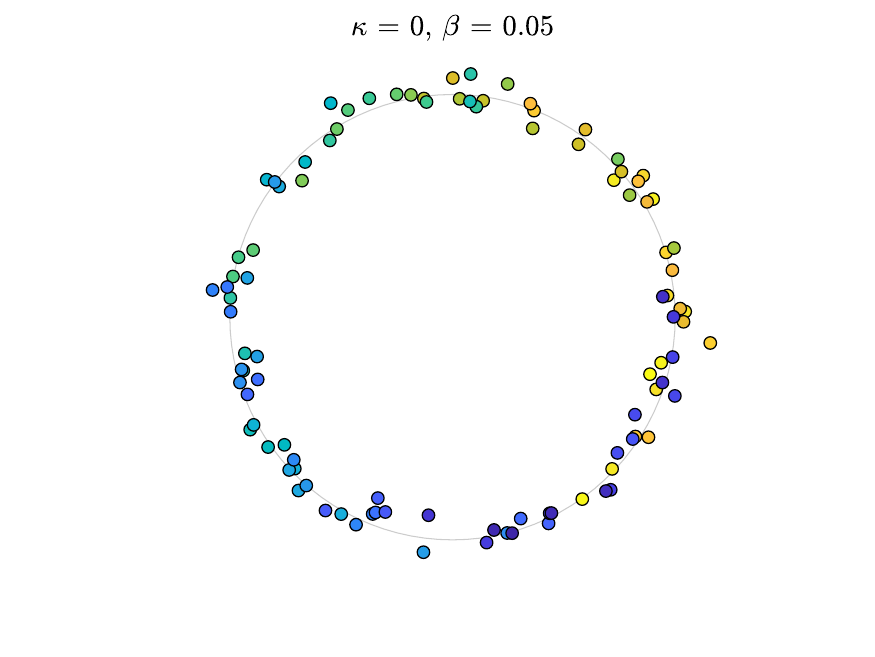}
  \includegraphics[trim = 3.3cm 1.5cm 2.6cm 0.7cm, clip, width=.496\linewidth]
  {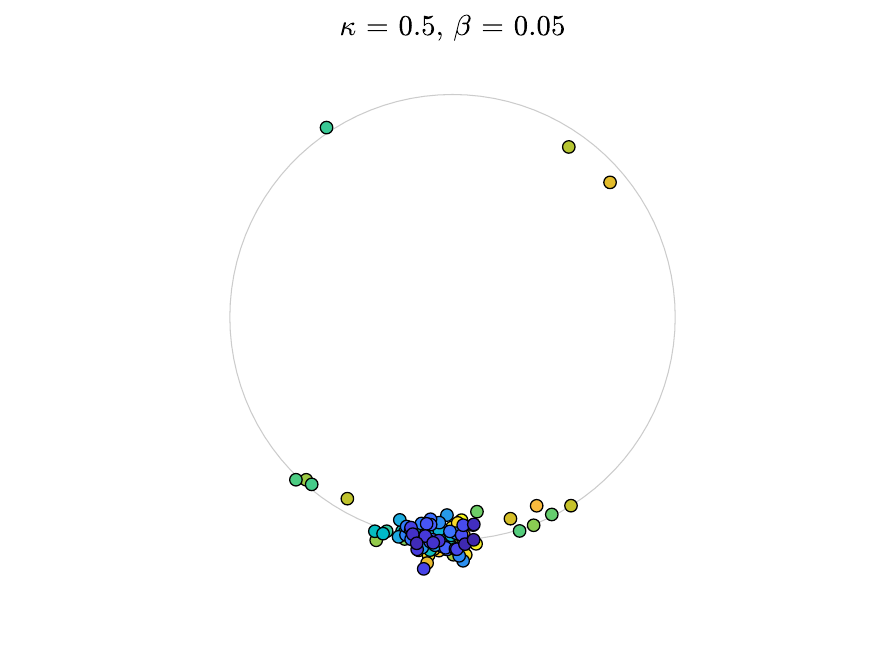}
  \caption{\textbf{Kuramoto model} (snapshots)\\ 
    Snapshots from animations of \autoref{fig:kuramotoDemo_video}.
    These snapshots (each row, one snapshot) demonstrate the dynamic of Kuramoto model consisting of 100 oscillators.
    Each dot represent an oscillator and the colors code for oscillator's intrinsic frequency.
    On the left, the oscillators do not interact with each other as the coupling parameter is set to zero ($\kappa = 0$).
    On the righ, the oscillators do interact with each other as the coupling parameter is non-zero ($\kappa = 0.5$).
    The first row is a snapshot from the initial condition of the simulation,
    the second row is a snapshot from an intermediate state of the simulation,
    and the last row is the last snapshot of this simulation.
    }
  \label{fig:kuramotoDemo_snapshors}
\end{figure}

Synchronization is not a genuine property of the individual units and there is no central coordinator in the system.
However, oscillators tend to synchronize their activity due to the presence of interactions between the units.
In this example, synchronization is considered an \emph{emergent} property of the system.

The brain can also be conceived as a complex system,
as it is made up of \emph{nested networks of interactions}
and demonstrates emergent-like behaviors such as oscillation.
Different constructing units or building blocks of the brain (from molecules to networks) interact with each other \cite[Chapter 1]{churchlandComputationalBrain1992}.
Indeed, this perspective toward the brain has been extensively articulated
\cite{siegelmannComplexSystemsScience2010,wernerConsciousnessViewedFramework2013,spornsConnectivityComplexityRelationship2000,singerBrainComplexSelforganizing2009,olbrichSleepingBrainComplex2011b,kochSystemsBiologyModular2012,bullmoreComplexBrainNetworks2009a,lynnHumanInformationProcessing2020,betzelMultiscaleBrainNetworks2017,bassettUnderstandingComplexityHuman2011,mitchellComplexityGuidedTour2011,buzsakiRhythmsBrain2011,chialvoEmergentComplexNeural2010c}.

\section{Complex system tools  in neuroscience}\label{sec:tools-used-neur}

Inspired by perspective introduced in the previous section,
various frameworks that stem from the field of complex systems
has been adapted to answer neuroscientific question.
Furthermore, various tools that have been developed for studying complex systems have also been customized to be applied to neural data.

The tools and frameworks adapted from the field of complex systems to address neuroscientific questions can be divided into four categories
(of courses, a subjective categorization):
1- Network science 2- Non-linear dynamics  3- Information theory and 4- Statistical physics.

\begin{description}
\item[Network science:]
  Network science is perhaps the most adapted tool from the filed of complex systems to be used in neuroscience.
  To use tools developed in network theory, 
  we abstract the object of interest as graphs,
  this includes defining the nodes and edges of the graph.
  Brain can also be abstracted as a graph in various levels of organization,
  from genes to behavior \cite{borsboomSmallWorldPsychopathology2011,bullmoreComplexBrainNetworks2009a,vandenheuvelSpotlightBridgingMicroscale2017,scholtensMultimodalConnectomicsPsychiatry2018,vandenheuvelMultiscaleNeurosciencePsychiatric2019,heuvelCrossdisorderConnectomeLandscape2019}.
\item[Non-linear dynamics:]
  Theory of dynamical systems has a broad application in neuroscience.
  The core idea is conceptualizing or modeling the dynamics of the brain at different scales as a [non-linear] dynamical system \cite{mckennaBrainDynamicPhysical1994,beerDynamicalSystemsPerspective1995,rabinovichDynamicalPrinciplesNeuroscience2006}.
  There have been various attempts to model single neuron \cite{izhikevichDynamicalSystemsNeuroscience2010}, neuronal populations \cite[Part 3]{gerstnerNeuronalDynamicsSingle2014}\cite{decoDynamicBrainSpiking2008}, large-scale brain networks \cite{izhikevichLargescaleModelMammalian2008,decoEmergingConceptsDynamical2011} and even brain-environment system as dynamical systems \cite{beerDynamicalSystemsPerspective1995}.
\item[Information theory:]
  Information-theoretic tools have been extensively used in neuroscience,
  for purposes, as simple as studying neural coding in a single neuron
  \cite{bialekReadingNeuralCode1991,steveninckReproducibilityVariabilityNeural1997,strongEntropyInformationNeural1998,borstInformationTheoryNeural1999,fredriekeSpikesExploringNeural1999}
  all the way to quantifying the level of consciousness
  \cite{tononiInformationIntegrationTheory2004,balduzziIntegratedInformationDiscrete2008b,oizumiPhenomenologyMechanismsConsciousness2014}
  and providing a mathematical framework to represent the content of the conscious experience \cite{balduzziQualiaGeometryIntegrated2009b}
  (for a review see \citet{tononiIntegratedInformationTheory2016}).
\item[Statistical physics:]
  Statistical physics is a branch of physics which seeks for simple behaviors in systems consisting of many interacting components \cite{sethnaStatisticalMechanicsEntropy2006}.
  Such systems can be atoms of water in a glass \cite{sethnaStatisticalMechanicsEntropy2006},
  all the way to collective activity of a flock of birds
  \cite{bialekStatisticalMechanicsNatural2012,bialekSocialInteractionsDominate2014}
  and pattern of tweets in the Twitter network \cite{hallStatisticalMechanicsTwitter2019}.
  One of the phenomena that has been central in statistical physics (and other fields as well),
  is criticality.
  which has also inspired theoretical frameworks in neuroscience \cite{munozColloquiumCriticalityDynamical2018} (will be briefly discussed in \autoref{cha:appr-thro-theo}).
\end{description}

\section{Novel complementary approaches}\label{sec:toward-neur-insp}

Certainly, using the approaches mentioned in the previous section (\autoref{sec:tools-used-neur})
has been tremendously insightful for understanding the brain as a complex system.
This is an important achievement, given their principled and foundational nature.
Nevertheless, they might also have some limitations when they are adapted for understanding the brain.
For instance,  information-theoretic measures are often difficult to apply to neural data in general settings due to the need for large amounts of data
\marginpar{Goal of the thesis}
(but also see innovative approaches such as \cite{zbiliQuickEasyWay2020}).
Such caveats become even more critical for functionally relevant information-theoretic measures such as integrated information \cite{oizumiPhenomenologyMechanismsConsciousness2014}.
Computing or estimating the amount of integrated information in a system for more than a handful of units is challenging \cite{tononiIntegratedInformationTheory2016}.
There are other kinds of limitation for the mentioned approaches,
but since the purpose of this thesis is introducing \emph{complementary} (not alternative) approaches 
I would rather focus on these complementary approaches and the motivation behind them.
In these complementary approaches, the goal is exploiting the development in the field of systems neuroscience to be close to the neuroscience side but still
remain related to the complex system perspective.

There are multiple examples in systems neuroscience,
in which a given function is attributed to a \emph{coordinated} activity of a group of neurons or neural units 
\eg a brain circuit or an area.
Just to name a few, we can mention population coding \cite{sangerNeuralPopulationCodes2003,shamirEmergingPrinciplesPopulation2014}, communication through coherence \cite{friesMechanismCognitiveDynamics2005,friesRhythmsCognitionCommunication2015},
and memory consolidation \cite[Chapter 7]{malsburgMalsburgDynamicCoordination2010}.
In these examples, the target function is implemented through the precise coordination of units;
In population coding, by the interaction between neurons; 
in communication through coherence through oscillatory interaction through neural populations; 
And in memory consolidation through interaction between multiple regions of hippocampal formation and neocortex.

Interestingly, some of the tools and notions that system neuroscientists used to understand the coordinated phenomenon can be closely related to perspectives inspired by 
or related to the field of complex systems.
For instance, various studies have investigated cross-scale relationships in neural activities such as
relationship between spikes and local field potentials (LFP) 
\cite{mitraObservedBrainDynamics2007} for understanding the mechanism involved in communication through coherence,
or considering simultaneously two successive scales such as neural event triggered fMRI (NET-fMRI)
studies to understand the memory consolidation mechanisms
\cite{logothetisIntracorticalRecordingsFMRI2012,logothetisHippocampalCorticalInteraction2012,ramirez-villegasDiversitySharpwaverippleLFP2015}.

In \autoref{cha:appr-thro-nda},
we introduce a set of methodologies for cross-scale and multi-scale analysis of neural data.
Developing these tools is motivated by a perspective that results from approaching the brain as a complex system. %toward the brain.
Every system, in particular, complex systems can be described at different scales.
Some systems (\eg our solar system) can be described, to a large degree, in \emph{isolated scales} and their behavior upon interacting with other systems can be predicted.
However, many systems wherein we are interested to understand are not well described in isolated scales.
To illustrate this important notion, we use a few intuitive examples adopted from
\citeAYt{bar-yamWhyComplexityDifferent2017}.
If we are interested in explaining the dynamics of the earth 
(orbits of the earth in the solar system),
and how it will change when a new planet is added to the solar system,
we do not need to know the details of processes happening inside the earth.
\marginpar{Approaching through neural data analysis}
Therefore, for this system, we can \emph{separate scales} without losing our descriptive and predictive power (to a large degree).
But if we are interested in the collective dynamics of a flock of birds \cite{bialekStatisticalMechanicsNatural2012},
we neither can focus on the micro-scale (motion of an individual bird) as it is too fine-grained,
nor the macro scale (average motion of the flock) as it is not sufficient to describe and predict the collective behaviour of the birds.
Generally speaking, understanding the complex behavior which is not completely independent (random) nor it is completely coherent requires investigation across scales \cite{bar-yamWhyComplexityDifferent2017}.
In \autoref{cha:appr-thro-nda}, we further elaborate on the motivation and necessity of investigating the brain by simultaneously considering  two successive scales and introduce our novel methodologies motivated by this mindset.

As mentioned earlier, the goal is establishing a bridge between systems neuroscience and a complex system perspective toward the brain.
In an effort toward achieving this goal,
in addition to developing analysis methods and generalizing the existing ones,
we also propose two other apertures in \autoref{cha:appr-thro-theo} and \autoref{cha:appr-thro-behav}.
Of course these new apertures also provide us new angles to build the bridge.

In \autoref{cha:appr-thro-theo} we provide a potential link between one of the most important theoretical frameworks in system neuroscience, \emph{efficient coding}, 
and one of the most important theoretical framework in the field of complex systems, \emph{criticality}. 
Efficient coding has different variants and many of them have been extensively investigated both experimentally and theoretically in systems neuroscience.
On the other hand, the theory of critical phase transition in complex systems 
has been successful in explaining many phenomena in nature \cite{mathisEmergenceLifeFirstOrder2017a,chialvoLifeEdgeComplexity2018}, 
and ``criticality hypothesis of the brain'' \cite{munozColloquiumCriticalityDynamical2018},
has been developed based on this solid foundation.
In nutshell, criticality hypothesis of the brain state that, the brain operates close to a critical state.
Being close to this state is beneficial for such an organ
\marginpar{Approaching through neural theories}
\cite{munozColloquiumCriticalityDynamical2018,tkacikInformationProcessingLiving2016,moraAreBiologicalSystems2011a},
as it has been shown that general information processing capabilities such as
sensitivity to input \cite{kinouchiOptimalDynamicalRange2006,brochiniPhaseTransitionsSelforganized2016}, 
dynamic range
\cite{kinouchiOptimalDynamicalRange2006,larremorePredictingCriticalityDynamic2011,nurProbingSpatialInhomogeneity2019},
and information transmission and storage
\cite{shewInformationCapacityTransmission2011,vanniCriticalityTransmissionInformation2011,vanniCriticalityTransmissionInformation2011,lukovicTransmissionInformationCriticality2014,marinazzoInformationTransferCriticality2014},
and various other computational characteristics are optimized in this state.
Certainly, being in a state with such optimized capabilities are relevant for the computations in the brain, 
but they are too abstract to provide a concrete explanation of the computations in the brain.
For instance, all the capabilities mentioned above are relevant for coding sensory information which is a relevant function for the brain and has been studied in systems neuroscience extensively,
however mere adjustment for being close to criticality cannot provide a neural implementation for the coding given resource constraints.
In \autoref{sec:efficient-coding-as} we provide more detail on both frameworks,
efficient coding and criticality hypothesis of the brain,
and provide evidence on the connection between them.

In \autoref{cha:appr-thro-behav}, we introduce another aperture for establishing the mentioned connection.
Perhaps, one of the most important goals of neuroscience is understating the machinery behind the cognitive capabilities of the human brain and  behavior.
In the first two approach we focused on method of neural data analysis and theories,
and in the third approach, the focus is on cognition.
We suggest bistable perception is a behavioral cognitive phenomenon that is relevant for the perspective we introduced.
This approach can be motivated, based on the fact that
bistable perception can be explained to some degree based on tools from complex systems
(see \autoref{sec:tools-used-neur}).
\marginpar{Approaching through behavior and cognition}
For instance, spontaneous transitory behavior that has been observed in bistable perception,
to some degree, can be explained based on principles of statistical physics \cite{bialekRandomSwitchingOptimal1995,atwalStatisticalMechanicsMultistable2014}
or the dynamics of the neural population can be explained by network models that are operating on the edge of a bifurcation \cite{theodoniCorticalMicrocircuitDynamics2011a,pastukhovMultistablePerceptionBalances2013a}.
In \autoref{cha:appr-thro-behav}, we introduce briefly the phenomenon of bistable perception,
then we justify its importance from the perspective of complex systems approach to the brain.
Perhaps this is the closest to one of the ultimate goals of systems and cognitive neuroscience,
and the most distant from the complex systems approach.
To minimize this gap we suggest and conduct novel experimental work,
namely, studying the phenomena on a mesoscopic scale which has not been done before.
I believe that this is just the very first step toward establishing the connection
such close to cognition and behavior.
Various theoretical and experimental steps need to be taken in the future studies to build a solid bridge between complex systems perspective toward the brain and cognition.

%%% Local Variables:
%%% mode: latex
%%% TeX-master: "../phdThesis_csb"
%%% TeX-master: "../phdThesis_csb"
%%% End:

 % Chapter 2
\chapter{Approaching through neural data analysis}\label{cha:appr-thro-nda}

Based on the motivation elaborated in \autoref{cha:brain-as-complex},
we believe multi-scale and cross-scale analysis of neural data is one of the important aspect of neural data analysis from the complex systems prospective toward the brain
and indeed is one of the apertures through which,
we can seek for  the complementary approaches mentioned in \autoref{sec:toward-neur-insp}.
In this chapter, after further elaboration on the need for multi-scale and cross-scale analysis of neural data,
very briefly we discuss some of the relevant cross-scale neural data 
analysis methodologies and then introduce two novel methodologies that has been developed as part of this thesis.

\section{Necessity of investigating across scales}\label{sec:necess-invest-across}

As it was briefly discussed in \autoref{sec:toward-neur-insp},
understanding behavior in a system whose components  are neither behaving completely independent nor completely coherent, 
requires investigation \emph{across scales} \cite{bar-yamWhyComplexityDifferent2017,einevollScientificCaseBrain2019}.
Certainly, the brain is a prominent example of such systems \cite{einevollScientificCaseBrain2019}.
Perhaps the most intuitive aspect of the brain which demonstrates this point is its oscillatory dynamics.
As \citeAYt{chialvoEmergentComplexNeural2010c} pointed out, 
\begin{displayquote}\textsl{
    ``Recent work on brain rhythms at small and large brain scales showed that spontaneous healthy brain dynamics is not composed by completely random activity patterns or by periodic oscillations\cite{buzsakiRhythmsBrain2011}''.
  }
\end{displayquote}

\begin{figure}[h]
  \centering
  \includegraphics[width = \linewidth]{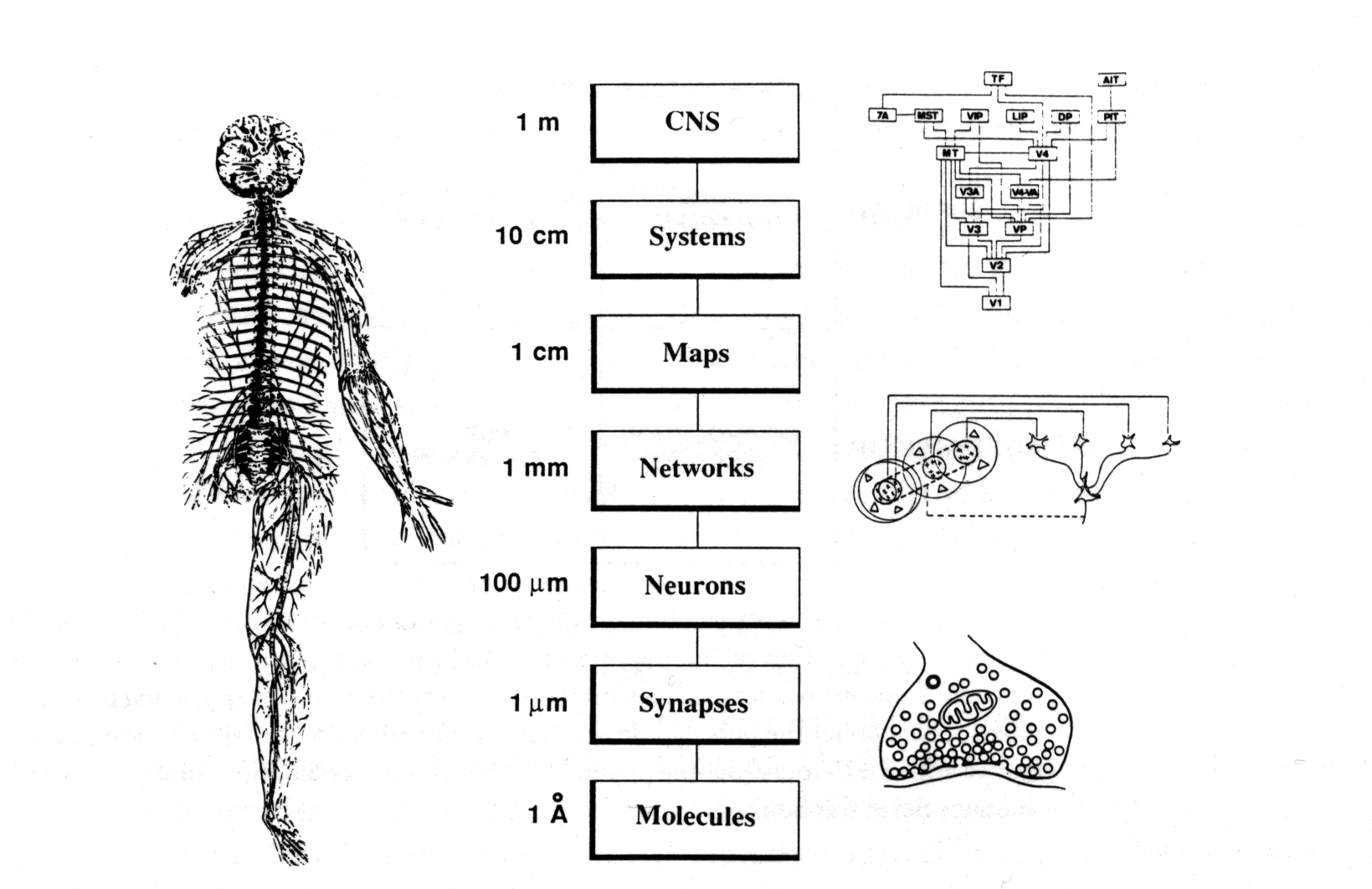}
  \caption{\textbf{Schematic depiction of levels of organization}\\
    Demonstrate extremely variable spatial scales at which anatomical organizations can be identified.
    Icons to the right represent structures at distinct levels:
    (top)
    a subset of visual areas in visual cortex;
    (middle)
    a network model of how ganglion cells could be connected to simple cells in visual cortex,
    and (bottom)
    a chemical synapse.
    Figure is adopted from \citet{churchlandPerspectivesCognitiveNeuroscience1988} with permission.}
  \label{fig:levelOfOrg}
\end{figure}

In order to investigate the brain across scales,
first we need to clarify what is considered as the scale.
In this thesis, we refer to different \emph{levels of organization} as scales.
Brain is organized in different \emph{levels}
(\autoref{fig:levelOfOrg}).

% figure
These levels range from scale of molecules all the way to large scale brain networks
\cite[Chapter 1]{churchlandComputationalBrain1992}.
Different phenomenon might primarily be explained in a limited range of these levels.
For instance, synaptic transmission, which is a basic form of communication in the brain, occurs at fairly small spatial scales, \ie level of molecules, synapses, and neurons.
Nevertheless, certain processes involve a broad range of levels. %\emph{levels}.
For instance in memory consolidation,
processes from gene expressions at the level of dendrites are involved,
all the way to larger-scale network \textbf{re}organization.
Therefore, one expects that process happening at different levels of organization to be related to each other.
It is worth to mention that, our understanding (especially from a theoretical perspective)
should be consistent across the levels of organization.
As elegantly described in \citet[Chapter 1]{churchlandComputationalBrain1992}:
\begin{displayquote}\textsl{
    ``... the theories on one level must mesh with the theories of levels both
    higher and lower, because an inconsistency or a lacuna somewhere in the tale
    means that some phenomenon has been misunderstood.
    After all, brains are assemblies of cells, and something would be seriously amiss if neurons under
    one description had properties incompatible with the same neurons under
    another description.''
  }
\end{displayquote}

Indeed, there are various empirical evidence on predictions across scales and relationships between scales:
From single neurons to microcircuits \cite{raschInferringSpikeTrains2008a,raschNeuronsCircuitsLinear2009},
from microcircuits to a single brain area \cite{liBurstSpikingSingle2009a},
from a single area to the whole brain \cite{schwalmCortexwideBOLDFMRI2017,zerbiRapidReconfigurationFunctional2019}.
In some cases, the cross-scale coupling is closely and causally related to a specific function,
such as global state changes that have been shown in a study by \citet{liBurstSpikingSingle2009a}.
They showed that burst spiking of a single
cortical neuron in somatosensory cortex can induce a global switch between the slow-wave sleep and Rapid-Eye-Movement (REM) sleep.
In some cases, cross-scale relationships are even mechanistically interpretable as well.
For instance, it has been demonstrated that spiking probability can be modulated by the underlying network oscillation.
Network oscillations modulate the membrane potential of the neuron and that leads to the different levels of excitability for the given neuron.
Depending on the phase of the underlying oscillation, this can lead to a higher or lower probability of spiking activity \cite{volgushevLongrangeCorrelationMembrane2011,hasenstaubInhibitoryPostsynapticPotentials2005}.
Based on these simple mechanisms, \emph{coordination by oscillation} has been hypothesized,
and this lends support to various cognitive functions such as attention. 
The hypothesis of ``Coordination by oscillation'' proposes that network
oscillations modulate differently the excitability of several target populations,
such that a sender population can emit messages during the window of time for which a selected target is active, 
while unselected targets are silenced
\citep{friesRhythmsCognitionCommunication2015,womelsdorfModulationNeuronalInteractions2007a,friesMechanismCognitiveDynamics2005}.
Overall, I believe, considering \emph{two successive scales simultaneously},
is a principled approach for understanding collective or coordinated organizations in neural systems.
Furthermore, as mentioned in  \autoref{sec:toward-neur-insp} this approach is also justified by empirical evidence.

Investigating across scales can also be motivated from a more abstract 
(and perhaps more fundamental) perspectives:
In dynamical systems with non-linear interaction there are various examples where
activity in different scales are related \cite{levanquyenBrainwebCrossscaleInteractions2011}.
One example for such non-linear dynamical systems is the Kuramoto model.
As described briefly in \autoref{sec:complex-systems},
Kuramoto model describes a system of multiple coupled oscillators
\cite{kuramotoSelfentrainmentPopulationCoupled1975,kuramotoChemicalOscillationsWaves2003} 
(for an integrative review see \cite{acebronKuramotoModelSimple2005}).
In this model, the activity of individual oscillators is related to quantities pertaining to the average or mean-field activity of the system as a whole.
More precisely, the phase of an individual oscillator can be related to the mean phase of oscillators and their phase coherence.
Such core ideas from the theory of dynamical systems went beyond mere conceptual connections,
but also inspired unifying formulations for neural oscillations in the brain (\eg see \cite{breakspearGenerativeModelsCortical2010}).
For more detailed elaboration on motivations from the theory of dynamical systems for cross scales investigation of the brain see works of Le Van Quyen and colleagues
\cite{levanquyenExploringNonlinearDynamics2003,levanquyenDisentanglingDynamicCore00,levanquyenBrainwebCrossscaleInteractions2011}.

The other abstract motivation for investigation across scales is the nature of computation in the brain.
The brain is a naturally evolved biological information processing system.
Therefore, the computational strategies or solutions served by the brain can be quite different from engineered information processing systems \cite[Chapter 1]{churchlandComputationalBrain1992}\cite{douglasRecurrentNeuronalCircuits2007}.
The main difference between commonly engineered information processing systems and natural information processing systems is that the latter is constrained by the existing form of evolving organisms.
As elaborately framed by \citet[Chapter 1]{churchlandComputationalBrain1992}:
\begin{displayquote}\textsl{
    ``Evolutionary modifications are
    always made within the context of an organization and architecture that are
    already in place. Quite simply, Nature is not an intelligent engineer. It cannot
    dismantle the existing configuration and start from scratch with a preferred
    design or preferred materials.
    It cannot mull the environmental conditions and construct an optimal device.''
  }
\end{displayquote}
Furthermore, there are other aspects that need to be taken into account in the process of thinking about the solution chosen by the brain.
For instance,
humans/animals are constrained by the response time (they need to be fast enough) to be able to survive in their natural environment.
Finding the solution for the required computation is expected to happen in a few hundred milliseconds.
This becomes even more puzzling if we take into account the computational machinery in the brain that is orders of magnitude slower than artificial information processing systems.
Events in neurons happen in range of milisecond ($10^{-3}$)
as opposed to  nano second ($10^{-9}$) in electronic computers \cite{douglasRecurrentNeuronalCircuits2007}.
Other such examples are, spatial constrains (limitation by available space), energy consumption, and metabolism
\cite[Chapter 1]{churchlandComputationalBrain1992}).
\textbf{All being said to minimize the surprise of mentioning novel proposals (in the following) on brain computational principle that pertains to cross-scale investigation.}
\citet{bellLevelsLoopsFuture1999,bellCrossLevelTheory2007} proposes that, 
the adaptive power of biological information processing systems comes from
the gating of information flows across levels, both upward and downward,
as \citet{bellCrossLevelTheory2007} stated:
\begin{displayquote}\textsl{
    ``There is thus no ``functionalist cut-off level'' anywhere in the biological hierarchy
    Nature does not seem to shield the macro from the micro in the way that a
    computer does.``
  }
\end{displayquote}
Although, to the best of my knowledge, this proposal is not yet formalized as a complete theoretical framework,
but perhaps it gains some empirical support through recent experimental and computational studies of \emph{ephaptic} interactions in the brain.
In recent years, we have experimental \cite{anastassiouEphapticCouplingCortical2011}
and modeling \cite{anastassiouEphapticCouplingCortical2011,ruffiniRealisticModelingMesoscopic2020,sheheitliMathematicalModelEphaptic2020}
on the possibility of having ephaptic interactions in the brain
(for a review also see \cite{anastassiouEphapticCouplingEndogenous2014}).
Indeed, this evidence that electrical fields in the brain can functionally modulate the activity of neurons is in line with \citet{bellLevelsLoopsFuture1999,bellCrossLevelTheory2007} proposal on the computational architecture of the brain.

Overall, I believe the arguments provided above,
justify the necessity of investigating brain activity across scales.
In spite of the importance of this need for understating the brain,
there are not sufficient methodologies for the multi-scale investigation of the brain activity
In the next two sections (sections \ref{sec:avail-tools-invest} and \ref{sec:need-new-tools}) 
we provide a brief overview of available tools and our contribution of novel methods for cross-scale investigation of brain dynamics.
\section{Available tools for investigating cross-scale relationships}\label{sec:avail-tools-invest}
\begin{figure}[h]
  \centering
  \includegraphics[width = \linewidth]{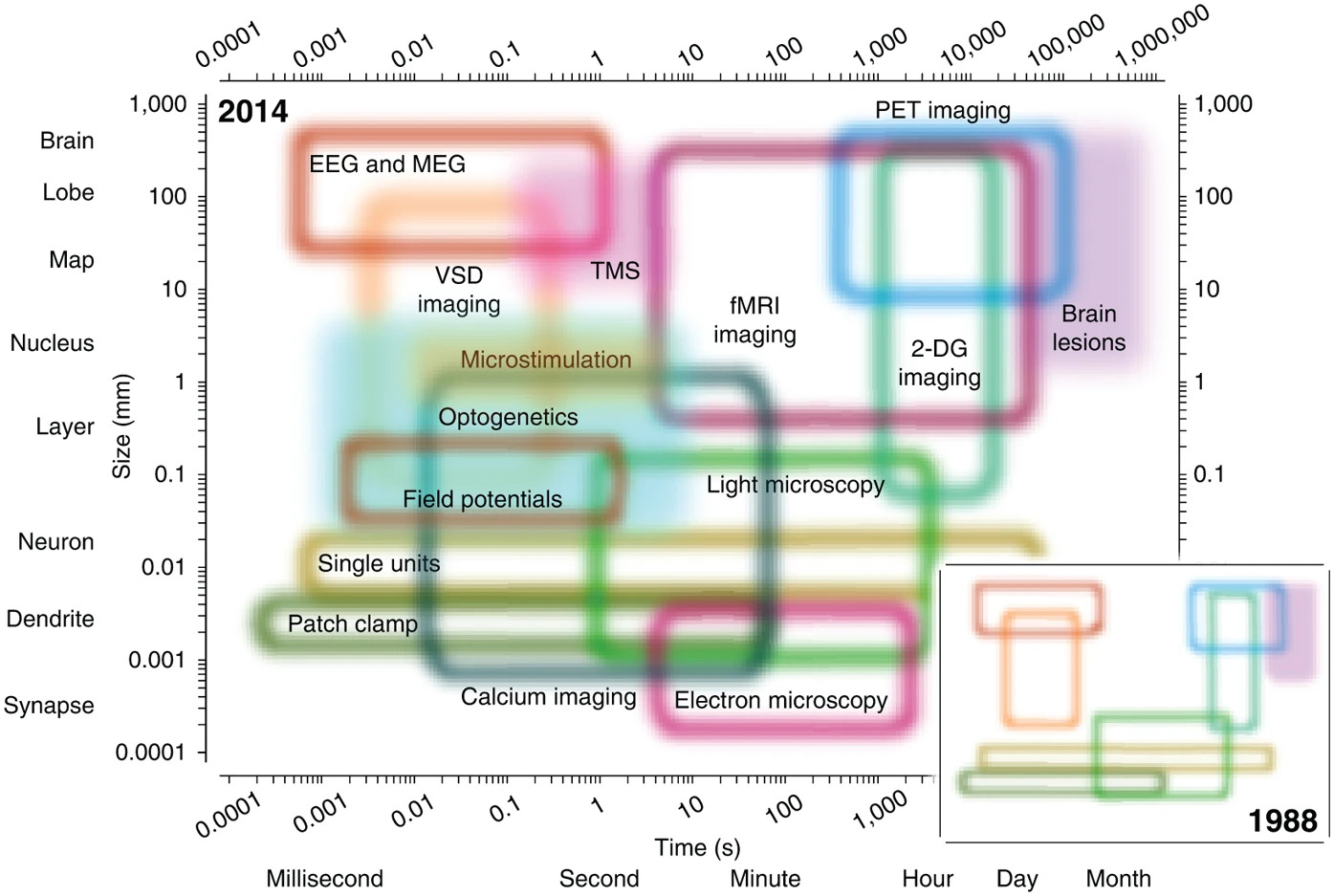}
  \caption{\textbf{Spatio-temporal resolution of measurement methods in neuroscience}\\
    Demonstrate the spatial and temporal resolution of measurement methods being used in neuroscience
    (up to 2014).
    Each box depict the spatial (y-axis) and temporal (x-axis) of one measurement method.
    Open regions represent measurement techniques and 
    filled regions, perturbation techniques.
    Inset, a cartoon rendition of the methods available in 1988.
    The regions allocated to each domain are somewhat arbitrary and represent the estimate of \citet{sejnowskiPuttingBigData2014}.
    Abbreviations used in the figure:
    EEG, electroencephalography;
    MEG, magnetoencephalography;
    PET, positron emission tomography;
    VSD, voltage-sensitive dye;
    TMS, transcranial magnetic stimulation; 
    2-DG, 2-deoxyglucose.
    Figure is adopted from \citet{sejnowskiPuttingBigData2014} with permission.}
  \label{fig:measureScale}
\end{figure}

Brain activity can be measured using various experimental methodologies at different scales
(\autoref{fig:measureScale}).
For instance, it can be spike trains from individual neurons, 
field potentials generated by small or large population of neurons or hemodynamic signals from the whole brain.
Our novel development for bridging scales pertains to the relationship between,
spiking activity
Local Field Potentials (LFPs) and 
Blood-Oxygen-Level Dependent (BOLD) signals.

A number of tools have  already been developed and applied to neural data, and they
gave us insight into the relationship between brain activity in different scales.
Here we mention very briefly a subset of such methods that are related to novel development that we introduce in the next section.

The relationship between spiking activity and LFP has been studied extensively in the context of mechanisms for coordination by oscillation in the brain.
Indeed, this was one of the examples briefly discussed in \autoref{sec:necess-invest-across} to motivate understanding cross-scale relationships.
Various techniques has been developed for investigating the relationship between spiking activity and LFP
\citep{zeitlerAssessingNeuronalCoherence2006,ashidaProcessingPhaseLockedSpikes2010,vinckPairwisePhaseConsistency2010,vinckImprovedMeasuresPhasecoupling2012,jiangMeasuringDirectionalityNeuronal2015,liUnbiasedRobustQuantification2016,zareiIntroducingComprehensiveFramework2018}.
Most of the approaches for investigating the spike-LFP coupling are restricted to pairwise first-order statistics of spike-LFP interactions.
Given the various experimental advances, there is a growing need for conceptual and methodological frameworks to investigate this relationship in multi-variate settings
(see further elaboration in \autoref{sec:relat-betw-micro}).

Another line of research pertaining to cross-scale relationships,
is investigating the relationship between LFP and fMRI BOLD signals.
In this branch,
extensive research has been done toward understanding the neural correlate
or neural activity underlying the BOLD signal
\cite{logothetisNeurophysiologicalInvestigationBasis2001,logothetisUnderpinningsBOLDFunctional2003,logothetisWhatWeCan2008,goenseNeurophysiologyBOLDFMRI2008a,zaldivarDopamineinducedDissociationBOLD2014}.
Methods used for exploring the relationship between these signals were
conventional correlation analsysis \cite{logothetisNeurophysiologicalInvestigationBasis2001},
system identification \cite{logothetisNeurophysiologicalInvestigationBasis2001},
Canonical Correlation Analysis (CCA) and its time-resolved kernelized version \cite{biessmannTemporalKernelCCA2009,murayamaRelationshipNeuralHemodynamic2010}.
Certainly, the mentioned investigation shed light on the basic nature of the coupling between LFP and fMRI BOLD, but more developments needed to get into functionally relevant couplings.

Mentioned developments construct the foundations and moreover led to important methodologies for addressing questions concerning functional implications of investigating the relationship between LFP and BOLD fMRI.
Along the same line of developments, Neural-Event-Triggered (NET) fMRI was also introduced recently. 
In NET-fMRI, characteristics neural activities of such as Sharp Wave-Ripple (SWR) are used as events to align and average the time course of large-scale brain activity to extract the global signature of the given events.
Indeed, ripple-triggered activities in macaque monkeys revealed important large-scale coordination involved in the process of memory consolidation \cite{logothetisHippocampalCorticalInteraction2012}.

NET-fMRI can be a very informative methodology if the \emph{event} is already well-defined.
Nevertheless, there are very few such well-characterized neural activity like SWR.
Therefore, we need novel methodologies to detect and characterize such distinct neural activities
(see further elaboration in \autoref{sec:relat-betw-meso}).

\section{Need for new tools for investigating cross-scale relationships} \label{sec:need-new-tools}

As was motivated in the previous section (\autoref{sec:avail-tools-invest}),
novel methodologies are needed for investigating the brain dynamics across the scale.
LFPs are signals at meso-scale  \cite{liljenstroemMesoscopicBrainDynamics2012}, 
which is an intermediate scale between micro- and macro-scale,
and they reflect a mesoscopic picture of the brain dynamics.
LFPs result from the superposition of the electric potentials generated by ionic currents flowing across the membranes of the cells located close to the tip of recording electrodes.
The LFP reflects neural cooperation due to the anisotropic cytoarchitecture of most brain regions,
allowing the summation of the extracellular currents resulting from the activity of neighboring cells and potentially remote populations. 
As such, a number of subthreshold integrative processes 
(\ie modifying the neurons' internal state without necessarily triggering spikes) 
contribute to the LFP signal
\cite{buzsakiOriginExtracellularFields2012,liljenstroemMesoscopicBrainDynamics2012,einevollModellingAnalysisLocal2013,herrerasLocalFieldPotentials2016,pesaranInvestigatingLargescaleBrain2018}.
As LFPs are rich and intermediary signals, they can be a pivotal point for bridging the scales. 
We can better illustrate the importance of LFP for cross-scale analysis with an example.
In LFPs, certain characteristics of neural activities, like
SWRs are detectable.
Interestingly, SWRs occur concurrently with well-coordinated activity at smaller scales (neurons and population of neurons),
and as well as a larger scale (entire brain).
For the connection to smaller scales (microscopic scale) various studies suggest SWRs emerge in the CA1 mainly due to afferent CA2- and CA3-ensemble \emph{synchronous} discharges \citep{csicsvariEnsemblePatternsHippocampal2000,csicsvariEnsemblePatternsHippocampal2000,olivaRoleHippocampalCA22016}.
For the larger scale (macroscopic scale), as briefly mentioned earlier, concurrent recording of BOLD signal of the entire brain and SWRs,
demonstrate large scale coordination of entire brain activity during SWRs \citep{logothetisHippocampalCorticalInteraction2012}.

Detecting characteristic activities like SWRs and finding such relationships across
scales (exemplified in the previous paragraph) was the result of years of experimental work and exploration in the data.
Developing new tools that allow us to find such characteristic
patterns (like SWRs) in an unsupervised fashion and finding their
relationship to measurement at other scales
[\eg with synchronization measures and NET-fMRI]
can be of paramount importance.

Based on the ideas and motivation elaborated above, we first focus on tools that allow us to explore
the relationship between spikes and LFPs (\autoref{sec:relat-betw-micro})
and then, a method for the detection of neural events in an unsupervised fashion (\autoref{sec:relat-betw-meso}).

\subsection{Tools to explore micro-meso relationsips}\label{sec:relat-betw-micro}
A prominent example of the relationship between micro- and meso-scale activity in the brain is the spike-field coupling.
Apart from its importance from the perspective discussed in \autoref{sec:necess-invest-across},
the synchronization between spiking activity and the phase of particular rhythms of LFP has been used as an important marker to reason about the underlying cooperative network mechanisms.
Nevertheless, there is not yet a systematic way to extract the coupling information from the largely multi-variate data available to state-of-the-art recording techniques
\cite{dickeySingleUnitStabilityUsing2009,junFullyIntegratedSilicon2017a,juavinettChronicallyImplantedNeuropixels2019}
with hundreds or even thousands of recording sites
\cite{pesaranInvestigatingLargescaleBrain2018,junFullyIntegratedSilicon2017a,buzsakiLargescaleRecordingNeuronal2004,fukushimaStudyingBrainFunctions2015}. 
We developed a multi-variate extension of phase-locking analysis 
and a statistical testing framework to assess the significance of the coupling strength.
With our method (which we call Generalized Phase Locking Analysis -- GPLA),
we can quantify, characterize, and statistically assess the interactions between population-level spiking activity and mesoscopic network dynamics (such as global oscillations and traveling waves).

We demonstrate the capability of the GPLA by applying the method to various simulated and experimental datasets.
For instance, the application of the method on simulation of hippocampal SWR can reveal various characteristics of hippocampal circuitry with minimal prior knowledge.
GPLA reveals CA1 and CA3 neurons are all coupled to the field activity in the gamma and ripple band
(in line with experimental and simulation results
\cite{buzsakiHighfrequencyNetworkOscillation1992,ramirez-villegasDissectingSynapseFrequencyDependent2018}),
suggesting this rhythm may support communication between CA1 and CA3 sub-fields during memory trace replay. 
Furthermore, it also allows us to tease apart the involved populations and provide hint on the communication flow from CA3 to CA1 based on label-free spike timing and LFP.
As another example, the application of the method on the experimental recordings from Prefrontal Cortex (PFC) suggests a non-trivial coupling between spiking activity and LFP traveling waves in this region of the PFC.
Assuming LFPs mostly reflect local and distal input post-synaptic currents to the underlying neural population,
analysis based on the GPLA accompanied by neural field simulations suggest that a connectivity structure consists of long excitatory horizontal connections and strong local recurrent inhibition as a plausible speculations for these PFC recordings
(in line with previous modeling and experimental studies \cite{safaviNonmonotonicSpatialStructure2018,sherfeyFlexibleResonancePrefrontal2018,sherfeyPrefrontalOscillationsModulate2020a}.

Notably,
an important component of our methodological contribution for investigating the relationship between micro- and meso-scale activity is the theoretical significance test for GPLA. 
We describe the theoretical foundation of the test in \citet{safaviUncoveringOrganizationNeural2023}
(\seealso, \nameref{cha:paper-5})
and the necessary development for practical applications on neural data
is described in \citet{safaviUncoveringOrganizationNeural2023}
% \shs{cite GPLA paper}
(\seealso, \nameref{cha:paper-gpla}).
In our theoretical investigation, we derive analytically the asymptotic distribution of Phase-Locking Value
(a uni-variate coupling statistics which is conventionally used for quantifying spike-LFP coupling),
which follows a Gaussian distribution.
The implication of these results for neural data is,
whitening of LFPs and normalization by the square root of the spike rate is necessary for the applicability of our theoretical results on neural data.
The asymptotic distribution for the uni-variate coupling was key for the development of the statistical test for the multivariate version of phase-locking analysis.
Based on Gaussianity of the uni-variate measure and random matrix theory we could derive the theoretical null distribution for the singular values of a matrix containing all pairwise coupling
that we call the coupling matrix.
Consequently, we show that singular values of such matrices converge to a
Marchenko-Pastur distribution
\citep{marcenkoDistributionEigenvaluesSets1967a}.
\footnote{\citet{marcenkoDistributionEigenvaluesSets1967a} in not
  written in English, but is the original publication.
  The reader can refer to \citet[Chapter
  2]{andersonIntroductionRandomMatrices2010} instead.}
This is a well-established asymptotic behavior in random matrix theory for matrices with independent normally distributed entries \cite{andersonIntroductionRandomMatrices2010}. 
The key is Marchenko-Pastur distribution has an upper bound, meaning that, 
under the null condition (no coupling between spike and LFP) largest singular value of the coupling matrix should not exceed this upper limit. 
If the singular values resulting from data are larger than this upper limit, 
then there is significant coupling between the population spikes and the multi-channel LFPs.
Developing a theoretical test is of paramount importance considering the constantly increasing dimensionality of modern recording techniques. 
\subsection{Tools to explore meso-macro relationsips}\label{sec:relat-betw-meso}

As pointed out in \autoref{sec:need-new-tools}, 
it is important to develop tools that allow us to find characteristic patterns of LFPs (such as SWRs) in an unsupervised fashion.
Such patterns are potentially very special, in the sense that,
they provide us a time window that meso-scale dynamics is closely related micro and macro scale dynamics.
In fact, this is of paramount importance for bridging the brain activity in different scales.

We developed an unsupervised methodology based on Non-negative Matrix Factorization (NMF) and dictionary learning to detect transient cooperative activities in a single channel LFP
(see \nameref{cha:paper-besserve2020ned} for more details).
Such activities were also introduced as \emph{neural events} in previous studies \cite{logothetisHippocampalCorticalInteraction2012,logothetisNeuralEventTriggeredFMRILargescale2014,ramirez-villegasDiversitySharpwaverippleLFP2015}.
With this method, is not only possible to detect well-established characteristic patterns such as sharp wave-ripples,
but also new characteristic neural activities that have not been identified and studied before.
We demonstrate the capability of our method by identifying neural events in Hippocampus and LGN and also and explored their brain-wide \emph{macro-scale} signatures using concurrent fMRI recordings from anesthetized monkey.
The result suggest that, similar to the previous study of \citet{logothetisHippocampalCorticalInteraction2012} that was focused on sharp wave-ripples,
the identified events in Hippocampus and LGN reflect a large scale coordinated dynamics,
namely a competition between cortical and subcortical regions.

Furthermore, neural events can also be informative for exploring micro-scale and meso-scale relationships.
By exploiting a simulation of thalamocortical circuitry  developed by \citet{costaThalamocorticalNeuralMass2016},
we demonstrate that such events have the potential of even relating meso-scale dynamics to \emph{micro-scale} dynamic, even at the cellular level.
With our methodology we identified different kinds of spindles in the activity of the thalamus module of the simulation (indeed, this is another demonstration for the capability of the method),
and demonstrate that different events co-occur with a characteristic activity pattern in cellular variables (such as membrane potentials and ionic currents) of the simulation.

%%% Local Variables:
%%% mode: latex
%%% TeX-master: "../phdThesis_csb"
%%% End:
 % Chapter 3
\chapter{Approaching through neural theories}\label{cha:appr-thro-theo}

As motivated in \autoref{cha:brain-as-complex},
in order to achieve the target bridge between complex systems and neuroscience, \ie
approaching the brain as a complex system by exploiting systems neuroscience tools and notions,
one of the apertures through which, we can seek for the complementary approaches is neural theories (see  \autoref{sec:toward-neur-insp}).
In this chapter we aim to explore two important theoretical frameworks,
one closely related to the field of neuroscience, and one to complex systems.
In order to establish the mentioned bridge, we explore the potential connection between them.
On the neuroscience side, we consider \emph{efficient coding} which is one of the most important theoretical frameworks in systems neuroscience,
and on the complex systems side, we reflect on the \emph{criticality hypothesis of the brain} that has strong roots in the field of complex systems.
We first provide a brief overview on each of them, and then their potential connection.

\section{Criticality hypothesis of the brain}\label{sec:crit-hypoth-brain}
In the course of studying the state of the matter
(\eg water, steam and ice as states of \ch{H2O})
and their phase transitions (\eg transition from water to vapor)
physicists discover some \emph{universal} behavior in a variety of phase transitions
(\eg freezing of water and  magnetization in metals \cite[Chapter 5]{schroederIntroductionThermalPhysics1999}
as well as in wider ranges of natural phenomenon such as human social behavior \cite{castellanoNonequilibriumPhaseTransition2000} 
(see \citet{mathisEmergenceLifeFirstOrder2017} and \citet{bar-yamWhyComplexityDifferent2017} for other examples).
Later on, in the process of examining the relationship between microscopic variables like speed of atoms and macroscopic variables like temperature,
it has been realized that, close to a critical point the usual methods fail to establish these relationships.
The critical point (for water) is the point where fluctuations between liquid-like and
vapor-like densities extend across the system so that the system is not smooth anymore and therefore averages are not well behaved.
Furthermore, this characteristic inharmonious behavior was observable at all scales \cite{bar-yamWhyComplexityDifferent2017}.
Indeed, the method of Renormalization Group (RG) has been developed to investigate mathematically such state of a system and has been applied on a wide range of systems.
It turns out, in spite of differences in details of various systems (\eg magnetic dipoles and molecules of water),
their behavior can be explained based on the RG method.
This important observation, led to the notion of \emph{universality},
that allow us to explain various systems with many interacting components with a small set of variables and some scaling relations.

Based on these fundamental ideas \emph{criticality hypothesis of the brain} has been proposed \cite{munozColloquiumCriticalityDynamical2018}.
Roughly speaking, criticality hypothesis of the brain states that,
brain operates close to a critical state, a state on the edge of transition between order and disorder.
The first experimental evidence on scale-freeness of the brain dynamics
(as one of the signatures of criticality -- see \autoref{sec:sign-crit-neur})
has been reported almost two decades ago by \citet{beggsNeuronalAvalanchesNeocortical2003}.
Later on such scale-free dynamics have been observed in various smaller and larger scales as well.
To name a few, see
\citet{bonilla-quintanaActinDendriticSpines2020} at the scale of actin in dendrites,
\citet{johnsonSinglecellMembranePotential2019} at the scale of neuronal membranes,
\citet{varleyDifferentialEffectsPropofol2020} at the scale of the entire brain
(for more references see \cite{munozColloquiumCriticalityDynamical2018,agrawalScaleChangeSymmetryRules2019}).
Moreover, being close to this state is beneficial for the brain
\cite{munozColloquiumCriticalityDynamical2018,tkacikInformationProcessingLiving2016,moraAreBiologicalSystems2011a},
as it has been shown that general information processing capabilities such as
sensitivity to input \cite{kinouchiOptimalDynamicalRange2006,brochiniPhaseTransitionsSelforganized2016}, 
dynamic range
\cite{kinouchiOptimalDynamicalRange2006,larremorePredictingCriticalityDynamic2011,nurProbingSpatialInhomogeneity2019},
or information transmission and storage
\cite{shewInformationCapacityTransmission2011,vanniCriticalityTransmissionInformation2011,lukovicTransmissionInformationCriticality2014,marinazzoInformationTransferCriticality2014},
and various other computational characteristics has been also considered to be relevant
\cite{turingComputingMachineryIntelligence1950,tanakaRecurrentInfomaxGenerates2008,hidalgoInformationbasedFitnessEmergence2014a,hidalgoCooperationCompetitionEmergence2016,medianoIntegratedInformationMetastability2016,khajehabdollahiEmergenceIntegratedInformation2019,hoffmannOptimizationSelfOrganizedCriticality2018a,michielsvankessenichPatternRecognitionNeuronal2019,wangHierarchicalConnectomeModes2019,finlinsonOptimalControlExcitable2020,zeraati2023intrinsic}
(also see  
\cite{beggsCriticalityHypothesisHow2008,shewFunctionalBenefitsCriticality2013,zeraatiSelfOrganizationCriticalitySynaptic2021}
for a reviews).

To summarize, multiple studies have reported signatures of criticality observed in various neuronal recordings at different scales,
and  theoretical investigations demonstrated various aspects of information processing are optimized at the second-order phase transition
(see references in \cite{munozColloquiumCriticalityDynamical2018,agrawalScaleChangeSymmetryRules2019}).

\section{Signatures of criticality in neural systems}\label{sec:sign-crit-neur}
As motivated in the previous section,
various empirical and theoretical investigations lend support to criticality hypothesis of the brain,
and signify the potential functional relevance of the criticality hypothesis of the brain.
Therefore, it has been motivating to search for diverse signatures of criticality in the brain.
These signatures can be categorized into three groups \cite{zeraatiStudyingCriticalityIts2017}:
scale-freeness neural activity (avalanche criticality),
dynamical regime of the neural system (edge of bifurcation criticality),
and thermodynamic of the neural data (maximum entropy criticality).
\begin{description}

\item[Avalanche criticality:]
  Scale-free cascade of activity is a ubiquitous type of dynamics in nature:
  For instance in interacting tectonic plates \cite{gutenbergEarthquakeMagnitudeIntensity1956},
  forest fires \cite{malamudForestFiresExample1998},
  nuclear chain reactions \cite{harrisTheoryBranchingProcesses1963},
  threshold-crossing events that appears as one unit (\eg a tree) exceeding a threshold (\eg a tree fires)
  and because the units of the system are coupled to each other,
  similar threshold-crossing events \emph{propagate} through other units of the system.
  Such propagating dynamics can lead to large \emph{avalanches} of activity.
  Almost two decades ago \citet{beggsNeuronalAvalanchesNeocortical2003} observed similar cascades in activity of in-vitro neural populations
  and later on others reported such scale-free cascades at various other neuronal recordings in various scales (see references in
  \cite{munozColloquiumCriticalityDynamical2018,agrawalScaleChangeSymmetryRules2019}).
  Truly critical systems, not only should show the mentioned scale free dynamics,
  but also they should follow the scaling laws introduced by \citet{sethnaCracklingNoise2001b},
  that were observed in neural data \cite{friedmanUniversalCriticalDynamics2012} as well
  \footnote{
    Indeed, scale-free neural avalanches without following scaling laws have been observed in neural models that are not operating close to a critical point 
  \cite{aitchisonZipfLawArises2016a,touboulPowerlawStatisticsUniversal2017}.}.

\item[Bifurcation criticality:]
  When a dynamical system has a transition from one dynamical regime to another
  (such as transition from order to chaos),
  it experiences a \emph{bifurcation} \cite{izhikevichDynamicalSystemsNeuroscience2010,breakspearDynamicModelsLargescale2017,cocchiCriticalityBrainSynthesis2017}.
  The point where the transition happens is also denoted as the critical point.
  There are various kinds of bifurcations (see \citet{izhikevichDynamicalSystemsNeuroscience2010}),
  but some of them have been particularly appealing for understating the dynamics of the brain as well as computation in the brain.
  Without getting into the theoretical details of these bifurcations and in very brief fashion,
  transitioning from order to chaos \cite{bertschingerRealtimeComputationEdge2004},
  and transitioning from an asynchronous to a synchronous state \cite{santoLandauGinzburgTheory2018}
  have been considered as two important bifurcations for the brain
  (for further elaboration see \citet{cocchiCriticalityBrainSynthesis2017,munozColloquiumCriticalityDynamical2018} and references therein).
  Avalanche criticality and bifurcation criticality can co-occur,
  when there is a continuous phase transition \cite{cocchiCriticalityBrainSynthesis2017}
  (for example see \cite{magnascoSelftunedCriticalAntiHebbian2009,pittorinoChaosCorrelatedAvalanches2017}),
  nevertheless, \citet{kandersAvalancheEdgeofchaosCriticality2017} 
  proposed that these two types of criticality do not necessarily co-occur and therefore should be attributed to two distinct phenomena.

\item[Thermodynamic criticality:]
  Statistical mechanic provides a powerful framework to study collective behavior in systems consisting of interacting units with many degrees of freedom \cite{sethnaStatisticalMechanicsEntropy2006}.
  Tools from statistical mechanic have been applied in neural networks in order to understand their collective dynamics \cite{amitModelingBrainFunction1992}.
  Along the same line \citet{tkacikThermodynamicsSignaturesCriticality2015} approached the activity of neurons from a thermodynamical perspective.
  They define a Boltzman-like distribution, derive various thermodynamic quantities such as heat capacity based on estimated Boltzman distribution, and ultimately define criticality based on thermodynamic quantities (like divergence of heat capacity).
  Moreover, in empirical data this novel framework is applicable and functionally relevant.
  This novel formulation introduces another signature or definition of criticality in neural system \cite{tkacikThermodynamicsSignaturesCriticality2015}
  (but also see \cite{nonnenmacherSignaturesCriticalityArise2017}).
\end{description}

\section{Seeking for a bridge: a complementary approach}\label{sec:seek-bridg-compl}

As mentioned earlier, over the last two decades, multiple
experimental and theoretical investigations lend support to criticality hypothesis of the brain. 
In particular, as it was briefly discussed in \autoref{sec:crit-hypoth-brain},
closeness to criticality has been suggested to be an optimal state for information processing.
To evaluate how closeness to criticality can be beneficial for the information processing in the brain,
the common approach is using a model
(\eg a branching network, a recurrent neural network)
that can attain various states (including critical and non-critical states),
 depending on control parameters (\eg branching ratio, connection strength) of the model.
Then by quantifying how general information processing capabilities such as
information transmission
depend on the control parameters, the advantages of being close to a critical state can be assessed.
For instance, if information transmission in the model under study is optimized exclusively close to the critical state of the model (defined based on the control parameter(s)),
then it can be considered as evidence for relevance of usefulness of criticality for the brain.

Indeed, one of the important reasons for the relevance of the criticality for the brain
is the optimized  information processing capabilities that operating close to this state offers.
Nevertheless, the \emph{optimized setting} implied by criticality hypothesis,
does not imply any specific computation that the brain may need to execute,
but rather \emph{general} capabilities for computation
\footnote{
  See also \citet{lizierLocalInformationDynamics2013} (in particular chapter 6) that argue
  closeness to criticality is a sate where [some] computing primitives (such as information storage, transfer and modification) are optimized.
  Furthermore, an complementary perspective is, non-critical states can be specifically advantageous for a particular computation,
  and therefore brain needs to be able to flexibly switch between them \cite{clawsonAdaptationScalefreeDynamics2017,zeraatiStudyingCriticalityIts2017}.
}.
For instance,
being in a state which is optimized to have the maximum sensitivity to input \cite{kinouchiOptimalDynamicalRange2006,brochiniPhaseTransitionsSelforganized2016}, 
and maximum dynamic range
\cite{kinouchiOptimalDynamicalRange2006,larremorePredictingCriticalityDynamic2011,nurProbingSpatialInhomogeneity2019}
are all relevant capabilities for coding sensory information,
but mere adjusting for the closeness to criticality cannot provide a neural coding algorithm and its implementation for coding given resource constraints.
In contrast, there are frameworks (such as efficient coding)
that  provide the functionally relevant objectives to be maximized or minimized
(which define the optimized computation),
the algorithm of computation (neural coding algorithm) and the neural implementation.
Therefore, we think we need complementary approaches to criticality
that can bridge the gap between criticality and frameworks which focus on \emph{functionally relevant} computations and their implementations.

\subsection{Efficient coding as the computational objective}\label{sec:efficient-coding-as}

We focus on \emph{coding}, as a functionally relevant computation
(and with the ultimate purpose of establishing the bridge to criticality). 
Efficiency of neural coding is particularly important, as sensory systems have evolved to transmit maximal information about incoming sensory signals,
given internal resource constraints (such as internal noise, and/or metabolic cost)
\cite[Chapeter 13]{rosenblithSensoryCommunication2012}\cite{fredriekeSpikesExploringNeural1999,quianquirogaPrinciplesNeuralCoding2013}.
Indeed, models using this simple principle made various verified predictions about neural responses
(\eg receptive field in in V1
\cite{olshausenEmergenceSimplecellReceptive1996,simoncelliNaturalImageStatistics2001}).

Several variants of efficient coding have been develop\-ed
(for a brief over\-view see \cite{chalkUnifiedTheoryEfficient2018}).
Depending on the answers to qualitative questions like,
``
\emph{What should be encoded?}
\emph{What sensory information is relevant?}
\emph{What can be encoded given the internal constraints?}
``,
the suitable variant of efficient coding can be determined
(see \citet{chalkUnifiedTheoryEfficient2018} for a quantitative elaboration).
For instance, one of the variants of efficient coding is based on \emph{redundancy reduction},
which has the  objective of encoding maximal information about \emph{all} inputs with statistically independent responses and it is applicable in low noise regime \cite{chalkUnifiedTheoryEfficient2018}.
Afterward, based on principles of efficient coding, a computational objective for a given neural system can be defined.
Our choice of efficient coding computational objective is the one introduced in \citet{boerlinPredictiveCodingDynamical2013}.
The objective of this coding schema is,
a network of Leaky-Integrate and Fire (LIF) neurons should encode the input through a pattern of spikes,
such that input stimulus can be reconstructed based on a linear readout of the spiking output.
Furthermore, the network should perform the coding with minimum number of spikes and as accurate as possible.
The same principle has been employed in \citet{chalkNeuralOscillationsSignature2016} in a more realistic network of LIF neurons and has been used in our investigation.

\subsection{Signature of criticality in efficient coding networks}

Following our motivation for the necessity of complementary approaches to criticality,
we study networks that implement efficient coding
(see \citet{boerlinPredictiveCodingDynamical2013} and \citet{chalkNeuralOscillationsSignature2016} for more details)
and we ask if any of the  criticality signatures
(discussed in \autoref{sec:sign-crit-neur})
are observable exclusively in the network that is optimized for performing  efficient coding.

We investigate the scale-freeness of neuronal avalanches \cite{beggsNeuronalAvalanchesNeocortical2003},
as a potential signature of the networks operating close to criticality.
A neuronal avalanche is defined as an uninterrupted cascade of spikes propagating through the network \cite{beggsNeuronalAvalanchesNeocortical2003}.
In a system operating close to criticality, the distribution of avalanche sizes (number of spikes in a cascade) follows a power law.
An event is an occurrence of at least 1 spike (among all neurons) within a small window of time.

Interestingly our analysis suggests that,
in the vicinity of the parameters that are optimized for efficient coding in the network %show scale-freeness in spiking activity
the distribution of avalanche sizes follow a power-law. 
When the noise amplitude is considerably lower or higher for efficient coding,
the network appears either super-critical or sub-critical, respectively
(see \nameref{cha:paper-safavi2020cribay} for more details). 
Certainly, this is only a preliminary step, but indeed, 
it might bring us a few  steps closer to bridging criticality and computational frameworks that complement the criticality.

%%% Local Variables:
%%% mode: latex
%%% TeX-master: "../phdThesis_csb"
%%% End:
 % Chapter 3
\chapter{Approaching through cognition} \label{cha:appr-thro-behav}

As motivated in \autoref{cha:brain-as-complex},
one of the apertures for approaching the brain as a complex system,
that let us remain close to the neuroscience side, is 
through behavior and cognition.
After providing a brief introduction to visual awareness and related phenomenon such as binocular rivalry, 
we argue that,
binocular rivalry is one of the important cognitive phenomenon,
that is particularly relevant for a complex system perspective toward the brain.
Based on this perspective toward binocular rivalry,
we suggest and conduct novel experimental works.
We study the phenomena of binocular rivalry on a mesoscopic scale which has not been done before.

\section{Visual awareness} %
Consciousness is one of the most challenging problems of science
\cite{christofQuestConsciousnessNeurobiological2004}.
However, during the last few decades, the vast technological and theoretical advancements brought consciousness research to an intense experimental phase. 
As a result, philosophical speculations on the nature and mechanisms of consciousness are slowly being replaced by empirical and theoretical approaches \cite{logothetisVisionWindowConsciousness2006,tononiNeuralCorrelatesConsciousness2008,christofkochConsciousnessConfessionsRomantic2012}.

There are various experimental paradigms in studying conscious\-ness.
We mention two example approaches and highlight our choice.
The first one is studying brain activity during various levels of conscious\-ness, \ie the differences between an awake, conscious state and various degrees of unconscious\-ness such as deep sleep, anesthesia, or coma.
The second one is studying how brain activity changes when a specific visual stimulus is subject\-ively perceived or supp\-ressed through experim\-ental paradigms like Binocular Rivalry (BR), Binocular Flash Suppression (BFS), masking etc. 

The first branch is about studying how brain activity changes in concert with changes in the overall level of consciousness, and indeed it is a fundamental approach. 
Nevertheless, it is extremely complex and it imposes a set of theoretical and experimental limitations. 
For example, it is technically difficult to monitor intracortical electrophysiological activity under conditions of coma. 
However, the second approach, \ie studying visual awareness 
(a "visual form of consciousness" \cite{crickVisualPerceptionRivalry1996a}), 
is an alternative approach to the problem with a more tractable framework,
especially at the neuronal level. 
In this approach, brain activity is monitored during changes in the \emph{content of} consciousness. 
For example, electrophysiological activity is studied when a visual stimulus becomes visible or invisible,
while everything else, including the overall level of consciousness as well as the sensory input, 
remains as constant as possible. 
Therefore, investigating various kinds of brain activity and their relation with the perception-related events ultimately might bring us steps closer toward an understanding of the neural mechanisms involved in visual awareness.

\subsection{Binocular rivalry}
One prominent example of such experimental paradigms that have been exhaustively exploited for
understanding the neural mechanisms involved in visual awareness is binocular
rivalry.
Binocular rivalry is one of the forms of ambiguous visual stimulation. 
It involves simultaneous stimulation of corresponding retinal locations
across the two eyes with incongruent visual stimuli.
It has been shown that different species experience this kind of ambiguous stimulation with some common characteristic \cite{carterPerceptualRivalryAnimal2020}.
When the subjects are presented with such visual stimuli, 
they typically experience fluctuations in perception between the two visual stimuli
(these fluctuations in perception are known as perceptual switches).

\subsection{Neural correlate of binocular rivalry}
In order to understand the neural correlate of phenomenon of binocular rivalry,
brain activity can be measured using various experimental methodologies at different scales. 
It can be spike trains from an individual neuron, field potentials or hemodynamic signals that reflect groups of neurons etc. 
Each measurement technique has its own limitations \cite{logothetisWhatWeCan2008}. 
For instance, non-invasive brain-imaging techniques are limited by their spatial and/or temporal resolution,
and electrophysiological recordings are limited in their coverage of cell populations.
Although all have their own limitations, 
they have provided us with a significant set of ideas about the neural mechanisms involved in conscious visual perception
that we briefly review in the following
(for detailed reviews, see for example \citet{blakeVisualCompetition2002a,tononiNeuralCorrelatesConsciousness2008,panagiotaropoulosSubjectiveVisualPerception2014a,kochNeuralCorrelatesConsciousness2016}).

Through single-unit recordings, we grasped a significant set of ideas and insights about the neural mechanisms underlying conscious visual perception on a local scale.
Specifically, through these studies, we learned that within each stage of visual hierarchy (from Lateral Geniculate Nucleus, V1 all the way to Profrontal Cortex (PFC)) 
there are a number of single units whose activity reflects the content of subjective perception of the animal.
The proportion of neurons which are modulated by the perception of the animal gradually increases across the visual hierarchy \cite{panagiotaropoulosSubjectiveVisualPerception2014a}.
From no modulated cell in Lateral Geniculate Nucleues (LGN) \cite{lehkyNoBinocularRivalry1996},
to superior temporal sulcus (STS) and inferotemporal cortex (IT) \cite{sheinbergRoleTemporalCortical1997},
and Lateral Prefrontal Cortex (LPFC) \cite{panagiotaropoulosNeuronalDischargesGamma2012,kapoorDecodingInternallyGenerated2022}
where 60-90\% of feature selective neurons are perceptually modulated.
But how does the activity of these distributed neurons relate to each other and also to other neurons (that are not involved in perception)?
How do they interact within their own population?
How is the activity of neuronal populations and large-scale networks organized, and how  are they related to perception-related events?
Single unit studies have potentially overlooked these important aspects of the underlying neural mechanisms,
Perhaps, such information is hidden in dynamic patterns of activity that are distributed over larger populations of neurons.

On the other side, imaging studies to some degree characterized the global network by revealing some specific large-scale interactions. 
For example, frequency-specific oscillatory interactions in the fronto-parieto-occipital \cite{hippOscillatorySynchronizationLargescale2011a}
and prefrontal-parietal networks \cite{doesburgRhythmsConsciousnessBinocular2009b}
and causal interactions in prefrontal-occipital \cite{imamogluChangesFunctionalConnectivity2012} network are involved in conscious perception.
However, these findings could not capture the \emph{neuronal} interactions due to their limited spatial and/or temporal resolution.
Indeed, such information is potentially available to multi-electrode recordings.

\section{Why is appealing from a complex system perspective} \label{sec:why-it-appealing}
An integrationist overview on the previous electrophysiology and imaging studies on the neural mechanisms involved in conscious visual perception implies that \emph{a global network of neuronal populations that interact with each other is involved in this phenomenon} \cite{blakeVisualCompetition2002a,panagiotaropoulosSubjectiveVisualPerception2014a}.
Therefore, visual awareness presumably is  a system property,
which is associated with a set of cooperative interactions within and between highly interconnected networks of neurons. 
These neurons are distributed within the entire thalamo-cortical system, 
mainly temporal, prefrontal,  occipital, parietal lobes and thalamus 
\cite{blakeVisualCompetition2002a,panagiotaropoulosSubjectiveVisualPerception2014a,wangBrainMechanismsSimple2013,lumerNeuralCorrelatesPerceptual1998,srinivasanIncreasedSynchronizationNeuromagnetic1999b,hippOscillatorySynchronizationLargescale2011a,doesburgRhythmsConsciousnessBinocular2009b,panagiotaropoulosNeuronalDischargesGamma2012,bahmaniNeuralCorrelatesBinocular2011,tononiNeuralCorrelatesConsciousness2008,kochNeuralCorrelatesConsciousness2016,imamogluChangesFunctionalConnectivity2012,safaviMultistabilityPerceptualValue2022}. 
The fact that, there is a large number of \emph{interacting} components (neurons and brain regions) involved in the phenomenon of  visual awareness,
is already one of the important characteristics that 
allows us to conceive perception as an \emph{emergent} property of a complex system.

Given this new conceptualization for visual awareness,
what are our  options to tackle it experimentally -- at least in terms of measuring the brain activity?
Almost all the previous studies of binocular rivalry
--in terms of spatial and temporal resolution-- 
are either single-unit recordings or whole-brain imaging (EEG/MEG, fMRI).
Such measurements can provide hints or evidence for the existence of such a distributed network (as indeed have been profoundly insightful), 
but they are not the most suitable  measurement techniques to characterize the \emph{neural interactions}
\footnote{
  With EEG/MEG and fMRI we can also characterize the interaction between the component of the neural system,
  but due to the nature of these measurement techniques, the picture they can provide about neural interactions is more ambiguous compare to what we can get from invasive recording techniques}.
Understanding the \emph{interaction} between units of a complex system is the key for characterizing collective behaviors and therefore it is important to observe the system at scales which give the clearest picture in this regard.
At first glance, we can realize that the phenomenon of binocular rivalry is poorly understood at the mesoscopic scale, 
which could not only reveal the phenomenon of coordinated activity 
within areas but also across areas in large-scale networks 
(see \autoref{sec:necess-invest-across}).
Therefore, a complex system perspective motivates observation at the mesoscopic scale as the first priority
and therefore motivates new experiments.
Studying at this scale, not only can inform about the involved cooperative mechanisms,
but also, it is the first step for bridging the studies based on single-unit recordings and imaging studies.

Conceiving perception as a system property or an emergent property
resulting from interactions within a large and distributed network of neurons, 
is not the only reason for the glamour of binocular rivalry from a complex system perspective.
Indeed, various models based on the theory of the dynamical system
(which is one of the most powerful frameworks to formalize a complex system)
can explain a range of characteristics of bistable perception (such as the distribution of dominance periods)
\cite{ditzingerOscillationsPerceptionAmbiguous1989b,braunAttractorsNoiseTwin2010,theodoniCorticalMicrocircuitDynamics2011a,pastukhovMultistablePerceptionBalances2013a}.
Perhaps, the most  appealing theoretical explanation is provided by \citet{pastukhovMultistablePerceptionBalances2013a}
that showed a network model operating on the edge of a bifurcation and can explain statistical characteristics of a wide range of multi-stable phenomenon.

Overall, based on available empirical and theoretical evidence we know, we need to deal with a large and distributed network of neurons;
Components of this network interact in a non-trivial way;
Phenomenon of binocular rivalry seems to be inherently multi-scale;
It seems, a neural network operating on an edge of bifurcation can explain various behavior-related statistical properties of the phenomena. 
Altogether, these finding make this phenomenon appealing from the perspective of complex systems.
We believe one of the very first steps for understating the cooperative neural mechanism pertaining to
binocular rivalry is \emph{measuring the mesoscopic neural activity},
\ie new experiments are needed which is the focus of the next sections.

\section{Experimental considerations} \label{sec:exper-cons}
In the previous section (\autoref{sec:why-it-appealing}) we argued that
meso-scale observations are necessary for understating the binocular rivalry  and
consequently, conducting new experiments are needed.
For conducting the  experimental work pertaining to binocular rivalry,
in addition to considerations pertaining to the level of observation,
some basic factors need to be considered as well.
These factors are briefly discussed in this section.

The first consideration pertains the recording area.
One of the target regions for new experiments is PFC for multiple reasons.
First, 
PFC is a central subnetwork (in a graph-theoretic sense)  \citep{modhaNetworkArchitectureLongdistance2010} that play a crucial role in cognitive computations \citep{millerIntegrativeTheoryPrefrontal2001},
especially due to an increase in the integrative aspect of information processing in higher-order cortical areas.
Second, ventro-lateral PFC (vlPFC), is reciprocally connected to Inferior Temporal (IT) coretex,
which contains the largest proportion of neurons that are perceptually modulated \citep{sheinbergRoleTemporalCortical1997} and neurons in PFC have been also shown to be perceptually modulated in similar tasks \citep{panagiotaropoulosNeuronalDischargesGamma2012,hesseNewNoreportParadigm2020}.
Third, PFC is outside of the core visual hierarchy.

For recording from PFC, we also need to be cautious with experimental design,
due to the ambiguous role of PFC in perception.
In a study by \citet{frassleBinocularRivalryFrontal2014},
it was suggested that ``frontal areas are associated with active report and introspection rather than with rivalry per se.''.
In \citet{safaviFrontalLobeInvolved2014} (\seealso, \nameref{cha:paper-safavi2014}), based on a broad set of evidence,
we argue that evidence provided by \citet{frassleBinocularRivalryFrontal2014} is not sufficient for this conclusion, and understating the role of PFC in visual awareness needs further investigation.
Due to potential confounding in activity of PFC that can happen due to behavioral report,
we needed to employ a no-report paradigm (decoding the perception of the animal using optokinetic nystagmus (OKN) responses \citep{leopoldMeasuringSubjectiveVisual2003}).

In this experiment, we particularly  needed to have the responses of neurons whose activities are modulated by features of a presented visual stimulus, 
and the visual stimulus had to induce OKN responses
(a certain pattern of eye movement in response to moving stimuli such as moving grating).
At the same time, as the core idea was monitoring the activity of neural population,
the recording had to be performed with Utah array 
(10 x 10 array of electrodes that need to be implanted chronically).
In contrast to previous similar experiments (\eg see \citet{panagiotaropoulosNeuronalDischargesGamma2012}) that used non-chronic recording with tetrods where the experimenter could explore to find the neuron by moving the electrodes,
Utah arrays are fixed and almost permanent.
In \citet{safaviNonmonotonicSpatialStructure2018} and \citet{kapoorDecodingInternallyGenerated2022}
(\seealsos, \nameref{cha:paper-safavi2018} and \nameref{cha:paper-kapoor2020})
we reported that such neurons are accessible with this recording technique (recording with Utah arrays) and under our experimental design.
Additionally, we also found that, similarly tuned neurons in this region of PFC
are correlated in large distances \cite{safaviNonmonotonicSpatialStructure2018} in contrast to most of sensory cortices
\cite{rothschildFunctionalOrganizationPopulation2010,cohenMeasuringInterpretingNeuronal2011,smithSpatialTemporalScales2008a,smithSpatialTemporalScales2013a,denmanStructurePairwiseCorrelation2014} (but also see \cite{rosenbaumSpatialStructureCorrelated2017}).
Interestingly, we also found that spatial structure of functional connectivity in ventro-lateral PFC is generally
\footnote{By generally, it is meant in presence and absence of visual stimulation, in awake and anesthetized state of the animal.}
different from  most sensory cortices.
In most  sensory cortices, noise correlation decay monotonically as a function of distance,
nevertheless, in ventro-lateral PFC we observed in both anesthetized and awake monkeys noise correlation rises again after an initial decay.
This observation is also compatible with anatomical differences between PFC and sensory areas
\cite{levittTopographyPyramidalNeuron1993,amirCorticalHierarchyReflected1993,lundComparisonIntrinsicConnectivity1993,kritzerIntrinsicCircuitOrganization1995,fujitaIntrinsicConnectionsMacaque1996,tanigawaOrganizationHorizontalAxons2005}.
The finding on the spatial structure of noise correlation in vlPFC was not relevant for the binocular rivalry experiment as the spatial structures were not take into account,
nevertheless, it was an important finding of the circuitry of PFC.

\section{Toward a meso-scale understanding}
The very first question that can be approached based on a mesosopic-level investigation,
is  what can population dynamics  reflect about the content of conscious perception.
Second question is what can we learn about the involved neural mechanism from micro-meso relationships in PFC.
Notably, both questions are approachable when we have observed the system in a mesoscopic scale
(level of neural populations), and are briefly discussed in the next sections (and associated papers).

\subsection{Meso-scale dynamics} %

The activity of the majority of PFC neurons that are responsive to visual attributes of sensory input
are correlated with conscious perception of animals as well.
In our case, we used vertically moving grating  -- upward or downward as stimuli \cite{safaviNonmonotonicSpatialStructure2018,kapoorDecodingInternallyGenerated2022}
(\seealsos, \nameref{cha:paper-safavi2018} and \nameref{cha:paper-kapoor2020})
and previously it was shown this is the case for face-selective neurons as well \cite{panagiotaropoulosNeuronalDischargesGamma2012}.
But additionally, the content of conscious perception is decodable from the spiking activity of neural \emph{populations} in ventro-lateral PFC.
This is the first confirmation of informativity of the meso-scale observation or measurement of the neural activity.
The next steps should focus on characterizing the coordinated dynamics and neural interactions 
(see the next section and the \autoref{part:outlook} for further elaboration on the next steps).

\subsection{Micro-Meso relationship} % 
Given the empirical evidence on the informativeness of population spiking of PFC neurons,
more specifically the fact that they reflect the content of conscious perception,
it is justified to consider more intricate aspects of mesoscopic dynamics.
Such aspect of mesoscopic dynamics includes signatures of neural coordination such as neural oscillation and spike-LFP relationship 
(also see \autoref{cha:appr-thro-nda} important aspect of neural coordination).
Furthermore, investigating the relationship between PFC [presumed] state fluctuations conjectured based on LFP oscillatory dynamics,
perceptual switches and spiking activity can hint at another aspect of the putative role of neural interactions in binocular rivalry.
Indeed, one of the important findings of our study was that,
spiking activity of population reflecting the dominant perception,
are coupled (relatively stronger than suppressed population) to LFP in range $25-45$ $Hz$ after the perceptual switch \cite{dwarakanathBistabilityPrefrontalStates2023}
(\seealso, \nameref{cha:paper-dwarakanath2020}).

This strong spike-LFP coupling can be a hint for an emphasized communication (or interaction) of PFC populations reflecting the conscious perception and other brain regions
(see \citet{buzsakiWhatDoesGamma2015} for the interpretation of spike-LFP coupling as a quantity to characterize the communication channel).
Further investigation is needed to characterize the interaction and functional role of this putative communication.
In particular, multiple experimental evidence should be taken into account for interpreting the functional role of the mentioned neuronal interaction.
First, we know that neural populations that monitor task-related activity exist in the same region of PFC in the absence of any behavioral report \citep{kapoorParallelFunctionallySegregated2018},
which is important given than various studies argue that PFC is strongly involved in task monitoring \cite{frassleBinocularRivalryFrontal2014}.
Second, we know that the activity of neural populations in IT cortex is also correlated with perception in the absence of behavioral reports \cite{hesseNewNoreportParadigm2020}.
On the other side, from studies with causal intervention, we know that the activity of PFC is needed for difficult object recognition tasks \cite{karFastRecurrentProcessing2020}.
Therefore, IT cortex might be a crucial component in this communication circuit and needed to be clarified in future studies.

%%% Local Variables:
%%% mode: latex
%%% TeX-master: "../phdThesis_csb"
%%% End:
 % Chapter 4

\cleardoublepage % Empty page before the start of the next part
% brain as a complex system
% 

% ------------------------------------------------
% Manuscripts

\ctparttext{
  In this part of the thesis,
  information of all manuscripts associated to this thesis is provided,
  which includes the title, list of authors, status of the manuscript
  and statement of contributions.
  For statement of contributions, the standard CRediT taxonomy \cite{brandAuthorshipAttributionContribution2015} has been used when it was available either in the published manuscript or its publicly available preprint,
  otherwise the ``author contributions'' stated in the published manuscript or its publicly available preprint has been used. 
  A summary --with emphasis on the relevant aspects to this thesis-- for each manuscript is provided as well.
  Summaries are written such that, redundancies between manuscripts are minimal.
  Furthermore, The reader is refereed to other relevant summaries or chapters of the synopsis (\autoref{part:synopsis}).
  Therefore,
  summaries remain brief and at the same time, convey the coherent picture of this thesis.
  Summaries are ordered such that earlier summaries provide backgrounds and foundations for later ones,
  making it possible to be more concise as we progress through them.
} % Text on the Part 2 page describing the content in Part 2

%%% Local Variables:
%%% mode: latex
%%% TeX-master: "../phdThesis_csb"
%%% End:

\part{Manuscripts Information}
\label{part:manuscr-inform}
% order of appearnace
\chapter{Paper \rom{1}} 
\label{cha:paper-5}
\section*{Paper information} % 

\begin{description}
\item[Title:]
  From univariate to multivariate coupling between continuous signals and point processes: A mathematical framework
\item[Authors:]
  Shervin Safavi, Nikos K. Logothetis, Michel Besserve
\item[Status:] 
  Published in Neural Computation, see \citet{safaviUnivariateMultivariateCoupling2021}
\item[Presentation at scientific meetings:]
  NeurIPS 2019 Workshop: Learning with Temporal Point Processes
  \cite{safaviMultivariateCouplingEstimation2019},
  Bernstein 2021 \cite{safaviGeneralizedPhaseLocking2021}  
\item[Author contributions: ]

  Conceptualization, S.S., and M.B.;
  Methodology, S.S., and M.B.;
  Software, S.S. and M.B.;
  Formal Analysis, S.S., and M.B.;
  Investigation, S.S., and M.B.;
  Resources, N.K.L.;
  Data Curation, S.S., and M.B.;
  Writing -- Original Draft, S.S., and M.B.;
  Writing -- Review \& Editing: S.S., M.B., and N.K.L.;
  Visualization, S.S., and M.B.;
  Supervision and Project administration, M.B.;
  Funding acquisition, N.K.L.
  
\end{description}

%%% Local Variables:
%%% mode: latex
%%% TeX-master: "../phdThesis_csb"
%%% End:

\section*{Summary} %
\subsection*{Motivation}
In various complex systems,
we deal with highly multi-variate temporal point processes,
that are corresponding to the activity of a large number of individuals.
They can be generated by the activity of neurons in brain networks \cite{johnsonPointProcessModels1996},
such as neurons' action potentials,
or by members in social networks \citep{daiRecurrentCoevolutionaryLatent2016,deLearningForecastingOpinion2016},
such as tweets in the Twitter network.
In practice, a limited number of events per unit are accessible experimentally or observable
(for instance numbers of spikes generated by neurons).
With such limitations, inferring the underlying dynamical properties of the studied system becomes challenging.
Nevertheless, in many cases, exploiting the coupling between the point processes and aggregate measure of the complex system (such as Local Field Potentials as an aggregate measure of population neural activity) can be insightful for understanding the underlying dynamics.

Meaningful and reliable estimates of coupling between such signals can be crucial for understanding many complex systems.
However, the statistical properties of many methods classically used remain poorly understood.
As a consequence, statistical assessment in practice largely relies on heuristics (\eg permutation tests).
While such approaches often make intuitive sense, they are computationally expensive and may be biased by properties of the data that are unaccounted for.
This is particularly relevant for quantities involving point processes and high-dimensional data, which have largely non-intuitive statistical properties, and yet are key tools for experimentalists and data analysts.
In this study, we establish a principled framework for statistical analysis of coupling between multi-variate point process and continuous signal.

\subsection*{Material and Methods}
First, we derive analytically the asymptotic distribution for a class of coupling statistics that quantify the correlation between a point process and a continuous signal.
The key to this theoretical prediction is expressing coupling statistics as stochastic integrals. 
Indeed, a general family of coupling measures can be expressed as stochastic integrals. 
The Martingale Central Limit Theorem allows us to derive analytically the asymptotic Gaussian distribution of such coupling measures. 
We show that these coupling statistics follow a Gaussian distribution. 
A commonly used example of such coupling statistics is Phase Locking Value (PLV) which typically is used for quantifying spike-LFP coupling in neuroscience.

We then go beyond uni-variate coupling measures and analyze the statistical properties of a family of multi-variate coupling measures taking the form of a matrix with stochastic integral coefficients.
We characterize the joint Gaussian asymptotic distribution of matrix coefficients,
and exploit Random Matrix Theory (RMT) principles to show that,
after appropriate normalization,
the spectral distribution of such large matrices under the null hypothesis
(absence of coupling between the point process and continuous signals),
follows approximately the Marchenko-Pastur law \citep{marcenkoDistributionEigenvaluesSets1967a}
\footnote{Referred paper \cite{marcenkoDistributionEigenvaluesSets1967a},
  is not written in English, but it is  the original publication.
  Reader can refer to \citet[Chapter
  2]{andersonIntroductionRandomMatrices2010} instead.}
(which is a well-characterized distribution in Random Matrix Theory),
while the magnitude of the largest singular value converges to a fixed value whose simple analytic expression depends only on the shape of the matrix.

\subsection*{Results}
We derive analytically the asymptotic distribution of Phase-Locking Value (PLV)
which is a coupling statistic conventionally used for quantifying the relationship between a pair of a point process (like spikes) and an oscillatory continuous signal (like LFPs).
We show that PLVs follow a Gaussian distribution with calculable mean and variance.

Based on the multi-variate extension,
we show how this result provides a fast and principled procedure to detect significant singular values of the coupling matrix, reflecting an actual dependency between the underlying signals.
This is of paramount importance for the analysis of empirical data given the ever-increasing dimensionality of datasets that need computationally efficient statistical tests.

\subsection*{Conclusion}
Our results not only construct a theoretical framework, which is valuable on its own
but also can have various applications for neural data analysis and beyond.
For instance, based on our theoretical framework we note realistic scenarios where the PLV can be a biased estimator of spike-LFP coupling, and in light of our framework, such biases can be treated.

%%% Local Variables:
%%% mode: latex
%%% TeX-master: "../phdThesis_csb"
%%% End:

     % 1
\chapter{Paper \rom{2}}\label{cha:paper-gpla}
\section*{Paper information} % 
\begin{description}
\item[Title:] Uncovering the organization of neural circuits with generalized phase locking analysis
\item[Authors:]
  Shervin Safavi,
  Theofanis I. Panagiotaropoulos,
  Vishal Kapoor,
  Juan F. Ramirez-Villegas,
  Nikos K. Logothetis,
  Michel Besserve
\item[Status:]
  Preprint is available online, see \citet{safaviUncoveringOrganizationNeural2020a}
\item[Presentation at scientific meetings:]
  ESI-SyNC 2017 \cite{safaviGeneralizedPhaseLocking2017},
  AREADNE 2018 \cite{safaviGeneralizedPhaseLocking2018},
  Cosyne 2019 \cite{besserveGeneralizedPhaseLocking2019},
  Cosyne 2020 \cite{safaviUncoveringOrganizationNeural2020b},
  Bernstein 2021 \cite{safaviGeneralizedPhaseLocking2021}
\item[Author contributions: ]

  Conceptualization, S.S., T.I.P., M.B.;
  Methodology, S.S., J.F.R.-V. and M.B.;
  Software, S.S. and M.B.;
  Formal Analysis, S.S. and M.B.;
  Investigation, S.S., T.I.P., V.K. and M.B.;
  Resources, N.K.L.;
  Data Curation, S.S., T.I.P., V.K., and M.B.;
  Writing -- Original Draft, S.S. and M.B.;
  Writing -- Review \& Editing: S.S., T.I.P., V.K., J.F.R.-V., N.K.L. and M.B.;
  Visualization, S.S. and M.B.;
  Supervision and Project administration, T.I.P. and M.B.;
  Funding acquisition, N.K.L.

\end{description}

%%% Local Variables:
%%% mode: latex
%%% TeX-master: "../phdThesis_csb"
%%% End:

\section*{Summary} %
\subsection*{Motivation}

The synchronization between spiking activity and the phase of particular rhythms of LFP has been suggested as an important marker to reason about the underlying cooperative network mechanisms; 
nevertheless, there is not yet a systematic way to extract concise coupling information from the largely multi-variate data available in current recording techniques.
We introduce Generalized Phase Locking Analysis (GPLA) which is a multi-variate extension of phase-locking analysis.
Phase-locking analysis is a common uni-variate method of quantifying the spike-LFP relationship.
With GPLA, we can quantify, characterize and statistically assess the interactions between pop\-ulation-level spiking activity and mesoscopic network dynamics
(such as global oscillations and traveling waves).

\subsection*{Material and Methods}

We collect the coupling information between spikes and LFP in a coupling matrix.
The coupling matrix, constructed by all the pairwise complex-value spike-field coupling coefficients, 
represents the population-level spiking activity and all LFP channels.
We use Singular Value Decomposition (SVD) to provide a low-rank representation of the coupling matrix.
Therefore, we summarize the information of the coupling matrix with the largest singular value and the corresponding singular vectors. 
Singular vectors represent the dominant LFP and spiking patterns and the singular value, called generalized Phase Locking Value (gPLV), characterizes the strength of the coupling between LFP and spike patterns.

We further investigate the statistical properties of the gPLV and develop an empirical and theoretical statistical testing framework for assessing the significance of the coupling measure gPLV.
For the empirical test, we synthesize surrogate data with spike jittering for the generation of the null hypothesis and use it to estimate the p-value for the gPLV calculated from the data.
For the theoretical test, we used Martingale theory and \cite{aalenSurvivalEventHistory2008}
Random Matrix Theory (RMT) \citep{andersonIntroductionRandomMatrices2010}
to approximate the distribution of singular values under the null hypothesis (see \citet{safaviUnivariateMultivariateCoupling2020} for the details and \autoref{cha:paper-5} for a summary).
This allows us to derive a computationally efficient significance test in comparison to the empirical one.

\subsection*{Results}
Firstly, if both GPLA and its uni-variate counterpart are applicable,
GPLA is superior as it can extract a more reliable  coupling structure in the presence
of an excessive amount of noise in LFP.
Furthermore, to demonstrate the capability of GPLA for mechanistic
interpretation of the neural data,
we apply GPLA to various simulated and experimental data.
Application of GPLA on simulation of hippocampal
Sharp-Wave-Ripples (SWR) can reveal various characteristics of hippocampal circuitry with minimal prior knowledge.
For instance, with GPLA we can show CA1 and CA3 neurons are all coupled to the field activity in the gamma and ripple band
(in line with experimental and simulation results 
\citep{buzsakiHighfrequencyNetworkOscillation1992,ramirez-villegasDissectingSynapseFrequencyDependent2018}),
suggesting this rhythm may support communication between CA1 and CA3 sub-fields during memory trace replay. 
Furthermore, it also allows us to tease apart the involved populations based on the label-free spike timing and LFP.
GPLA can also provide hints on the propagation of activity between the populations (propagation from CA3 to CA1).
Application of the method on the experimental recordings from monkey PFC suggests a \emph{global} coupling between spiking activity and LFP traveling waves in this region of PFC.
Overall, exploiting the phase distributions across space and frequencies captured by GPLA combined with neural field modeling help to untangle the contribution of inhibitory and excitatory recurrent interactions to the observed spatio-temporal dynamics.

\subsection*{Conclusion}
GPLA is a multi-variate method to quantify, characterize and statistically assess the interactions between population-level spiking activity and mesoscopic network dynamics such as global oscillations, traveling waves, and transient neural events.
Spike and LFP vectors compactly represent the dominant LFP and spiking patterns and  generalized Phase Locking Value (gPLV),
characterizes the strength of the coupling between LFP and spike patterns.
Our theoretical statistical testing framework allows a computationally efficient assessment of the significance of coupling measure gPLV.
This is of paramount importance for neural data analysis given the ever-increasing dimensionality of modern recording techniques that need computationally efficient statistical tests.

%%% Local Variables:
%%% mode: latex
%%% TeX-master: "../phdThesis_csb"
%%% End:

   % 2
\chapter{Paper \rom{3}}\label{cha:paper-besserve2020ned}
\section*{Paper information} % 

\begin{description}
\item[Title:] The complex spectral structure of transient LFPs reveals subtle aspects of network coordination across scales and structures
\item[Authors:]
  Michel Besserve, Shervin Safavi, Bernhard Sch\"olkopf, Nikos Logothetis
\item[Status:]
  Work-in-progress; a preliminary  manuscript is available in the appendix,
  see \hyperref[pdf:besserve2020ned]{Paper 3}.
\item[Presentation at scientific meetings:]
  Machine Learning Summer School \cite{besservePracticalMachineLearning2016}
\item[Author contributions: ]

  Conceptualization, M.B. and N.K.L;
  Methodology, M.B. and S.S.;
  Software, S.S. and M.B.;
  Formal Analysis, M.B.;
  Investigation, S.S. and M.B.;
  Resources, B.S. and N.K.L.;
  Data Curation, M.B. and N.K.L;
  Writing - Original Draft, M.B. and S.S.;
  Writing - Review \& Editing: M.B., S.S., B.S. and N.K.L;
  Visualization, M.B. and S.S.;
  Supervision and Project administration,  M.B.;
  Funding acquisition, B.S. and N.K.L.

\end{description}

%%% Local Variables:
%%% mode: latex
%%% TeX-master: "../phdThesis_csb"
%%% End:

\section*{Summary} %
\subsection*{Motivation}
LFPs are intermediary signals, and as such, they
reflect a mesoscopic picture of the brain dynamics \cite{liljenstroemMesoscopicBrainDynamics2012}.
As LFPs are rich signals \cite{buzsakiOriginExtracellularFields2012,liljenstroemMesoscopicBrainDynamics2012,einevollModellingAnalysisLocal2013},
they can be a pivotal point for bringing the brain dynamics at different scales together. 
In particular, certain transient activities of LFPs reflect cooperative dynamics (we call them \emph{neural events}).
A prominent example of such neural events are sharp wave-ripples (SWRs),
and it has been observed they co-occur with well-coordinated activity at smaller scales (neurons and populations of neurons) \cite{csicsvariEnsemblePatternsHippocampal2000,csicsvariEnsemblePatternsHippocampal2000,olivaRoleHippocampalCA22016},
as well as larger scale (entire brain) \cite{logothetisHippocampalCorticalInteraction2012,karimiabadchiSpatiotemporalPatternsNeocortical2020}.
In spite of the importance of such characteristic neural activities (neural events),
there are not many principled methods for identifying them in a single channel LFP.
We introduce a principled method for identifying neural events in a single channel LFP.

\subsection*{Material and Methods}
We detect the neural events by isolating transient characteristic neural activities.
We first compute the spectrograms of the LFP signals by applying short-term Fourier transform (STFT) on LFPs in order to exploit the spectral content of the LFPs.
To identify the frequent transient neural activity with similar spectral content we apply non-negative Matrix Factorization (NMF).
Notably, due to scale-invariant nature of LFPs (similar to other extracellular field potential \cite{buzsakiOriginExtracellularFields2012}) \cite{freemanScalefreeNeocorticalDynamics2007a,heScalefreeBrainActivity2014},
we used Itakura-Saito divergence in the optimization procedure of NMF \cite{fevotteNonnegativeMatrixFactorization2009}
in order to avoid under-weighting of high-frequency components due to their low power in the spectrum.
The components result from NMF, provide the information on the spectral content of the neural events.
In order to temporally isolate the neural events and characterize their temporal profile, 
we apply a shift-invariant dictionary learning
(a modified version of dictionary learning provided by \citet{mailheShiftinvariantDictionaryLearning2008a}).
The latter step, allows us to temporally locate the neural events and also identify the time-domain profiles of events that their spectral content are characterized by the NMF step.

We demonstrate the capability of our method by identifying neural events and their brain-wide signatures in Hippocampus and LGN recorded from anesthetized monkeys.
Furthermore, in order to demonstrate that neural events have the potential of  relating the meso-scale dynamics even to cellular dynamics,
we investigate the neural events in the simulation of thalamocortical circuitry  developed by \citet{costaThalamocorticalNeuralMass2016} where allow us to access both meso-scale dynamics and also some level of cellular dynamics.
The simulation consists of neural mass models with two modules,
one for the thalamus and one for the cortex, and mimics the behavior of these circuits during different stages of sleep.

\subsection*{Results}
We developed a novel methodology for detecting neural events (transient cooperative neural activities) such as sharp wave-ripples.
With our method, neural events can be detected with minimal prior knowledge about the structure under study.
Namely, the spectral content is automatically identified by the method,
and various other attributes of neural events such as the number of neural event clusters  can also be identified by the method in an unsupervised fashion.

Furthermore, we demonstrate the capability of the method by identifying neural events in Hippocampus and LGN and also explore their brain-wide \emph{macro-scale} signatures using concurrent fMRI recordings from anesthetized monkeys.
The results suggest that similar to the previous study of \citet{logothetisHippocampalCorticalInteraction2012} that was focused on sharp wave-ripples,
the identified events in Hippocampus and LGN reflect a large-scale coordinated dynamics.
Indeed, this demonstrates the insightfulness of neural events for bridging the meso-scale and macro-scale brain dynamics.

Our results also suggest that neural events can be insightful for establishing a bridge between meso-scale and micro-scale brain dynamics, even at the cellular level.
We demonstrate this aspect, by investigating a simulation of the thalamocortical system developed by \citet{costaThalamocorticalNeuralMass2016}.
With our methodology, we identified different kinds of spindles in the activity of the thalamus module of the simulation,
and demonstrate that different events co-occur with characteristic activity patterns in the cellular variables (such as membrane potentials and ionic currents) of the simulation.

\subsection*{Conclusion}

With this method, we can find characteristic patterns of LFPs in an unsupervised fashion.
This methodology not only allows us to detect well established neural events such as SWRs in a principled fashion,
it also identifies characteristic patterns in a single channel LFP that have not been explored, and they can be insightful about cooperative and multi-scale dynamics of the brain.
Such patterns are potentially very special in the sense that,
they provide us a time window at which meso-scale dynamics are closely related to micro- and macro-scale dynamics.
In fact, as pointed out in \autoref{sec:necess-invest-across} and \autoref{sec:need-new-tools}, this is of paramount importance for bridging the scales of neural dynamics,
in particular when combined with GPLA introduced in \nameref{cha:paper-gpla} and NET-fMRI \cite{logothetisNeuralEventTriggeredFMRILargescale2014}.

%%% Local Variables:
%%% mode: latex
%%% TeX-master: "../phdThesis_csb"
%%% End:

     % 3
\chapter{Paper \rom{4}}\label{cha:paper-safavi2020cribay}
\section*{Paper information} % 

\begin{description}
\item[Title:]
  Signatures of criticality in efficient coding networks
\item[Authors:]
  Shervin Safavi,
  Matthew Chalk,
  Nikos K. Logothetis,
  Anna Levina

\item[Status:]
  Work-in-progress; a preliminary  manuscript is available in the appendix,
  see \hyperref[pdf:safavi2020sce]{Paper 4}.
\item[Presentation at scientific meetings:]
  Conference on Complex Systems (CCS 2018) Satellite: Complexity from Cells to Consciousness: Free Energy, Integrated Information, and Epsilon Machines
  \cite{safaviOptimalEfficientCoding2018},
  DPG-Fr\"uhjahrstagung 2019
  \cite{safaviSignaturesCriticalityEfficient2019a},
  Cosyne 2020
  \cite{levinaSignaturesCriticalityObserved2020a}  
\item[Author contributions: ]

  Conceptualization, S.S., and A.L.;
  Methodology, S.S., M.C., A.L.;
  Software, S.S. and M.C;
  Formal Analysis, S.S., M.C and A.L.;
  Investigation, S.S., M.C and A.L.; 
  Resources, N.K.L. and A.L.;
  Data Curation, S.S., M.C and A.L.;
  Writing -- Original Draft, S.S.;
  Writing -- Review \& Editing, not applicable
  (this letter has not been communicated with other co-authors so far);
  Visualization, S.S.;
  Supervision and Project administration, A.L.;
  Funding acquisition, N.K.L. and A.L.
\end{description}

%%% Local Variables:
%%% mode: latex
%%% TeX-master: "../phdThesis_csb"
%%% End:

\section*{Summary} %
\subsection*{Motivation}

Understanding the computations that the brain needs to implement (neural computation)
and the dynamics of the brain activity (neural dynamics) are two important goals of computational neuroscience \cite[Chapter 1]{churchlandComputationalBrain1992}.
Ideally, we need a framework that can accommodate both aspects of the brain in one framework
\cite{churchlandComputationalBrain1992,eurichNeuralDynamicsNeural2003}.
Nevertheless, to the best of my knowledge, no framework has been developed to satisfy this important need.

An intermediate step toward developing such a framework is exploiting the frameworks and models that are either centered around neural computation or neural dynamics \emph{with implications for the other aspect}.
Indeed, there are normative models that have implications for neural dynamics
\cite{lengyelMatchingStorageRecall2005,deneveBayesianSpikingNeurons2008a,deneveBayesianSpikingNeurons2008,tanakaRecurrentInfomaxGenerates2008,buesingNeuralDynamicsSampling2011,boerlinPredictiveCodingDynamical2013,billDistributedBayesianComputation2015,chalkNeuralOscillationsSignature2016,zeldenrustEfficientRobustCoding2019,echevesteCorticallikeDynamicsRecurrent2020}
and also models of neural dynamics with implications for neural computation
\cite{bertschingerRealtimeComputationEdge2004,eliasmithUnifiedApproachBuilding2005,sussilloNeuralCircuitsComputational2014,hidalgoInformationbasedFitnessEmergence2014a,shrikiOptimalInformationRepresentation2016,maassSearchingPrinciplesBrain2016,kimLearningRecurrentDynamics2018,chenComputingModulatingSpontaneous2019,michielsvankessenichPatternRecognitionNeuronal2019,finlinsonOptimalControlExcitable2020}
We suggest seeking for ``bridges'' between such frameworks can be a first step.
Neural coding is of particular interest for building such bridges
as there have been various studies that suggest potential connections between neural coding and neural dynamics 
\cite{ermentroutRelatingNeuralDynamics2007c,boerlinPredictiveCodingDynamical2013,shrikiOptimalInformationRepresentation2016,chalkNeuralOscillationsSignature2016,alamiaAlphaOscillationsTraveling2019,kadmonPredictiveCodingBalanced2020,roethEfficientPopulationCoding2020,echevesteCorticallikeDynamicsRecurrent2020}.
In particular, multiple recent studies provide qualitative or quantitative evidence on the usefulness of operating close to a phase transition for coding
\cite{shrikiOptimalInformationRepresentation2016,chalkNeuralOscillationsSignature2016,kadmonPredictiveCodingBalanced2020,roethEfficientPopulationCoding2020}.
Interestingly, the phase transition is also one of the pillars of the criticality hypothesis of the brain
\cite{munozColloquiumCriticalityDynamical2018,tkacikInformationProcessingLiving2016,moraAreBiologicalSystems2011a}.
In spite of this apparent and exciting connection,
networks implementing neural coding have never been investigated for signatures of criticality.
In this study, we investigate networks that can be optimized for neural coding for signatures of criticality.

\subsection*{Material and Methods}

In this study, we investigate a network of Leaky-Integrate and Fire (LIF) neurons whose connectivity and dynamics can be optimized for coding a one-dimensional sensory input \cite{chalkNeuralOscillationsSignature2016}.
This network can be optimized to encode the input efficiently
(\ie with a minimal number of spikes) and accurately (\ie with minimal reconstruction error).
The input is reconstructed by performing a linear readout of spike trains
(see \cite{boerlinPredictiveCodingDynamical2013}).
Given an idealized network with instantaneous synapses, the optimal network could be derived analytically from first principles \cite{boerlinPredictiveCodingDynamical2013}.
In this case, neurons that receive a common input avoid communicating redundant information via instantaneous recurrent inhibition.
However, adding realistic synaptic delays leads to network synchronization, which impairs coding efficiency.
\citet{chalkNeuralOscillationsSignature2016} demonstrated that, in the presence of synaptic delays, a network of LIF neurons can nonetheless be optimized for efficient coding by adding noise to the network.
The network's performance depends non-monotonically on the noise amplitude,
with the optimal performance achieved for an intermediate noise level. 
We investigate potential signatures of criticality such as the scale-freeness of neuronal avalanches \cite{beggsNeuronalAvalanchesNeocortical2003} in the spiking activity of the network.

\subsection*{Results}

In this study, we introduce a new approach to better connect neural dynamics and neural computation.
Here we search for a potential connection between models of neural dynamics with implications on neural computation,
and normative models of neural computation with implications for neural dynamics.
We search for signatures of criticality in neuronal networks that can be optimized based on objectives of efficient coding. 
We investigate
efficient coding networks for signatures of criticality.
Interestingly, almost exclusively in the optimized network, we observe the signatures of criticality
and when the noise amplitude is too low or too high for efficient coding, the network appears either super-critical or sub-critical, respectively.
In both cases, the noise level that was optimal for coding also resulted in a scale-free avalanche behavior.

\subsection*{Conclusion}

Our results suggest that coding-based optimality might co-occur with closeness to criticality.
This result has important implications, as it shows how two influential,
and previously disparate fields --- efficient coding, and criticality --- might be intimately related.
This work proposes several promising avenues for future research on the computation and dynamics of the neural system.

%%% Local Variables:
%%% mode: latex
%%% TeX-master: "../phdThesis_csb"
%%% End:

 % 4
\chapter{Paper \rom{5}} \label{cha:paper-safavi2014}
\section*{Paper information} % 

\begin{description}
\item[Title:]
  Is the frontal lobe involved in conscious perception?
\item[Authors:]
  Shervin Safavi$^*$,
  Vishal Kapoor$^*$,
  Nikos K. Logothetis,
  Theofanis I. Panagiotaropoulos
  ($^*$ indicate equal contribution) 
\item[Status:]
  Published in Frontiers in Psychology, see \citet{safaviFrontalLobeInvolved2014}
\item[Author contributions: ]
  Conceptualization, S.S., V.K., N.K.L. and T.I.P.; 
  Methodology, not applicable;
  Software, not applicable;
  Formal Analysis, not applicable;
  Investigation, S.S., V.K. and T.I.P.; 
  Resources, N.K.L.;
  Data Curation, not applicable;
  Writing -- Original Draft, S.S., V.K. and T.I.P.; 
  Writing -- Review \& Editing, S.S., V.K., N.K.L. and T.I.P.; 
  Visualization, not applicable;
  Supervision and Project administration, T.I.P.;
  Funding acquisition, N.K.L.
\end{description}

%%% Local Variables:
%%% mode: latex
%%% TeX-master: "../phdThesis_csb"
%%% End:

\section*{Summary}

PFC as part of the subsystem that serves the goal-directed character of behavior 
\cite{logothetisStudiesLargeScaleNetworks2014},
needs to closely interact with two other subsystems. One is responsible for sensory representation and the other reflects the internal states of the organism,
such as arousal or motivation \cite{logothetisStudiesLargeScaleNetworks2014}.
Moreover, PFC is also a central sub-network [in a graph-theoretic sense]  \cite{modhaNetworkArchitectureLongdistance2010} that plays a crucial role in various cognitive functions \cite{millerIntegrativeTheoryPrefrontal2001}.
Therefore, it is expected to behave differently compared to sensory-related networks in various tasks (\eg binocular rivalry).

In recent years, novel paradigms have been used to dissociate the activity related to
conscious perception from the activity reflecting its prerequisites and consequences \cite{aruDistillingNeuralCorrelates2012,degraafCorrelatesNeuralCorrelates2012,tsuchiyaNoReportParadigmsExtracting2015}.
In particular, one of these studies focused on resolving the role of frontal lobe in conscious perception \cite{frassleBinocularRivalryFrontal2014}.
In this study, \citet{frassleBinocularRivalryFrontal2014} through a novel experimental design,
concluded that
``frontal areas are associated with active
report and introspection rather than with
rivalry per se.''
Therefore, activity in prefrontal regions could be considered as a consequence rather than a neural correlate of conscious perception.

However, based on both fMRI and electrophysiological studies we suspect that PFC is indeed involved in conscious visual perception.
Regarding the fMRI studies, \citet{zaretskayaIntrospectionAttentionAwareness2014}, in response to \citet{frassleBinocularRivalryFrontal2014},
reviewed the experimental evidence based on fMRI BOLD activity in frontal lobe which suggests even with contrastive analysis (similar to \citet{frassleBinocularRivalryFrontal2014}), some regions of frontal lobe are engaged and therefore play a role in conscious perception.
Electrophysiological studies  also provided evidence on involvement of some regions of frontal lobe in the absence of behavioral reports (\ie using no-report paradigms),
namely lateral PFC, in visual awareness \cite{panagiotaropoulosNeuronalDischargesGamma2012,kapoorDecodingInternallyGenerated2022,dwarakanathBistabilityPrefrontalStates2023}.
In particular, two recent studies  \cite{kapoorDecodingInternallyGenerated2022,dwarakanathBistabilityPrefrontalStates2023},
(which were carried out as a part of this thesis, see \autoref{cha:appr-thro-behav})
used a similar paradigm to the one used in \citet{frassleBinocularRivalryFrontal2014}.
Moreover, a recent study by \citet{kapoorParallelFunctionallySegregated2018}
based on analysis of a wider range of single units in vlPFC (not just feature selective neurons)
suggests that, both task-related and perception-related neurons co-exist in the same region of PFC.

Last but not least, the last decade witnessed a similar disagreement but on the role of primary visual cortex instead of frontal lobe
\cite{leopoldActivityChangesEarly1996,maierDivergenceFMRINeural2008a,kelirisRolePrimaryVisual2010,leopoldPrimaryVisualCortex2012}.
Ultimately, measuring both electrophysiological activity and the BOLD signal in the
same macaques engaged in an identical
task of perceptual suppression settled the debate \cite{maierDivergenceFMRINeural2008a,leopoldPrimaryVisualCortex2012}.
Therefore, to address such discrepancies we can benefit from multiple measurement techniques simultaneously or in the same animal along with a careful experimental design.

In this opinion paper, we advocate that 
formulating our conclusions related to prerequisites, consequences and true correlates of conscious experiences,
we need to have an \emph{integrative} view on the in hand collection of new evidence.
Our investigations and conclusions about the neural correlates of
consciousness must not only entail better designed experiments
but also diverse experimental techniques (e.g., BOLD fMRI, electrophysiology)
that could measure brain activity at different spatial
and temporal scales.
Moreover, different measurement techniques can reflect complementary information on the brain activity.
Therefor, such a multi-modal approach holds great promise in refining our current
understanding of conscious processing (and understating the brain in a broader sense).

%%% Local Variables:
%%% mode: latex
%%% TeX-master: "../phdThesis_csb"
%%% End:

      % 5
\chapter{Paper \rom{6}} \label{cha:paper-safavi2018}

\section*{Paper information} % 

\begin{description}
\item[Title:]
  Nonmonotonic spatial structure of interneuronal correlations in prefrontal microcircuits
\item[Authors:]
  Shervin Safavi$^*$,
  Abhilash Dwarakanath$^*$,
  Vishal Kapoor,
  Werner Joachim,
  Nicholas Hatsopoulos,
  Nikos K. Logothetis,
  Theofanis I. Panagiotaropoulos 
  ($^*$ indicate equal contributions) 
\item[Status:]
  Published in PNAS, see \citet{safaviNonmonotonicSpatialStructure2018}
\item[Presentation at scientific meetings:]
  NeNa 2015 \cite{dwarakanathTemporalRegimesStateDependent2015},
  AREADNE 2016 \cite{safaviNonMonotonicCorrelationStructure2016}
\item[Author contributions:]
  Conceptualization, T.I.P.; 
  Methodology, S.S., A.D., V.K. and T.I.P.;
  Software, S.S., A.D., T.I.P. and J.W.;
  Formal Analysis, S.S., A.D. and T.I.P.;
  Investigation, V.K., A.D., T.I.P., S.S. and N.G.H.;
  Resources, N.K.L.;
  Data Curation, A.D., T.I.P., V.K., and S.S.;
  Writing -- Original Draft, T.I.P., S.S., and A.D.;
  Writing -- Review \& Editing: V.K., A.D., T.I.P., N.G.H., and N.K.L.;
  Visualization, S.S., A.D, V.K. and T.I.P.;
  Supervision and Project administration, T.I.P.;
  Funding acquisition, N.K.L.
\end{description}

%%% Local Variables:
%%% mode: latex
%%% TeX-master: "../phdThesis_csb"
%%% End:

\section*{Summary} %
\subsection*{Motivation}

It has been suggested that mammalian's neocortex follow certain canonical features
\cite{douglasCanonicalMicrocircuitNeocortex1989,douglasNeuronalCircuitsNeocortex2004,douglasMappingMatrixWays2007,harrisCorticalConnectivitySensory2013}.
One of the features is in the spatial pattern of connectivity.
Indeed, there is a large body of evidence suggesting that functional connectivity, inferred based on spike count correlations \cite{cohenMeasuringInterpretingNeuronal2011},
rapidly decay as a function of lateral distance in most of the sensory areas of the brain
\cite{constantinidisCorrelatedDischargesPutative2002,rothschildFunctionalOrganizationPopulation2010,cohenMeasuringInterpretingNeuronal2011,smithSpatialTemporalScales2008a,smithSpatialTemporalScales2013a,denmanStructurePairwiseCorrelation2014}.
Nevertheless, there are functional and anatomical evidence,
that hint at deviations from these canonical features in PFC.
PFC is a central sub-network [in a graph-theoretic sense]  \cite{modhaNetworkArchitectureLongdistance2010} that play a crucial role in cognitive computations \citep{millerIntegrativeTheoryPrefrontal2001},
especially due to an increase in the integrative aspect of information processing in higher-order cortical areas.
Moreover, anatomical studies have shown that in contrast to early visual cortical areas
where we have a limited spread of lateral connections, in later stages of cortical hierarchy like PFC
\cite{amirCorticalHierarchyReflected1993,kritzerIntrinsicCircuitOrganization1995,angelucciCircuitsLocalGlobal2002,tanigawaOrganizationHorizontalAxons2005,vogesModelerViewSpatial2010}
lateral connections are considerably expanded 
\cite{levittTopographyPyramidalNeuron1993,amirCorticalHierarchyReflected1993,lundComparisonIntrinsicConnectivity1993,kritzerIntrinsicCircuitOrganization1995,fujitaIntrinsicConnectionsMacaque1996,tanigawaOrganizationHorizontalAxons2005}.
In this study, we investigate the functional connectivity ventro-lateral PFC (vlPFC) as a function of lateral distance.

\subsection*{Material and Methods}\label{sec:papers-pnas2018-material-methods}
In this study, we investigate the correlated fluctuations of single-neuron discharges in a mesoscopic scale.
Electrophysiology data was recorded from 4 macaque monkeys,
two in anesthetized state, and two in awake state.
Spiking activity was recorded from a Utah array chronically implanted in vlPFC.
For the awake experiments, monkeys were trained to fixate for 1000 ms on
moving grating in 8 different directions distributed randomly across multiple trials.
Tasks were started with the appearance of a red dot as a fixation point (with the size of $0.2^\circ$) on the screen for $\sim$300 ms (followed by a moving grating in one of the 8 directions).
The moving grating was only presented if the monkey maintains the fixation for the $\sim$300 ms period.
Moving grating had the size of $8^\circ$, speed of 12-13 degrees per second, and spatial frequency of 0.5 cycles per degree.

In anesthetized experiments, monkeys were exposed with 10 s of stimulation with natural movies.
Both awake and anesthetized experiments also included,
spontaneous sessions where neural activities recorded in the absence of any behavioral task.

Tuning curves were computed based on conventional procedures \cite{cohenMeasuringInterpretingNeuronal2011} by averaging the firing rate across trials for each of the eight presented directions of motion.
Signal correlations were defined as the correlation coefficient between the tuning curves of a neuronal pair.

Noise correlations for anesthetized data were computed by dividing the period of visual stimulation into 10 periods, each being 1000 ms long, and considered these periods as different successive stimuli.
The same procedure was used for the intertrial periods as well.
In the awake data, visual stimulation and intertrial periods were 1000 ms long each;
therefore, no additional procedure was required.
In the spontaneous data (both anesthetized and awake),
the entire length of the recording period was divided into periods of 1000 ms bins and they were treated as a trial.

The spike count correlation coefficients were computed similarly to previous classical studies
\cite{bairCorrelatedFiringMacaque2001a}
First, for each condition (either presentation of each moving grating in awake experiment or a single bin of movie clip in the anesthetized experiment),
we normalized the spike counts across all trials by converting them into z scores.
For each pair, we computed the Pearson's correlation coefficient for normalized spike counts and averaged across conditions to obtain the correlation value.

\subsection*{Results}

We found that the spatial structure of functional connectivity
(measured based on noise correlations) in vlPFC is different from most of the sensory cortices.
In most sensory cortices, noise correlations decay monotonically as a function of distance;
nevertheless, in vlPFC we observed in both anesthetized and
awake monkeys noise correlation rises again after an initial decay.
Moreover, we showed that the characteristic non-monotonic spatial
structure in vlPFC,
is pronounced with structured visual stimulation.

\subsection*{Conclusion}

Our results suggest that spatial inhomogeneities in the functional
architecture of the PFC arise from strong local and long-range lateral
interactions between neurons.
These characteristic patterns of interactions among PFC neurons lead
to a non-monotonic spatial structure of correlations in vlPFC.
Moreover, the mentioned spatial inhomogeneities are pronounced during structured
visual stimulation in the awake state which can be instrumental for
distributed information processing in PFC.

%%% Local Variables:
%%% mode: latex
%%% TeX-master: "../phdThesis_csb"
%%% End:

       % 6
\chapter{Paper \rom{7}}\label{cha:paper-kapoor2020}
\section*{Paper information} % 

\begin{description}
\item[Title:]
  Decoding the contents of consciousness from prefrontal ensembles
\item[Authors:]
  Vishal Kapoor$^*$,
  Abhilash Dwarakanath$^*$,
  Shervin Safavi,
  Joachim Werner,
  Michel Besserve,
  Theofanis I. Panagiotaropoulos,
  Nikos K. Logothetis
  ($^*$ indicate equal contributions) 
\item[Status:]
  Accepted for publication in Nature Communication (preprint is available online, see \citet{kapoorDecodingInternallyGenerated2022})
\item[Presentation at scientific meetings:]
  FFRM 2015 \cite{antoniouPerceptualModulationPupillary2015a},
  SfN 2018 \cite{panagiotaropoulosModulationNeuralDischarges2018a},
  FENS 2018 \cite{kapoorSpikingActivityPrefrontal2018},
  ASSC 2019 \cite{kapoorNeuronalDischargesPrefrontal2019}
\item[Author contributions: ]
  V.K., A.D. and T.I.P. designed the study. V.K., A.D. and S.S. trained animals. V.K. and A.D.
  performed experiments and collected data, with occasional help from S.S. V.K. and A.D.
  analyzed the data. S.S. contributed to spike sorting and selectivity analysis of control
  experiments. M.B. contributed to the decoding analysis. V.K. prepared and arranged the figures
  in the final format. S.S. provided the MATLAB generated version of the figures displayed in
  figure 3D, S12, S13 and S14 A. T.I.P. and N.K.L. supervised the study. N.K.L. and J.W.
  contributed unpublished reagents/analytical tools. N.K.L. provided the support to the group. V.K.
  and T.I.P. wrote the original manuscript draft. All authors participated in discussion and
  interpretation of the results and editing the manuscript.
\end{description}

%%% Local Variables:
%%% mode: latex
%%% TeX-master: "../phdThesis_csb"
%%% End:

\section*{Summary} %
\subsection*{Motivation}
The role of prefrontal cortex (PFC) has been controversial in recent consciousness studies.
Different frameworks of consciousness attribute different, even contradictory roles for PFC in generation of conscious experience.
Several frameworks, namely,
frontal lobe hypothesis \cite{crickConsciousnessNeuroscience1998},
higher order theory \cite{lauEmpiricalSupportHigherorder2011} and
global neuronal workspace framework \cite{baarsGlobalWorkspaceTheory2005,dehaeneExperimentalTheoreticalApproaches2011}
consider PFC play a mechanistic role in generation of conscious experience.
On the opposite side, another important framework of studying consciousness,
integrated theory of consciousness \cite{tononiInformationIntegrationTheory2004,balduzziIntegratedInformationDiscrete2008b,balduzziQualiaGeometryIntegrated2009b,oizumiPhenomenologyMechanismsConsciousness2014}
(for a review see \citet{tononiIntegratedInformationTheory2016}),
does not consider a similar role for PFC in generation of conscious experience,
rather attribute the role of PFC to prerequisites and consequences of  consciousness
\cite{aruDistillingNeuralCorrelates2012,degraafCorrelatesNeuralCorrelates2012}.

There are various differences between the aforementioned studies that support each of the two hypothesis.
For instance, studies that support attributing the role of PFC to prerequisites and consequences of consciousness,
used fMRI as the primary measurement technique, which can potentially lead to discrepancies.
In contrast, studies that support the opposite conclusion use electrophysiology
(see \autoref{cha:paper-safavi2014} for a short discussion).
Second, a large portion of studies that support a mechanistic role for PFC in conscious perception,
use externally induced perceptual switches such as Binocular Flash Suppression (BFS) \cite{panagiotaropoulosNeuronalDischargesGamma2012}.
Third, the majority of the experiments used behavioral reports by the subject in order to know the content of conscious experience (for a review see \cite{tsuchiyaNoReportParadigmsExtracting2015,kochNeuralCorrelatesConsciousness2016}).
This study was an effort, to bring this controversy one step closer to the resolution by
recording the neural activity from monkey ventro-lateral PFC (vlPFC) during a no-report Binocular Rivalry (BR) paradigm.

Focus of investigations on phenomenon of BR, in terms of spatio-temporal scales of measurements,
was mainly micro-scale (level of individual neurons) and macro-scale (level of large-scale networks)
Almost all the previous studies either focus on the activity of feature selective neurons measured based on single unit recordings
\cite{lehkyNoBinocularRivalry1996,sheinbergRoleTemporalCortical1997,kelirisRolePrimaryVisual2010,bahmaniNeuralCorrelatesBinocular2011,panagiotaropoulosNeuronalDischargesGamma2012},
or the whole-brain dynamics measured with imaging techniques (EEG/MEG, fMRI)
\cite{wangBrainMechanismsSimple2013,lumerNeuralCorrelatesPerceptual1998,srinivasanIncreasedSynchronizationNeuromagnetic1999b,hippOscillatorySynchronizationLargescale2011a,doesburgRhythmsConsciousnessBinocular2009b,tononiNeuralCorrelatesConsciousness2008,imamogluChangesFunctionalConnectivity2012}
(for reviews see \cite{blakeVisualCompetition2002a,panagiotaropoulosSubjectiveVisualPerception2014a,kochNeuralCorrelatesConsciousness2016}).
A complex system perspective to binocular rivalry phenomenon, motivates observation of the system in a mesoscopic scale as a very first step to understand the role of neural interactions (see \autoref{sec:why-it-appealing} for further elaboration).
In this study, we address this need, by measuring spiking activity of neural populations in vlPFC with multi-electrode recording techniques.

\subsection*{Material and Methods}\label{sec:papers-biorxiv2020-material-methods}
In this study, we investigate the neural correlate of visual awareness in mesoscopic scale.
Recording procedure is similar to awake experiments of \nameref{cha:paper-safavi2018} explained earlier (see \matmet of \autoref{sec:papers-pnas2018-material-methods}).
The core behavioral paradigm used in this study was a passive ambiguous stimulation, 
and consist of two tasks, Binocular Rivalry (BR) and Physical Alternation (PA).
Both tasks consist of fixation period similar to fixation task explained earlier in \autoref{sec:papers-pnas2018-material-methods},
and followed by presentation of 1 or 2 seconds  upward or downward moving gratings
(presented only to one eye -- half of the trials for each eye).
After the phase of stimulus presentation,
in PA trials, the first stimulus was removed and a moving grating in the contralateral eye was presented in the opposite direction.
BR trials had the identical structure of the stimulus presentation,
but with the difference that, the second stimulus was presented without removing the first stimulus.
In BR trials that two opposite moving grating were presented simultaneously, 
the perception of the monkey spontaneously switches between the stimulus
(\ie upward and downward grating) across the the entire length of trial (8-10 seconds).
Whereas, in PA trials, there are no perceptual switches,  but perception of the animal changes by the alternation of the presented stimuli (upward and downward grating).
Parameters of the visual stimulus (moving gratings) are identical to the experiment explained in \autoref{sec:papers-pnas2018-material-methods}.
Furthermore, Optokinetic Nystagmus (OKN) reflexes
\footnote{
  OKN reflexes are characteristic patterns of eye movements in response to moving stimuli,
  that consist of smooth pursuit and fast saccadic eye movements.}
has been used to determine the perception of the animal.

In addition to the main experiment that consist of BR and PA tasks,
we additionally have a control experiment for controlling eye movement as a confounding factor.
Given that determining the animal perception is based on eye movements (OKN reflexes),
to rule out the eye movement as a confounding factor,
we perform a passive fixation experiment similar to the awake experiment of \nameref{cha:paper-safavi2018} explained earlier (see \matmet of \autoref{sec:papers-pnas2018-material-methods}), but without eye movement.
In this experiment, the eye movement during presentation of moving grating were suppressed by instructing the animal to maintain the fixation during the task
(by overlaying a fixation point with size of $1$-$2^\circ$ on top of the moving grating).

\subsection*{Results}
Firstly, the perpetual dominance periods detected based on OKN reflexes follow a gamma distribution which is compatible with previous studies \cite{leveltNoteDistributionDominance1967}.
This indicates that using no-report paradigms of BR lead to compatible results with human studies.
Given the availability of neurons [recorded by Utah array] that respond to direction of motion of moving grating stimuli in PFC (see \nameref{cha:paper-safavi2018}),
we can quantify the proportion of perceptual modulation of neurons in our experiment that use upward and downward moving gratings as rivaling patterns. 
Interestingly, compatible with previous studies that used different tasks and visual stimuli \cite{panagiotaropoulosNeuronalDischargesGamma2012},
majority of sensory modulated units were also perceptually modulated.
Moreover, in the population level, the content of conscious perception of the animals was decodable from spiking activity of neural populations in vlPFC.
Lastly, the decoding algorithm that we used for decoding the content of the perception \cite{meyersNeuralDecodingToolbox2013},
could also reliably decode the content of the presented visual stimulus
(in the passive fixation experiment) both in presence and absence of eye movement
\ie training the decoder with responses in presence of eye movement,
and test when the eye movement are suppressed (fixation-on task) and vice versa.
Therefore, our control analysis suggest that
eye movements are not a confounding factor for our perceptual modulation.

\subsection*{Conclusion}
In this study, we showed that activity of the majority of sensory modulated neurons of vlPFC is correlated with conscious perception in a no-report binocular rivalry task,
and the content of conscious experience is decodable from mesoscopic dynamics of PFC.
Moreover, this study has an important implication for the neural correlate of visual awareness.
This study adds another piece of evidence for the involvement of PFC in conscious perception
which has been an important debate in the field of consciousness research in the last few years (also see \nameref{cha:paper-safavi2014}).

%%% Local Variables:
%%% mode: latex
%%% TeX-master: "../phdThesis_csb"
%%% End:

    % 7
\chapter{Paper \rom{8}}\label{cha:paper-dwarakanath2020}
\section*{Paper information} % 

\begin{description}
\item[Title:]
  Prefrontal state fluctuations gate access to consciousness
\item[Authors:]
  Abhilash Dwarakanath$^*$,
  Vishal Kapoor$^*$,
  Joachim Werner,
  Shervin Safavi
  Leonid A. Fedorov,
  Nikos K. Logothetis,
  Theofanis I. Panagiotaropoulos
  ($^*$ indicate equal contributions) 
\item[Status:]
  Preprint is available online, see \citet{dwarakanathBistabilityPrefrontalStates2023}
\item[Presentation at scientific meetings:]
  FFRM 2015 \cite{antoniouPerceptualModulationPupillary2015a},
  SfN 2018 \cite{panagiotaropoulosModulationNeuralDischarges2018a},
  AREADNE 2018 \cite{dwarakanathPerisynapticActivityPrefrontal2018}
\item[Author contributions: ]
  
  Conceptualisation: A.D., V.K., T.I.P. (lead), N.K.L.;
  Data curation: A.D. (lead), V.K. and J.W.;
  Formal analysis: A.D. (lead), V.K., J.W., L.A.F.;
  Funding acquisition: N.K.L.;
  Investigation: A.D. (equal), V.K. (equal), T.I.P. (supporting);
  Methodology: A.D. (equal), V.K. (equal), J.W. \& S.S. (supporting), T.I.P. (equal);
  Project administration: T.I.P.;
  Resources: J.W., N.K.L. (lead);
  Software: A.D. (lead), V.K., J.W., L.A.F. \& S.S. (supporting);
  Supervision: T.I.P.;
  Visualisation: A.D. (lead), T.I.P. (supporting);
  Writing -- original draft: A.D., T.I.P. (lead);
  Writing -- review \& editing: A.D., V.K., L.A.F., S.S., T.I.P. (lead), N.K.L.

\end{description}

%%% Local Variables:
%%% mode: latex
%%% TeX-master: "../phdThesis_csb"
%%% End:

\section*{Summary} %
\subsection*{Motivation}
In \autoref{sec:why-it-appealing} we elaborated on the motivations for studying the phenomenon of binocular rivalry (BR) in a mesoscopic scale
and in \nameref{cha:paper-kapoor2020} we showed that content of conscious experience is decodable from mesoscopic dynamics of PFC. %which is the
This was the first confirmation on the usefulness of the meso-scale observation.
This allows us to go one step further in studying the mesoscopic dynamics of PFC.
One of the most important markers of coordination in mesoscopic dynamics of the brain,
is neural oscillations \cite{buzsakiRhythmsBrain2011,buzsakiScalingBrainSize2013a}.
In this study we investigate oscillatory dynamics in ventro-lateral PFC (vlPFC) and its connection to conscious visual perception.

\subsection*{Material and Methods}
Most of the experimental details for this study was explained in summaries of the other papers
(\nameref{cha:paper-safavi2018}, and \nameref{cha:paper-kapoor2020}).
Recording procedure is similar to awake experiment of
\nameref{cha:paper-safavi2018} explained earlier (see \matmet of \autoref{sec:papers-pnas2018-material-methods}).
The behavioral paradigm used in this study is also explained earlier
(see \matmet of \autoref{sec:papers-biorxiv2020-material-methods}).
In this study, Continuous Wavelet Transform (CWT) \cite{mallatWaveletTourSignal1999} has been used to extract spectral content of LFPs
and Chronux toolbox \cite{bokilChronuxPlatformAnalyzing2010} for quantifying spike-LFP coupling by computing Spike-Field-Coherence (SFC).

\subsection*{Results}
This study reveals various characteristic oscillatory activities which are happening in the vicinity of the perceptual switches detected based on Optokinetic Nystagmus (OKN) reflexes.
The frequency of these transient oscillatory activities are covering low and intermediate ranges (namely 1-9 Hz and 20-40 Hz).
In addition to presence of these coordinated dynamics in the mesoscopic activity of PFC neural populations and their relationship to perceptual events,
the statistics and spatio-temporal patterns of some of these transitory events lend support to important frameworks of studying the consciousness.

\subsection*{Conclusion}
This study adds another piece of evidence for the involvement of PFC in conscious perception, in addition to the one discussed earlier in \nameref{cha:paper-kapoor2020}.
In particular, it reveals signatures of neural coordination reflected in the oscillatory dynamics (see \autoref{sec:relat-betw-meso}) of neural populations involved in conscious visual perception.
Revealing these signatures could not be possible without investigating the system in meso-scale 
(see more elaborating in \autoref{sec:why-it-appealing}).
Lastly similar to \nameref{cha:paper-kapoor2020}, this study has an important implication for the neural correlate of visual awareness.
This study highlights the involvement of PFC in conscious perception
which has been an important debate in the field of consciousness research in the last few years (also see \nameref{cha:paper-safavi2014}).

%%% Local Variables:
%%% mode: latex
%%% TeX-master: "../phdThesis_csb"
%%% End:

 % 8

\cleardoublepage % Empty page before the start of the next part

% ------------------------------------------------
% Outlook

\ctparttext{
  This part is dedicated to a subjective perspective on how the research line of this thesis can or should be extended.
  In this thesis, we sought for \emph{principled} ways of approaching the brain.
  Although this thesis touched on various such aspects,
  but I believe it misses an important aspect of the brain which is its \emph{adaptivity}.
  In the end, brain, presumably the most ``complex system'', needs to survive in the environment.
  Indeed, in the field of \emph{complex adaptive systems}, the endeavor is understanding very similar 
  questions in the nature.
  Inspired by some ideas discussed in the field of complex adaptive systems,
  we suggest a set of new research directions that intend to incorporate the adaptivity aspect of the brain as one of the principles. 
  Of course, these research directions, remain close to the neuroscience side, similar to the intention of the research presented in previous parts.

} % Text on the Part 2 page describing the content in Part 2

%%% Local Variables:
%%% mode: latex
%%% TeX-master: "../phdThesis_csb"
%%% End:

\part{Outlook}\label{part:outlook}
\chapter{Brain as a complex \emph{\&} adaptive system}\label{cha:brain-as-complex-adaptive}
In \autoref{cha:brain-as-complex}, we argue that brain can be approached as a complex system.
Certainly, this is a valuable perspective toward the brain and was the pivotal idea of this thesis.
Nevertheless, an important aspect of the brain, as a biological information processing system,
is not taken into account in the approach we followed and discussed in this thesis.
This important aspect is \emph{adaptivity} of humans/animals.
They need to be \emph{adaptive} in order to survive.
That being said, perhaps we should consider humans/animals as \emph{adaptive agents} and the brains as a complex \emph{and} adaptive system.
Indeed, Complex Adaptive Systems (CAS) have been an independent field of research 
(see \citeAYt{hollandStudyingComplexAdaptive2006} for a brief review).

\marginpar{Approaching the brain as a complex \textbf{and} adaptive system}

Inspired by general properties and mechanisms introduced for CAS
(that are briefly discussed in \autoref{sec:compl-adapt-syst}), 
again, new questions can be asked in various domains of neuroscience, 
and moreover, even old questions can be revisited based on this perspective.
In this chapter, we introduce a set of new research directions that we believe are complementary to the ideas that motivated and shaped this thesis.

Conceiving the brain as a CAS implies that certain computations are needed to satisfy the adaptivity of the agent (see \autoref{sec:comp-object} for further elaboration).
Moreover, as we discussed earlier (see \autoref{cha:brain-as-complex}),
conceiving the brain as a complex system has implications on the dynamics of the brain.
More generally, on one hand, behavior is a rich source for seeking and understanding the computational objectives
(pertaining to adaptivity of humans and animals)
\marginpar{Through behavior we can understand computation needed to be adaptive and through multi-scale dynamics of the brain we can understand the brain's biophysical machinery}
On the other hand, multi-scale dynamics of the brain, as briefly discussed in \autoref{cha:appr-thro-nda},
is a rich source for understanding the biophysical machinery of this adaptive agent implementing the computation.
For instance, concerning the adaptivity of the humans and animals, focusing on behavior have led us to various developments in
ecological psychology \cite{reedEncounteringWorldEcological1996},
reinforcement learning \cite{nivReinforcementLearningBrain2009},
and even understanding the emotion \cite{bachAlgorithmsSurvivalComparative2017}
that all inform us about the brain computations \cite{nivPrimacyBehavioralResearch2020}. 
Concerning the multi-scale dynamics,
studying the brain across scales,
has helped us to understand the emergent properties of this biophysical machinery
(for further elaboration, see \citet[Chapter 1]{pesensonMultiscaleAnalysisNonlinear2013} and \citet{siettosMultiscaleModelingBrain2016}).

From a broader perspective, particularly in terms of Marr's levels of understating \cite{marrUnderstandingComputationUnderstanding1979},
it can  be argued that, understanding the brain dynamics,
brings us closer to  the implementation level and perhaps to some degree to the algorithmic level;
and understating the behavior brings us closer to understanding the computation and more explicitly the algorithm.
With no doubt, both of these aspects are utterly important for understating the brain.
\marginpar{An \textbf{integrative} understating of the brain need a bridge}
Therefore, it is import to establish a connection between these two, in order gain an \emph{integrative} understating of the brain
(see \citet[Chapter 2, Section 2]{churchlandComputationalBrain1992} for a broad perspective on the importance of this bridge and \citet{stephanTranslationalPerspectivesComputational2015} and \citet[Chapter 8]{forstmannIntroductionModelbasedCognitive2015} for showcases of their importance in  translational neuroscience).
Motivated by the importance of establishing this bridge,
in \autoref{sec:relat-behav-multi} we outline various approaches we can take for relating behavior to multi-scales brain dynamics.

\section{Complex adaptive systems}\label{sec:compl-adapt-syst}

Complex adaptive systems (CAS) can be broadly defined as a system composed of multiple elements, called agents,
\emph{"that learn or adapt in response to other agents"} \cite[Chapter 3]{hollandComplexityVeryShort2014}.
CAS have been studied for decades (see \citet{morowitzMindBrainComplex1995} for historical note),
and there have been efforts to explain the behavior of various natural and artificial systems based on the CAS formalism;
They include adaptive behavior of the immune system \cite{chowdhuryImmuneNetworkExample1999},
finical market \cite{hollandComplexityVeryShort2014}
and even language \cite{ellisLanguageComplexAdaptive2009}.

Different sets of properties and mechanisms which are considered to be common between different CAS have been suggested
\cite{brownleeComplexAdaptiveSystems2007}.
We outline the 4 features proposed by \citet{hollandStudyingComplexAdaptive2006}.
Although, some of the core ideas are common among most of the other proposals and indeed those commonalities are the foundations for ideas presented in the following,
but readers are also encouraged to refer to properties and mechanism proposed by others as well (for example see \citet{gell-mannComplexAdaptiveSystems1994} and \citet[Chapter 1]{arthurEconomyEvolvingComplex1997}).

\citet{hollandStudyingComplexAdaptive2006} introduces 4 major features or characteristics that CAS have in common in spite of their substantial differences:
\begin{enumerate}
\item Parallelism:
  Complex systems (also briefly discussed in \autoref{cha:brain-as-complex}) are constructed with many \emph{intently interacting} components.
  Due to the need for tight coordination, simultaneous communications between components of the system are inevitable.
\item Conditional actions:
  In CAS, agents need to act conditionally as the required action is defined by the agent's internal state (condition) and actions of external agents.
\item Modules and hierarchies:
  CAS are often organized in a modular and hierarchical fashion (for the latter see \cite[Chapter 7]{hollandComplexityVeryShort2014} and \cite[Chapter 8]{hollandSignalsBoundariesBuilding2012a}).
\item Adaptation and evolution:
  Agents in CAS need to change over time in order to gain a better performance.
  Adaptation requires solutions to two important problems, namely \emph{credit assignment} and \emph{rule discovery}.
\end{enumerate}
Features or characteristics mentioned in the number two and four of Holland's idea are particularly pertaining to \emph{computations} that CAS need to perform.
Interestingly, some of these computations are already a focus of research in the field of neuroscience as well
(but not necessarily based on a similar foundation we motivate by CAS ideas).
In section \autoref{sec:comp-object} we briefly discuss some of these computational objectives that can be closely connected to the brain.

\section{Brain computational objectives}\label{sec:comp-object}
As briefly  discussed earlier, humans/animals as information processing systems,
are adaptive agents, and need to interact with a complex environment.
We can conceive the brain as a CAS, and based on CAS notions introduced earlier,
we can argue that due to their adaptivity they need to perform certain computations.
Indeed, \citet[Chapter 12]{mitchellComplexityGuidedTour2011} argue that,
\begin{displayquote}\textsl{
    "At a very general level, one might say that computation is what
    a complex system does with information in order to succeed or adapt in its
    environment."
  }
\end{displayquote}

To emphasize conceiving the brain as a CAS and the computations it implies, 
we highlight some of the computational objectives of the brain that are under active investigation \emph{and}
and are closely related to general properties of CAS discussed in \autoref{sec:compl-adapt-syst}.
The need for \emph{conditional actions}, solving the \emph{credit assignment} problem and \emph{discovering rules} in the environment
that were mentioned in \autoref{sec:compl-adapt-syst} as general properties of CAS,
are closely related to \emph{representation}, \emph{decision making} and \emph{reinforcement learning}
which are actively investigated in neuroscience.

One of these computational objectives is efficient representations.
The ability of an agent to act upon actions and states of external agents relies on \emph{efficient representation} of information pertaining to external agents.
The other computational objective is credit assignment and rule discovery that are both premises of reinforcement learning \cite{woergoetterReinforcementLearning2008}.

Certainly, this section, by no means, provides a comprehensive list of computational objectives of the brain that have been already studied in neuroscience.
Rather, it highlights examples that are closely related to the ones CAS should have in a general sense.
In the next step, we need to find the connections between these computational objectives and their biophysical machinery by investigating the relationship between behavior and multi-scale dynamics of the brain. 

\section{Relating behavior to multi-scale brain dynamics}\label{sec:relat-behav-multi}
As argued earlier, behavior is a rich source for understating such computational objectives in human/animals and 
multi-scale dynamics is a rich source for understating the biophysical machinery behind it.
This is the motivation for relating the behavior to multi-scale brain dynamics.
In this section, we introduce potential approaches that we think can relate these two facets of the brain.

Certainly, establishing this connection is challenging.
Therefore, we need to decompose it into smaller but complementary steps that can be supported by the existing models and/or empirical evidence.
In the next sections 
(\autoref{sec:relat-neur-dynam}, \autoref{sec:expl-models-pivot}, and \autoref{sec:princ-framw-data}),
we propose various approaches that are more or less accessible and can potentially bring us a few steps closer to establishing a bridge between behavior and multi-scale brain dynamics.

\subsection{Relating neural dynamics and neural computation}\label{sec:relat-neur-dynam}

As discussed earlier, neural computation and dynamics are both important aspects of the brain.
There are various frameworks and models in neuroscience which are either centered around
neural computation
\cite{lengyelMatchingStorageRecall2005,deneveBayesianSpikingNeurons2008a,deneveBayesianSpikingNeurons2008,tanakaRecurrentInfomaxGenerates2008,buesingNeuralDynamicsSampling2011,boerlinPredictiveCodingDynamical2013,billDistributedBayesianComputation2015,chalkNeuralOscillationsSignature2016,zeldenrustEfficientRobustCoding2019,echevesteCorticallikeDynamicsRecurrent2020}
or neural dynamics
\cite{bertschingerRealtimeComputationEdge2004,eliasmithUnifiedApproachBuilding2005,sussilloNeuralCircuitsComputational2014,hidalgoInformationbasedFitnessEmergence2014a,shrikiOptimalInformationRepresentation2016,maassSearchingPrinciplesBrain2016,kimLearningRecurrentDynamics2018,chenComputingModulatingSpontaneous2019,michielsvankessenichPatternRecognitionNeuronal2019,finlinsonOptimalControlExcitable2020}
but also have some implications for the other one
(also see \citet{maassSearchingPrinciplesBrain2016} for a brief review).
These models are not necessarily well connected to \emph{behavior} and \emph{multi-scale} dynamics of the brain,
but still can fill some space in this large gap between behavior and multi-scale.
Further investigation in such frameworks and models, that are outlined in the next sections,
can potentially help us to accomplish the mentioned goal,
which is relating behavior to multi-scale brain dynamics.

\subsubsection{Normative models with implications for neural dynamics}\label{sec:norm-models-with}
There have been various efforts to relate neural computation to neural dynamics by introducing normative models of neural computation (\eg based on sampling theories, Bayesian inference algorithms) which can explain some aspects of observed dynamics of the brain such as irregular spiking and neural oscillations
\cite{ermentroutRelatingNeuralDynamics2007c,buesingNeuralDynamicsSampling2011,boerlinPredictiveCodingDynamical2013,billDistributedBayesianComputation2015,chalkNeuralOscillationsSignature2016,mastrogiuseppeLinkingConnectivityDynamics2018,echevesteCorticallikeDynamicsRecurrent2020,dubreuilComplementaryRolesDimensionality2020}.
More generally there have been efforts to relate the state of the machinery implementing a given neural computation to a putative dynamical regime of the neural circuits.
For instance, \citet{echevesteCorticallikeDynamicsRecurrent2020} and \citet{lengyelMatchingStorageRecall2005} have developed neuronal networks which implement Bayesian inference
that are attractor networks as well.
Neural coding, in particular, is one of the well established computations that brain needs to accomplish \cite{quianquirogaPrinciplesNeuralCoding2013}
and there have been various efforts to connect neural coding and neural dynamics \cite{ermentroutRelatingNeuralDynamics2007c,boerlinPredictiveCodingDynamical2013,chalkNeuralOscillationsSignature2016,echevesteCorticallikeDynamicsRecurrent2020}.
In most of such normative models, we optimize or train a network of neurons based on a specific computational objective (such as reconstruction error),
and the features of the neural dynamics appear in the resulting network activity automatically.

All the features of neural dynamics that have been explained by the previous normative models are among the important ones and some  of them are even considered computationally relevant
(like oscillations \cite{chalkNeuralOscillationsSignature2016,petersonHealthyOscillatoryCoordination2018}).
Nevertheless, the brain dynamics has been shown to be more complex than the reach of normative models so far \cite{decoDynamicBrainSpiking2008,breakspearDynamicModelsLargescale2017}.
Not only in terms of complexity of the observed dynamics,
but also in terms of scale, particularly large scale dynamics
and multi-scale dynamics \cite{freemanScalefreeNeocorticalDynamics2007a,agrawalScaleChangeSymmetryRules2019}.
Next steps should include developing normative models with richer neural dynamics, in particular, the large scale and multi-scale dynamics.

\subsubsection{Models of neural dynamics with implications  for neural computation}

One of the frameworks for explaining the neural dynamics with connection to neural computation is the ``criticality hypothesis of the brain'' 
(for a review see \cite{munozColloquiumCriticalityDynamical2018} -- also briefly discussed in \autoref{sec:crit-hypoth-brain}).
Certainly, frameworks like criticality are insightful for brain dynamics
\cite{munozColloquiumCriticalityDynamical2018}
in particular because they provide explanations for observed multi-scale dynamics of the brain \cite{agrawalScaleChangeSymmetryRules2019}.

One approach to better connect the criticality hypothesis of the brain to neural computation could be the one we used in \autoref{cha:appr-thro-theo},
which is searching for signatures of criticality in neuronal networks that can be optimized based functionally relevant computational objectives
(in \autoref{cha:appr-thro-theo}, we used efficient coding objectives).
Of course, this is not necessarily informative on a mechanistic level,
rather is an indication of \emph{potential} connections.
Presence of signatures of criticality may or may not hint for more mechanistic approaches.
Nevertheless, some clues can guide us toward more formal investigations.
For instance, for the particular case discussed in \autoref{cha:appr-thro-theo},
Fisher information can be a candidate quantity that both frameworks -- efficient coding \cite{weiMutualInformationFisher2015} and criticality \cite{prokopenkoRelatingFisherInformation2011,danielsQuantifyingCollectivity2016,kalloniatisFisherInformationCriticality2018,kueblerOptimalFisherDecoding2019} -- use to assess the closeness to their optimal point.

Another potential approach is seeking 
for other kinds of functionally relevant attributes for notions established in criticality hypothesis of the brain.
For instance, it has been suggested that neural avalanches are related to cell assemblies  \cite{plenzOrganizingPrinciplesNeuronal2007} and
indeed the notion of cell assemblies are closely connected to computations implemented in the brain 
\cite{singerFormationCorticalCell1990a,harrisOrganizationCellAssemblies2003,harrisNeuralSignaturesCell2005a,buzsakiNeuralSyntaxCell2010b,tetzlaffUseHebbianCell2015}.

\subsection{Exploiting models of pivotal tasks}\label{sec:expl-models-pivot}
For the purpose expressed in \autoref{sec:relat-behav-multi}, we can also exploit behavioral tasks which have been comprehended from a wide range of perspectives.
To the best of my knowledge, not so many such tasks are identified and exhaustively explored.
Nevertheless, we believe this small number is sufficient to  make further exploration in this direction justified, given the potential insight that we can get from them.
For instance, \citet{cavanaghCircuitMechanismIrrationalities2019} studied perceptual decision-making through interventional experimentation, and multi-scale computational modeling.
Indeed, such theory-experiment hybrid approaches can be insightful,
both for understanding the multi-scale dynamics of the phenomenon (in this case from synapse to behavior) and also the computations involved in the task (in this case evidence accumulation process).
\citet{frankLinkingLevelsComputation2015} and colleagues also studied the decision making and cognitive control through reinforcement learning models and biophysical modeling of a single cortico-basal ganglia circuit and
similarly, they could gain an integrative understating of the involved computation and also biophysical and dynamical characteristics that have been observed during such tasks.
A key in both examples was exploiting the tasks that have been comprehended from a wide range of perspectives
(normative modeling, biophysical modeling, measuring electrophysiological activity of involved circuits).

One example of such tasks that has been studied from a wide range of perspectives and wide range of tools is the \emph{bistable perception}.
On one hand, a large body of computational studies focus on explaining the dynamics of bistable perception \cite{moreno-boteNoiseinducedAlternationsAttractor2007a,shpiroBalanceNoiseAdaptation2009a,pastukhovMultistablePerceptionBalances2013a,vattikutiCanonicalCorticalCircuit2016,cohenDynamicalModelingMultiscale2019}; 
On the other hand, another class of computational models which
tried to explain the phenomenon with normative approaches centered around the computation that the brain might need to perform pertaining to perception \cite{bialekRandomSwitchingOptimal1995,dayanHierarchicalModelBinocular1998,hohwyPredictiveCodingExplains2008a,atwalStatisticalMechanicsMultistable2014a,samuelgershmanPerceptualMultistabilityMarkov2014a}.
Notably, most of these studies are centered around Bayesian model of the brain \cite{knillBayesianBrainRole2004,doyaBayesianBrainProbabilistic2007}.

Next to this extensive computational models
(which include both normative and biophysical models)
there is a large body of psychophysical (for review see \cite{klinkUnitedWeSense2012b}), electrophysiological and imaging
(for review see \cite{blakeVisualCompetition2002a,panagiotaropoulosSubjectiveVisualPerception2014a}), pharmacological \cite{carterPsilocybinSlowsBinocular,mentchGABAergicInhibitionGates2019}, and genetic studies \cite{millerGeneticContributionIndividual2010,ngoPsychiatricGeneticStudies2011a,lawEffectStimulusStrength2017,chenGenomicAnalysesVisual2018}.
Particularly, as briefly discussed in \autoref{cha:appr-thro-behav}, from electrophysiological and imaging we learn that a distributed network of neurons is involved in the phenomenon and therefore this is inherently a multi-scale problem.

We believe a wide range of perspectives toward the phenomenon of bistable perception,
that led to this immense range of studies and their resulting insight,  
justify bistable perception as one of the ideal tasks to be studied with the purpose of relating behavior (and their accompanied computation) to multi-scales brain dynamics \cite{safaviMultistabilityPerceptualValue2022}.
In this thesis, we approach the phenomenon of binocular rivalry differently from the conventional approaches (see \autoref{cha:appr-thro-behav}),
and our initial results (see \nameref{cha:paper-kapoor2020} and \nameref{cha:paper-dwarakanath2020}) justified the usefulness of our proposed mesoscopic scale observation of the brain during a binocular rivalry task.
Indeed, a meso-scale observation can also be the first step for understanding the multi-scale dynamics of binocular rivalry.
In \autoref{cha:appr-thro-nda} we introduced a set of novel methodologies for cross-scale and multi-scale analysis of neural data, in particular mesoscopic signals like LFPs.
Transient and cooperative neural activities in hippocampus (such as sharp wave-ripples) have been studied extensively.
As exemplified in \autoref{sec:need-new-tools}, such characteristic events
can co-occur with well-coordinated activity in smaller scales (scale of neurons and population of neurons),
and a larger scale (whole brain) as well.
Therefore, investigating the presence of such events in the mesoscopic activity of neurons during binocular rivalry [assuming their existence] and
the relationship between these neural events and behavior can potentially bridge the multi-scale dynamics of the brain and behavior (which is binocular rivalry in this case). 

Indeed, recent electrophysiological studies in the cortex also revealed neural activities with cooperative and transient nature that are involved in cognitive functions other than memory consolidation. 
For instance, \citet{womelsdorfBurstFiringSynchronizes2014} reported burst firing events in Prefrontal Cortex accompanied with particular large-scale synchronization patterns and attention switches.

What has been discussed can be a potential road map to bridge the multi-scale dynamics of the brain and behavior in binocular rivalry,
but still the connection to computation remains elusive.
Regarding the computations that brain presumably needs to perform, as mentioned earlier,
there are already computational models
\cite{bialekRandomSwitchingOptimal1995,dayanHierarchicalModelBinocular1998,hohwyPredictiveCodingExplains2008a,atwalStatisticalMechanicsMultistable2014a,samuelgershmanPerceptualMultistabilityMarkov2014a}.
Some of these models can even explain many aspects of binocular rivalry psychophysics and some aspects of neural dynamics \cite[Chapter 3]{leptourgosDynamicalCircularInference2018}.
Certainly, bridging the multi-scale dynamics and computations explicitly, should be investigated in the next steps. 

\subsection{A principled framework for data fusion}\label{sec:princ-framw-data}
One of the core components of the proposed goal, \emph{relating behavior to multi-scale brain dynamics},
is relating dynamics of the brain across scales even independent of behavior and computation.
Indeed, in \autoref{cha:appr-thro-nda}, we introduced novel methodologies for the very same purpose -- bridging the scales.
Nevertheless, most of such methodologies (including the ones introduced in this thesis) are designed for particular choices of data modalities (\eg spike-LFP coupling, LFP-BOLD relationship).
This implies, for each pair of modalities, we tend to develop a set of tools accustomed to the nature of that particular type of data (which is a reasonable choice for the first try).
Of course, such modality-specific methodologies have been insightful and certainly will be,
but having a general framework which is capable of embedding or allowing the investigation of different datasets in a common space
can potentially bring a wider range of opportunities for investigating brain dynamics across scales and ultimately relate them to the behavior and computation.

Indeed, a few frameworks exploiting kernel-based methods \cite{biessmannTemporalKernelCCA2009,murayamaRelationshipNeuralHemodynamic2010,biessmannAnalysisMultimodalNeuroimaging2011,fazliLearningMoreOne2015} and topological data analysis \cite{zhangTopologicalPortraitsMultiscale2020}  have been proposed,
that are potentially capable of fusing multi-modal data in a principled fashion.
Next steps should include broad investigation of such frameworks for various modalities including the ones accessible via invasive recording techniques such as spikes and extracellular field potentials
(as they are less explored compared to non-invasive ones).
In particular, data modalities that can be better represented by point processes (such as spike trains)
are more challenging to be fused with the other kinds of neural data which are continuous in nature 
(should be noted that there have been some efforts in this direction based on kernel-based methods \cite{shpigelmanSpikernelsPredictingArm2005,paivaReproducingKernelHilbert2009b,paivaInnerProductsRepresentation2010,liTensorproductkernelFrameworkMultiscale2014a}, and for a review see \citet{parkKernelMethodsSpike2013a}). 

\section{Understating the neuro-principles through dysfunctions}
Understanding the brain dysfunctions, in addition to its humanistic aspects and potential societal impacts can also be insightful for gaining a mechanistic understanding of the brain.
In particular, understanding cognition and behavior is one of the most important goals of the brain science,
and among brain dysfunctions, psychiatric disorders are specifically connected to the malfunctioned cognition and disorders of behavior \cite{huysAdvancesComputationalUnderstanding2020}.
A window for understanding the machinery behind cognitive capabilities and neural correlates of behavior can happen through the understanding of when and why they malfunction \ie \emph{mechanistically} to understand the syndromes we observe in psychiatric disorders.

Furthermore, Psychiatry is unique from various other perspectives.
Approaches used for understanding the psychiatric disorders are extremely diverse.
In terms of scales or levels of organization \cite[Chapter 1]{churchlandComputationalBrain1992}, 
psychiatric disorders have been studied from their genetic basis 
\cite{burmeisterPsychiatricGeneticsProgress2008,isslerDeterminingRoleMicroRNAs2015,smelandPolygenicArchitectureSchizophrenia2020}
all the way to their roots in the social interactions 
\cite{schilbachSecondpersonNeuropsychiatry2016,leongPromiseTwopersonNeuroscience2019,sevgiSocialBayesUsing2020}
In terms of [Marr] levels of understanding \cite{marrUnderstandingComputationUnderstanding1979}, psychiatric disorders have been attacked in all three levels
\cite[Chapter 5]{redishComputationalPsychiatryNew2016}\cite{huysAdvancesComputationalUnderstanding2020}.

The mentioned diversity of approaches goes beyond the conventional research in the systems neuroscience. 
As the last example, %for the diversity of the approaches, 
it is worth mentioning the research on psychiatric disorders for establishing the connection between the nervous system and the immune system. 
Recently, a peculiar connection between psychiatric disorders
(in particular depression and schizophrenia)
and dysfunctions of the immune system has been established 
\cite{khandakerInflammationImmunitySchizophrenia2015,bullmoreInflamedMindRadical2018,teixeiraImmunopsychiatryClinicianIntroduction2019,yuanInflammationrelatedBiomarkersMajor2019,mayerOptimalImmuneSystems2017,schillerNeuronalRegulationImmunity2020,haddadMaternalImmuneActivation2020,khandakerNeuroinflammationSchizophrenia2020}
and more generally the interaction between the immune system and the brain has been receiving more attention and support recently
(\cite{bullmoreInflamedMindRadical2018, deabreuPsychoneuroimmunologyImmunopsychiatryZebrafish2018,badimonNegativeFeedbackControl2020,pfeifferBrainImmuneCells2020,shieldsPsychosocialInterventionsImmune2020,mSocialIsolationAlters2020,heineTransdiagnosticHippocampalDamage2020,korenRememberingImmunityNeuronal2020,kolMemoryOrchestraRole2021}).

Despite this diversity, there are also potential connections and bridges between them.
For instance, in many brain dysfunctions we have clues about both impaired computation and  brain dynamics.
Whether there is a connection between them, it needs to be thoroughly investigated.
However, at least the current state of [Computational] Psychiatry
is not clueless about integration of neural computation and neural dynamics.
For instance, \cite{deneveCircularInferenceMistaken2016}, based on their implementation of circular inference, 
have suggested that pathological inference attributed in schizophrenia can be mapped into excitation-inhibition imbalance in the neural circuit implementing the inference.

Overall, we believe, understating the brain dysfunction is an intriguing window for gaining an integrative understating of the brain function given the richness and diversity of the empirical data in the field.

%%% Local Variables:
%%% mode: latex
%%% TeX-master: "../phdThesis_csb"
%%% End:

%----------------------------------------------------------------------------------------
%	THESIS CONTENT - APPENDICES
%----------------------------------------------------------------------------------------

\appendix

%----------------------------------------------------------------------------------------
%	POST-CONTENT THESIS PAGES
%----------------------------------------------------------------------------------------

\cleardoublepage% Table of Contents - List of Tables/Figures/Listings and Acronyms

\addtocontents{toc}{\protect\vspace{\beforebibskip}} % Place the bibliography slightly below the rest of the document content in the table of contents
\pdfbookmark[0]{\listfigurename}{lof} % Bookmark name visible in a PDF viewer

\listoffigures

\vspace*{8ex}
\newpage

% ----------------------------------------------------------------------------------------
%	Acronyms
% ----------------------------------------------------------------------------------------

\refstepcounter{dummy}

\markboth{\spacedlowsmallcaps{Acronyms}}{\spacedlowsmallcaps{Acronyms}}

\addcontentsline{toc}{chapter}{\tocEntry{Acronyms}}

\chapter*{Acronyms}

\begin{acronym}[UML]
  \acro{BOLD}{Blood-Oxygen-Level Dependent}
  \acro{BFS}{Binocular Flash Suppression }
  \acro{CAS}{Complex Adaptive System}
  \acro{CCA}{Canonical Correspondence Analysis}
  \acro{fMRI}{functional Magnetic Resonance Imaging}
  \acro{LFP}{Local Field Potential}
  \acro{LGN}{Lateral Geniculate Nucleues}
  \acro{LIF}{Leaky-Integrate and Fire}
  \acro{LPFC}{Lateral Prefrontal Cortex}
  \acro{vlPFC}{ventro lateral Prefrontal Cortex}
  \acro{PFC}{Prefrontal Cortex}
  \acro{PLV}{Phase Locking Value}
  \acro{MUA}{Multi Unit Activity}
  \acro{NET-fMRI}{Neural-Event-Triggered functional Magnetic Resonance Imaging}
  \acro{NMF}{Non-negative Matrix Factorization}
  \acro{OKN}{Optokinetic Nystagmus}
  \acro{REM}{Rapid-Eye-Movement}
  \acro{RG}{Renormalization Group}
  \acro{STFT}{Short-Term Fourier Transform }
  \acro{SUA}{Single Unit Activity}
  \acro{SFC}{Spike Field Coherence}
  \acro{SNR}{Signal to Noise Ratio}
  \acro{SVD}{Singular Value Decomposition}

\end{acronym}

%%% Local Variables:
%%% mode: latex
%%% TeX-master: "../phdThesis_csb"
%%% End:
 % List of Figures etc.
\cleardoublepage% Bibliography

\label{app:bibliography} % Reference the bibliography elsewhere with \autoref{app:bibliography}

\manualmark
\markboth{\spacedlowsmallcaps{\bibname}}{\spacedlowsmallcaps{\bibname}} % work-around to have small caps also
\refstepcounter{dummy}
\addtocontents{toc}{\protect\vspace{\beforebibskip}} % to have the bib a bit from the rest in the toc
\addcontentsline{toc}{chapter}{\tocEntry{\bibname}}

% \bibliographystyle{APA}

% \IfFileExists{/home/ssafavi/Nextcloud/libraries/zotlib.bib}{\bibliography{/home/ssafavi/Nextcloud/libraries/zotlib,locallib,tmplib}}{\bibliography{locallib,tmplib}}

% % bib source setting (this might need to go to a seperate bib file)
% \IfFileExists{/home/ssafavi/Nextcloud/libraries/zotlib.bib}{\bibliography{/home/ssafavi/Nextcloud/libraries/zotlib.bib,locallib,tmplib.bib}}{\bibliography{locallib.bib,tmplib.bib}}%,tmplib.bib}}

\printbibliography

%%% Local Variables:
%%% mode: latex
%%% TeX-master: "../phdThesis_csb"
%%% TeX-master: "../phdThesis_csb"
%%% End:
 % Bibliography

% ********************************************************************
% Backmatter
%*******************************************************
\cleardoublepage
\ctparttext{
  This appendix of the full thesis \cite{safaviBrainComplexSystem2022} includes the PDF of all the published papers, preprints and in-preparation manuscripts.
  They appear as they appeared in \autoref{part:manuscr-inform},
  with the following order:

  \begin{enumerate}
  \item \citet[Neural Computation 2021]{safaviUnivariateMultivariateCoupling2021}
  \item \citet[PLoS Computational Biology 2023]{safaviUncoveringOrganizationNeural2023} 
  \item Besserve et al.;
    preliminary  manuscript is available in the appendix,
    (\hyperref[pdf:besserve2020ned]{Paper 3})
  \item \citet[BioRxiv 2023]{safaviSignaturesCriticalityEfficient2023}
  \item \citet[Front. Psychol. 2014]{safaviFrontalLobeInvolved2014}
  \item \citet[PNAS 2018]{safaviNonmonotonicSpatialStructure2018}
  \item \citet[Nature Communcations 2022]{kapoorDecodingInternallyGenerated2022} %ref 224: 25, 39, 47).
  \item \citet[Neuron 2023]{dwarakanathBistabilityPrefrontalStates2023}
  \end{enumerate}
}

%%% Local Variables:
%%% mode: latex
%%% TeX-master: "../phdThesis_csb"
%%% TeX-master: "../phdThesis_csb"
%%% TeX-master: "../phdThesis_csb"
%%% End:

\part{Manuscripts}\label{part:manuscripts}

%----------------------------------------------------------------------------------------
%	MANUSCRIPTS
%----------------------------------------------------------------------------------------

% \input{chapters/app_manuscriptPDFs.tex}

%----------------------------------------------------------------------------------------
%	COLOPHON
%----------------------------------------------------------------------------------------

\cleardoublepage% Colophon (a brief description of publication or production notes relevant to the edition)

\pagestyle{empty}

\hfill

\vfill

\pdfbookmark[0]{Colophon}{colophon}

\section*{Colophon}

This document was typeset using the typographical look-and-feel \texttt{classicthesis} developed by Andr\'e Miede, and modified by Shervin Safavi for the purpose of this thesis. The style was inspired by Robert Bringhurst's seminal book on typography ``\emph{The Elements of Typographic Style}''. \texttt{classicthesis} is available for both \LaTeX\ and \mLyX: 

\begin{center}
\url{http://code.google.com/p/classicthesis/}
\end{center}

\bigskip

%%% Local Variables:
%%% mode: latex
%%% TeX-master: "../phdThesis_csb"
%%% End:
 % Colophon

%----------------------------------------------------------------------------------------

\end{document}